\documentclass[prb,twocolumn,showpacs,amsmath,amssymb,superscriptaddress,longbibliography]{revtex4-1}
\usepackage{graphicx}
\usepackage{breqn}
\usepackage{xcolor}
\usepackage{color}
\usepackage{booktabs}
\usepackage{float}
\usepackage{longtable}
\usepackage{epsfig}
\usepackage{bm}
\usepackage{dcolumn}
\usepackage{lipsum, babel}
\usepackage{soul}

\usepackage{rotating} % for sidewaystable

\usepackage[normalem]{ulem}
\usepackage{epstopdf}
\usepackage{natbib}
\usepackage[bookmarksopen, colorlinks,linkcolor=blue,citecolor=blue,urlcolor=blue]{hyperref}

\makeatletter
\let\cat@comma@active\@empty
\makeatother
\begin{document}

\title{Symmetry, microscopy and spectroscopy signatures of altermagnetism}
%\title{Distinctive symmetry, microscopy and spectroscopy signatures of altermagnetism}

\author{Tomas~Jungwirth}
\email{jungw@fzu.cz}
\affiliation{Institute of Physics, Czech Academy of Sciences, Cukrovarnick\'a 10, 162 00, Praha 6, Czech Republic}
\affiliation{School of Physics and Astronomy, University of Nottingham, NG7 2RD, Nottingham, United Kingdom}

\author{Jairo Sinova}
\affiliation{Institut f\"ur Physik, Johannes Gutenberg Universit\"at Mainz, D-55099 Mainz, Germany}

\author{Rafael M. Fernandes}
\affiliation{Department of Physics, The Grainger College of Engineering, University of Illinois Urbana-Champaign, Urbana, IL 61801, USA}
\affiliation{Anthony J. Leggett Institute for Condensed Matter Theory, The Grainger College of Engineering, University of Illinois Urbana-Champaign, Urbana, IL 61801, USA}

\author{Qihang Liu}
\affiliation{Department of Physics, State key laboratory of quantum functional materials and Guangdong Basic Research Center of Excellence for Quantum Science, Southern University of Science and Technology, Shenzhen, 518055, China}

\author{Hikaru Watanabe}
\affiliation{Department of Physics, University of Tokyo, Bunkyo-ku, Tokyo 113-0033, Japan}
\affiliation{Max Planck Institute for the Physics of Complex Systems, N\"othnitzer Str. 38, 01187 Dresden, Germany}

\author{Shuichi Murakami}
\affiliation{Department of Applied Physics, University of Tokyo, 7-3-1 Hongo, Bunkyo-ku, Tokyo 113-8656, Japan}
\affiliation{ International Institute for Sustainability with Knotted Chiral Meta Matter (WPI-SKCM$^2$), Hiroshima University, Higashi-hiroshima, Hiroshima
739-0046, Japan}

\author{Satoru Nakatsuji}
\affiliation{Department of Physics, University of Tokyo, Bunkyo-ku, Tokyo 113-0033, Japan}
\affiliation{Institute for Solid State Physics, University of Tokyo, Kashiwa, Chiba 277-8581, Japan}
\affiliation{Institute for Quantum Matter and Department of Physics and Astronomy, Johns Hopkins University, Baltimore, Maryland 21218, USA}
\affiliation{Trans-scale Quantum Science Institute, University of Tokyo, Bunkyo-ku, Tokyo 113-8654, Japan}

\author{Libor \v{S}mejkal}
\email{lsmejkal@pks.mpg.de}
\affiliation{Max Planck Institute for the Physics of Complex Systems, N\"othnitzer Str. 38, 01187 Dresden, Germany}
\affiliation{Max Planck Institute for Chemical Physics of Solids, N\"othnitzer Str. 40, 01187 Dresden, Germany}
\affiliation{Institute of Physics, Czech Academy of Sciences, Cukrovarnick\'a 10, 162 00, Praha 6, Czech Republic}

\date{\today}

\begin{abstract}
\noindent The recent discovery of altermagnetism was in part motivated by the research of compensated magnets towards highly scalable spintronic technologies. Simultaneously, altermagnetism shares the anisotropic higher-partial-wave nature of ordering with unconventional superfluid phases which have been at the forefront of research for the past several decades. These examples illustrate the interest in altermagnetism from a broad range of science and technology perspectives. Here we review the symmetry, microscopy and spectroscopy signatures of altermagnetism. We describe the spontaneously broken and retained symmetries which delineate altermagnetism as a distinct  phase of matter with $d$, $g$ or $i$-wave compensated collinear spin ordering. In materials ranging from weakly-interacting metals to strongly correlated insulators, the microscopic crystal-structure realizations of the altermagnetic symetries feature a characteristic ferroic order of anisotropic higher-partial-wave components of atomic-scale spin densities. These symmetry and microscopy signatures of altermagnetism are directly reflected in spin-dependent electronic spectra and responses. We review salient band-structure features originating from the altermagnetic ordering, and from its interplay  with spin-orbit coupling and topological phenomena. Throughout the review we compare altermagnetism to traditional ferromagnetism and N\'eel antiferromagntism, and to magnetic phases with symmetry-protected compensated non-collinear spin orders. We accompany the theoretical discussions  by references to relevant experiments.
\end{abstract}

\maketitle

% Word count without equations and figure captions: 2835
% figures: 5x aspsect ratio 1 gives: 5 x 170=850
% equations: 2x32=64
% captions: 384
% total: 4133
% PRL total limit 3750
%\subsection{Research context}
%Comparison of conventional (s-wave) FM with the higher-partial-wave AM in fig. 1. in unconventional d-wave SCs even parity order paramenter sim kx^2 - ky^2 like in AMs but scalar - no spin-space ordering. In p-wave SF 3He the order parameter is a vector in spin space (0, ikx-ky,0) which is kollinear and nodal in momentum like in AMs, but its magnitude is k-dependent unlike spin in AM having k-independent quantization axis, and odd-parity unlike even-parity AM ordering. Most importantly, SF of 3He at low T and high P originates from a subtle momentum space FS instability of a homogeneous fluid, while the microscopic ordering mechanism in AM is based on robust interplay of single-particle crystal potential and many-body (exchange) interaction applicable to metals and insulators and enabling ordering at ambient conditions. More in Ref. [Newton].

\begin{figure}[h!]
	\centering
	\includegraphics[width=.5\linewidth]{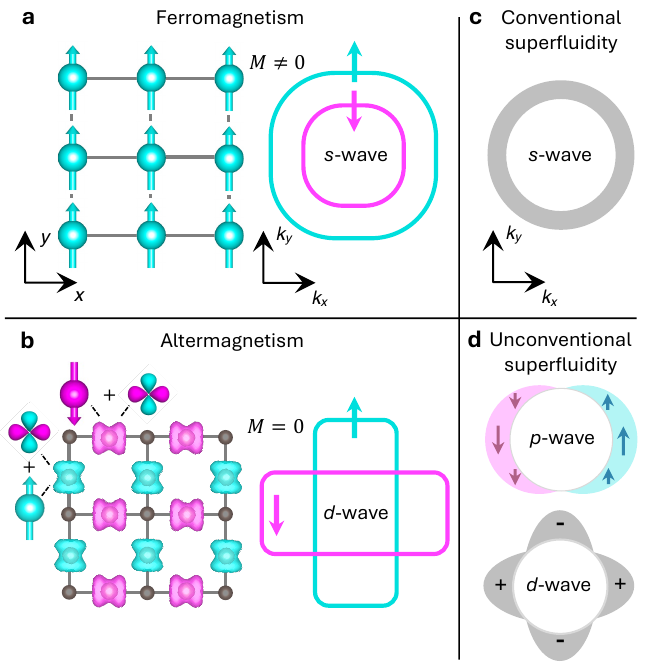}
	\vspace{-.2cm}
	\caption{\textbf{Ferromagnetism and altermagnetism vs. conventional and unconventional superfluidity.}
{\bf a,} Cartoon of uncompensated ($M\neq 0$) ferromagnetism  with a parallel (ferroic) alignment of atomic magnetic dipoles in the position space of the crystal  and corresponding majority-spin  and minority-spin  energy iso-surfaces in the momentum space preserving the crystallographic point-group rotation symmetry ($s$-wave). {\bf b,}  Cartoon of compensated ($M=0$) altermagnetism with depicted decomposition of the local anisotropic spin density\cite{Smejkal2020,Smejkal2021a,Smejkal2022a}   into a dipole (circle with arrow) and a higher-partial-wave ($d$-wave) spin-density component. Cyan and magneta colors mark opposite spin polarizations. 
The higher-partial-wave  ($d$-wave) component is ferroically ordered on the crystal. 
The depicted model is a 2D Lieb lattice\cite{Mazin2023a,Brekke2023,Antonenko2025,Kaushal2024}.  The momentum-space spin-up and spin-down energy iso-surfaces show the corresponding anisotropic ($d$-wave) order breaking the $T$-symmetry\cite{Smejkal2020,Smejkal2021a,Smejkal2022a}.  They intersect at spin-degenerate nodes. {\bf c,} Cartoon of the isotropic scalar order parameter and the quasiparticle gap function (filled area) on the Fermi surface for conventional  $s$-wave superfluidity (superconductivity). {\bf d,} Top: Anisotropic order parameter (arrows) and the quasiparticle gap (filled area) for unconventional $p$-wave superfluidity. The  $p$-wave order parameter is a collinear vector in the spin space, reminiscent of collinear altermagnetism, but has odd parity.  Bottom:  Anisotropic order parameter ($\pm$) and the quasiparticle gap (filled area) for unconventional  $d$-wave superfluidity. The  $d$-wave order parameter has even parity, reminiscent of  altermagnetism, but is a scalar in the spin space (no spin order). For a detailed comparison between altermagnetism and unconventional superfluidity see Ref.~\onlinecite{Jungwirth2025}.
}
\label{fig:AM-FM}
\end{figure}

\noindent Altermagnetism is a newly identified ordered phase of electrons and their spins in condensed matter\cite{Smejkal2021a,Smejkal2022a}. The distinctive signatures of altermagnetism  can be illustrated on simplified model structures shown in Fig.~\ref{fig:AM-FM}. As a reference, ferromagnetism, carrying  a finite net magnetization, is represented by a parallel (ferroic) alignment of atomic magnetic dipoles formed by electron spins in the position space of the  crystal. The corresponding electronic spectrum in the momentum space is  split into majority and minority channels with opposite spin. In contrast, altermagnetism with its compensated spin ordering is illustrated in Fig.~\ref{fig:AM-FM} by a model highlighting the presence of an anisotropic component of the local spin density whose sign alternates on the atomic scale. The  ferroic crystal order of these local anisotropic spin-density components  has its counterpart in the alternating-sign spin polarization in the momentum space. The corresponding electronic spectrum has a form of equal-size, anisotropically-distorted and mutually-rotated energy iso-surfaces of the opposite spin channels, intersecting at spin-degenerate nodes\cite{Smejkal2020,Smejkal2021a,Smejkal2022a}.

The microscopic mechanism of the spontaneous spin ordering in magnetic ground states, including altermagnetic, is due to an interplay of the spin-independent electron-electron Coulomb interaction  in the many-body Hamiltonian with the Pauli exclusion principle. The mechanism is commonly referred to as the exchange interaction. Due to a large energy scale of typical exchange interactions, the spin orders tend to be robust. In contrast, microscopic mechanisms by which the isotropic electron-electron Coulomb interaction can lead to ordered ground states with a spontaneously developed anisotropy  have been considered to be of a more subtle correlated nature\cite{Jungwirth2025}. Here the prominent examples are the unconventional $d$-wave superconductivity in cuprates or  the $p$-wave superfluidity of  $^3$He (Figs.~\ref{fig:AM-FM}c,d). Altermagnetism stands apart from these anisotropic ordered phases driven by the subtle correlated instabilities in metallic Fermi fluids \cite{Jungwirth2025}. The anisotropic spin ordering in altermagnets  is commonly stabilized by a robust microscopic mechanism. It originates from a direct interplay of, on one hand, the exchange interaction and, on the other hand, the single-particle electron interaction  with the  ionic potential of the underlying crystal lattice with a suitable symmetry\cite{Smejkal2021a,Smejkal2022a,Jungwirth2025}. As a result, altermagnetism can form at ambient conditions and in  a variety of metallic, as well as insulating materials. This has opened a range of new research directions exploring the extraordinary properties and responses enabled by altermagnetism, and the synergies of altermagnetism with other condensed-matter phases 
\cite{Smejkal2021a,Smejkal2022a,Mazin2025,Beenakker2023,Mazin2023a,Brekke2023,Li2023,Zhu2023b,Sumita2023,Ghorashi2024,Papaj2023,Wei2024,Zhang2023,Cheng2024,Zhao2024,Maeland2024,Chakraborty2024b,Banerjee2024a,Jeschke2024,Verbeek2023,Bernardini2025,Zyuzin2024,Sim2025,Hu2025,Smejkal2024,Duan2025,Gu2025,Parthenios2025}. (For more details on the research context and additional references see Supplementary information.)

In the following sections we review the distinctive symmetry, microscopy and spectroscopy signatures of the altermagnetic ordering which underpin the emerging broad research of this newly identified phase of matter. For other review, perspective or commentary articles focusing on altermagnetism in the context of the symmetries of compensated magnets, the anomalous Hall effect and other spintronic responses, the initial experimental research, or discussing distinctions between altermagnetism and correlated metallic and superfluid instabilities of Fermi liquids, we point to Refs.~\onlinecite{Smejkal2022AHEReview,Smejkal2022a,Bai2024,Jungwirth2025,Liu2025,Song2025,Jungwirth2025b}.

\bigskip
\noindent{\large\bf Symmetry signatures}
\bigskip

\noindent The single-particle interaction term of the electronic Hamiltonian due to the ionic potential of the crystal lattice has no explicit dependence on the electron spin. It is, therefore, invariant under the symmetry group SO(3)  of all continuous spin-space rotations. In the real space, the crystal potential obeys symmetries of the crystallographic group {\bf G}. The many-body term in the Hamiltonian due to the electron-electron interactions has also no explicit dependence on spin, i.e. has the spin-space SO(3) symmetry, plus it is isotropic in the real space. The normal phase of the electronic system determined by these interaction terms retains the full symmetry, ${\bf Z}_2^{T}\times{\rm SO(3)}\times {\bf G}$, of the corresponding Hamiltonian. Here we also explicitly included the $T$-symmetry via the ${\bf Z}_2^{T}$ group containing $T$ and the identity. 

Conventional collinear ferromagnetism, represented by the model spin arrangement on the crystal in Fig.~\ref{fig:AM-FM}a, spontaneously breaks (some) spin-space SO(3) symmetries and  the $T$-symmetry. Simultaneously, the retained symmetries in the ordered ground state are given by the  spin group ${\bf Z}_2^{C_2T}\ltimes{\rm SO(2)}\times {\bf G}$. Here the $C_2T$ symmetry in ${\bf Z}_2^{C_2T}$ combines $T$ with a two-fold spin-space rotation $C_2$ around an axis orthogonal to the collinearity axis of spins, and SO(2) is a group of continuous spin-space rotations around the collinearity axis. 

The altermagnetic ordering\cite{Smejkal2021a} (Fig.~\ref{fig:AM-FM}b) also spontaneously lowers the Hamiltonian's ${\bf Z}_2^{T}\times{\rm SO(3)}$ symmetry to the ground-state's ${\bf Z}_2^{C_2T}\ltimes{\rm SO(2)}$ symmetry. (This part of the spontaneous symmetry lowering is common to all collinear spin-ordered phases\cite{Smejkal2021a,Liu2021,Litvin1974}.) In contrast to ferromagnetism, however, the altermagnetic ordering retains only symmetries of a halving subgroup {\bf H} of the crystallographic group {\bf G}. The other half of the symmetries contained in ${\bf G}-{\bf H}$ are spontaneously broken on their own, while the altermagnetic ground state retains symmetries  combining the crystallographic transformations from ${\bf G}-{\bf H}$  with the spin-space rotation $C_2$ around an axis orthogonal to the spins. In the case of the altermagnetic order, ${\bf G}-{\bf H}$ contains real-space rotation transformations (proper or improper and symmorphic or non-symmorphic) and does not contain the real-space inversion (parity $P$)  or a translation ($t$). With this constraint, the altermagnetic-ordering class is unambiguously and exclusively delineated by  the spin groups of the form\cite{Smejkal2021a},  ${\bf Z}_2^{C_2T}\ltimes{\rm SO(2)}\times ([E\parallel {\bf H}] +[C_2\parallel {\bf G}-{\bf H}])$. (Here $E$ is the spin-space identity and the spin-space/real-space symmetry transformations are on the left/right from the double-bar in the square bracket.) 

The 	$[C_2\parallel {\bf G}-{\bf H}]$ symmetries imply that the altermagnetic spin ordering is compensated, i.e., that the integrated spin-up and spin-down densities have the same magnitude. The  $[E\parallel {\bf H}]$ term  implies that in each spin channel, the spin density is anisotropic. The altermagnetic spin ordering retains the symmetries of the halving subgroup {\bf H}, while spontaneously breaking the remaining ${\bf G}-{\bf H}$ rotation symmetries.

Before we turn to the discussion in the following sections on how the altermagnetic spin-group symmetries are reflected in the position-space microscopy and momentum-space spectroscopy signatures, we conclude this section with a brief comment on the physical distinction between symmetry breaking by the spin ordering and the spin-orbit coupling. 

The spin ordering can occur in ground states of many-body systems of interacting electrons by spontaneously breaking the  spin-space SO(3) symmetry of the spin-independent crystal-potential and the electron-electron interaction terms in the Hamiltonian\cite{Andreev1984,Gorkov1992,Leggett2003,Smejkal2021a,Moessner2021}.  We note here that the corresponding spin-ordered ground state is degenerate since different ordered states related by any spin-rotation transformation from the SO(3) symmetry group of the Hamiltonain have the same energy. 

The spin-orbit coupling refers to single-particle interaction term(s) in the relativistic Hamiltonian which explicitly depend on the electron spin and which couple the spin-space and the real-space degrees of freedom\cite{Strange1998}. In the case of the spin-orbit coupling, the spin-space SO(3) symmetry is thus already broken in the Hamiltonian of the electronic system. This makes the physical nature of the symmetry breaking by the spin-orbit coupling principally distinct from the spontaneous symmetry breaking by the spin ordering. Albeit typically weak compared to the crystal potential and electron-electron interaction terms in materials without heavy elements, the spin-orbit coupling can have important consequences for equilibrium and non-equilibrium properties of the electronic systems. For example, in normal states of systems with broken $P$-symmetry it can lift the spin degeneracy of the electronic spectra,  or in magnets it can lift the degeneracy of the spin-ordered ground states\cite{Winkler2003,Landau1984}. It can also facilitate a range of topological band structure and transport phenomena\cite{Sinova2015,Nagaosa2010,Franz2013,Murakami2017a,Bradlyn2017,Armitage2018,Elcoro2021}. 

The primary focus of this review is on the novel physics associated with the altermagnetic spin ordering, but we also give examples of the interplay of the altermagnetic ordering with the spin-orbit coupling. On the symmetry level, effects of the spin-orbit coupling are disentangled from the spin-ordering physics by employing the spin-group and the magnetic-group formalisms. In the spin-group formalism\cite{Smejkal2021a,Smejkal2022a,Litvin1974,Mazin2021,Liu2021,Reichlova2024,Smejkal2022GMR,Betancourt2021,Smejkal2023,Chen2025,Hariki2023,Hellenes2023,Hellenes2023a,McClarty2024,Smolyanyuk2024,Shinohara2024,Watanabe2024,Jiang2024,Zhu2025,Chen2024,Schiff2025a,Xiao2024}, focusing on the spin ordering alone, the symmetry transformations applied simultaneously in the spin space and in the real space can be different, reflecting the absence of any explicit spin dependence in the crystal-potential and the electron-electron interaction terms in the Hamiltonian. In contrast, the magnetic-group formalism\cite{Landau1984,Litvin2013} only considers a simultaneous application of the same symmetry transformations in both the spin space and the real space, reflecting the explicit coupling of the spin-space and the real-space degrees of freedom in the relativistic spin-orbit coupled Hamiltonian. The complementary merits of the two symmetry formalisms are illustrated in Box.

\bigskip
{\color{brown}
\noindent{\large\bf Box} 
\bigskip

\noindent In Fig.~Box, we compare the spin groups\cite{Smejkal2021a,Liu2021} of a collinear uncompensated spin ordering in Fe, a collinear compensated spin ordering in (altermagnetic) NiF$_2$, and a non-collinear coplanar compensated spin ordering in Mn$_3$Sn, with magnetic groups\cite{MAGNDATA} of the three structures. The spin group is a direct product of the so-called spin-only group and the so-called non-trivial spin group\cite{Smejkal2021a,Liu2021,Litvin1974}. The spin-only group distinguishes the collinear (Fe and NiF$_2$) from the non-collinear coplanar (Mn$_3$Sn) spin orders, where in the former case the spin-only group is given by 
${\bf Z}_2^{C_2T}\ltimes{\rm SO(2)}$ and in the latter case by ${\bf Z}_2^{C_2T}$. The non-trivial spin group for the uncompensated spin ordering (Fe) has no term combining a crystallographic transformation with a spin rotation. In contrast, such a combined transformation is present  in the case of the compensated  spin ordering (NiF$_2$, Mn$_3$Sn). 

Unlike the spin groups, the magnetic groups for all three structures are the same. This illustrates, among other features, that the magnetic groups do not generally distinguish between the collinear and the non-collinear (coplanar or non-coplanar) magnets. In addition, the specific magnetic group of the three structures belongs to the so-called ferromagnetic groups which allow for a net magnetization. All three structures are thus rendered as uncompensated by the magnetic group.  A comparison with the respective spin groups shows that in the case of NiF$_2$ and Mn$_3$Sn, the net magnetization is of the relativistic spin-orbit coupling origin, while the spin ordering alone yields a zero net magnetization in the non-relativistic limit protected by the spin-group symmetry. In the case of  Fe, a non-zero net magnetization is allowed  without the spin-orbit coupling by the spin-group symmetry.  All three magnetic structures in Fig.~Box have spin-group symmetries that allow for spin-split electronic spectra in the absence of the spin-orbit coupling.  In general, spin-group symmetries (and not magnetic-group symmetries) determine whether and at what momenta the non-relativistic band structure can be spin split, or whether spin-degeneracy of the non-relativistic bands is symmetry-enforced  across the whole Brillouin zone.
}

\bigskip
\noindent{\large\bf Microscopy signatures}
\bigskip

\begin{figure}[h!]
	\centering
	\includegraphics[width=.6\linewidth]{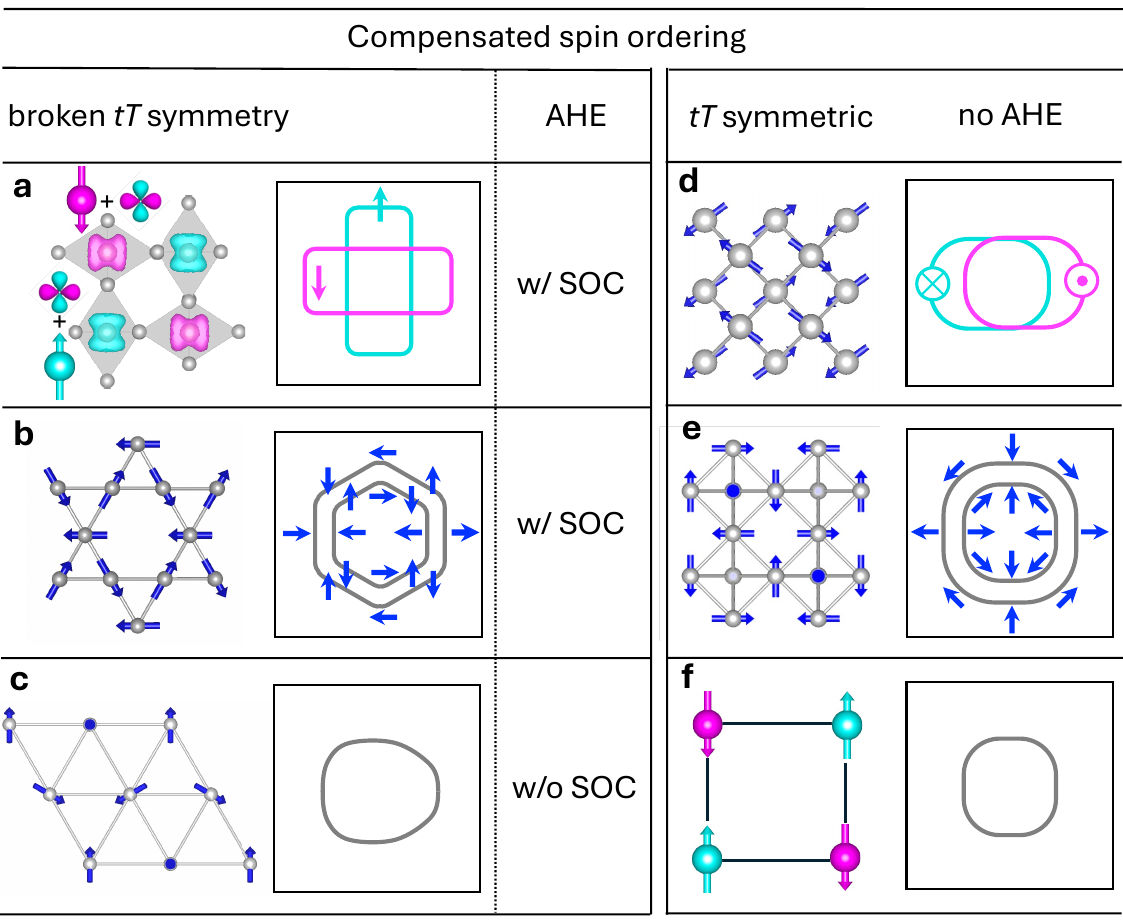}
	\vspace{-.0cm}
	\caption{
	\textbf{Magnets with compensated spin ordering.}
{\bf a,} Left: Altermagnetic ordering on a rutile crystal (RuO$_2$, FeF$_2$, etc.)\cite{Smejkal2020,Smejkal2021a,Smejkal2022a} with anisotropic atomic spin densities schematically decomposed into antiparallel isotropic $s$-wave (dipole)  and ferroically-ordered anisotropic $d$-wave  components. Right: Corresponding $d$-wave order on the schematic momentum-space energy iso-surfaces. {\bf b,} Position-space (left) and momentum space (right) cartoons of a non-collinear coplanar even-parity magnetic order represented by Mn$_3$Sn\cite{Nakatsuji2015,Zelezny2017a}. {\bf c,} Non-coplanar magnetic order in position space with a corresponding spin-degenerate electronic structure in the momentum space represented, e.g., by CoM$_3$S$_6$ (M = Ta or Nb)\cite{Martin2008,Feng2020,Takagi2023,Watanabe2024,Chen2024}. {\bf d,} Non-collinear coplanar magnetic order in the position space with corresponding odd-parity ($p$-wave) collinear spin order in the momentum space represented by CeNiAsO\cite{Hellenes2023}. {\bf e,} Non-collinear non-coplanar magnetic order in the position space with corresponding  odd-parity non-collinear spin order in the momentum space represented by Ce$_3$InN\cite{Hellenes2023a}. {\bf f,} Conventional collinear N\'eel antiferromagnetism with the antiferroic order ($tT$-symmetry) of opposite atomic dipole moments on an even number of crystal sublattices\cite{Neel1971}, and corresponding spin-degenerate electronic structure in the momentum space. Cartoons in all panels are for non-relativistic electronic structures. Whether or not the anomalous Hall effect  is allowed (AHE or no AHE), and whether with or without the relativistic spin-orbit coupling (w/ or w/o SOC) is also indicated for all panels.
		}
\label{fig:split_AHE}
\end{figure}

\noindent In this section we look how the altermagnetic spin-group symmetries are realized on the microscopic level in representative  crystal-structures, and discuss the microscopic physics of ordering in altermagnets. Fig.~\ref{fig:split_AHE}a shows the arrangement of anisotropic atomic spin densities in an altermagnetic rutile crystal obtained from microscopic theory\cite{Smejkal2020}. These local spin densities can be decomposed into $s$-wave (dipole) components with antiparallel alignment between neighboring atoms, and $d$-wave  components aligned ferroicly on the crystal lattice, as schematically indicated in Fig.~\ref{fig:split_AHE}a. The $d$-wave spin density components correspond to the $d$-wave symmetry of anisotropic exchange interactions in the rutile crystal lattice\cite{Smejkal2023}. The ferroic order of the local $d$-wave spin-density components generates the $T$-symmetry breaking electronic structure and responses, reminiscent of the effects of the ferroic order of local dipoles in conventional ferromagnets\cite{Smejkal2020}. In contrast to the ferroicly ordered dipoles, however, the higher-partial-wave ferroic order does not generate a net magnetization and leads to the characteristic anisotropy of the phenomena observed in altermagnets.

We note that experimental signatures of the magnetic ordering in RuO$_2$, a metallic rutile crystal considered among this class of altermagnetic candidates from the early theoretical studies\cite{Smejkal2020}, are presently a matter of debate \cite{Berlijn2017a,Zhu2018,Lovesey2022,Occhialini2021,Feng2022,Bose2022,Bai2022,Karube2022,Lovesey2023c,Liu2023,Fedchenko2024,Smolyanyuk2023,Lin2024,Kessler2024,Li2024a,Wenzel2025,Jeong2024,Hiraishi2024}.  These seemingly controversial reports can be possibly explained by the magnetic order in ruthenates, including RuO$_2$, being fragile with respect to impurities, disorder or strain\cite{Smolyanyuk2023}.
In other insulating rutile crystals\cite{Smejkal2022a,Lopez-Moreno2016,Noda2016,Hayami2019,Yuan2020,Bhowal2024}, such as altermagnetic NiF$_2$ mentioned in Fig.~Box, the experimental signatures of the compensated collinear magnetic order are well established.

%\newpage

\begin{figure}[h!]
	\centering
	\includegraphics[width=1\linewidth]{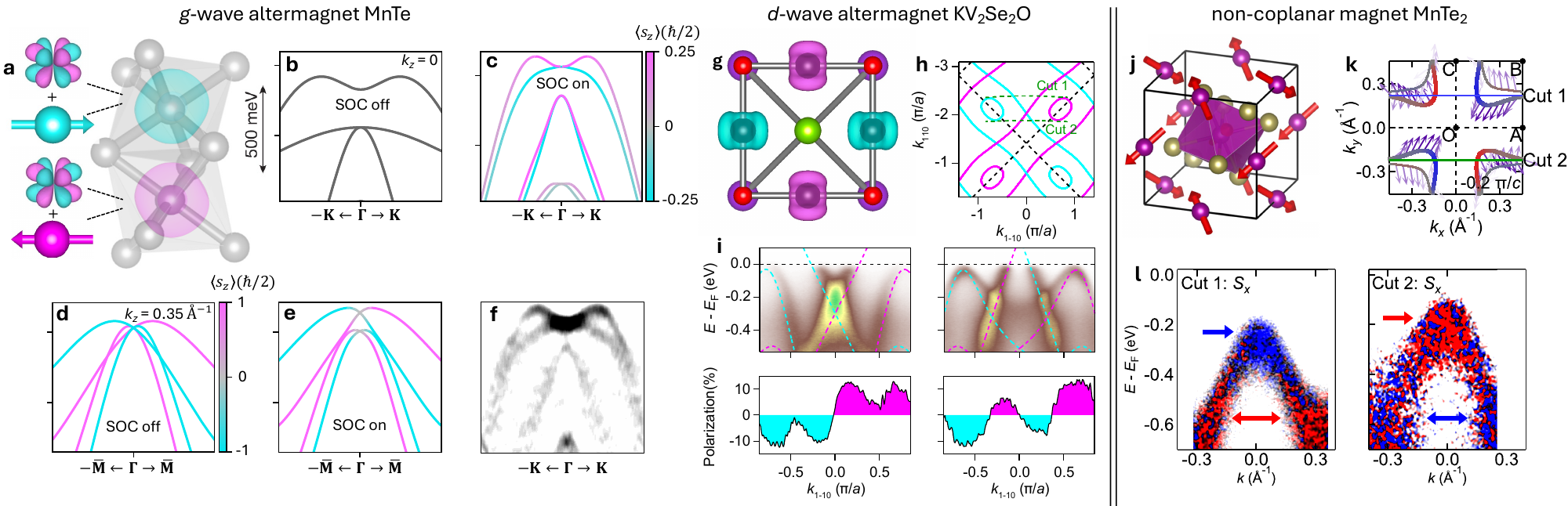}
	\vspace{.0cm}
	\caption{
	\textbf{{\bf\em g} - wave and {\bf\em d} - wave altermagneta and a non-coplanar magnet.}
{\bf a,} MnTe magnetic crystal structure with grey spheres representing Te atoms.  Cyan/magenta local spin densities on Mn atoms inside the Te octahedra are schematically  decomposed into isotropic dipole (sphere with arrow) and anisotropic $g$-wave components. Cyan and magenta colors mark opposite spin polarizations. {\bf b,} DFT spin-polarized band structure along a $k_z=0$ path in the Brillouin zone in the $k_z=0$ nodal plane with the spin-orbit coupling (SOC) turned off.  {\bf c,} Same as panel b with SOC turned on. {\bf d,e,} Same as panels b,c along a $k_z \neq 0$ path in the Brillouin zone away from the four nodal planes of the $g$-wave altermagnetic order.  {\bf f,} Angle-resolved photoemission measurements of the spin-split band structure corresponding to the DFT panel c. Adapted from Ref.~\onlinecite{Krempasky2024}. {\bf g,} KV$_2$Se$_2$O magnetic crystal structure with local anisotropic spin densities on V atoms. (K atoms are purple, O red, and Se green spheres.) {\bf h,} Calculated spin-resolved Fermi surface at $k_z=0$ with green dashed lines indicating the momentum locations of cuts 1 and 2 in the Brillouin zone. {\bf i,} ARPES intensity plots showing the band dispersion along cut 1 and 2. Magneta and cyan dashed curves are calculated spin-up and spin-down bands. Magneta and cyan filled areas highlight the momentum-dependent spin-up and spin-down polarizations calculated by the asymmetry of the measured spin-up and spin-down signals. Adapted from Ref.~\onlinecite{Jiang2025}. {\bf j,} Crystal structure of MnTe$_2$. The compensated non-coplanar magnetic configuration on Mn atoms is indicated by the red arrows (yellow spheres represent Te atoms forming octahedra around Mn atoms). {\bf k,}  DFT-derived, spin-resolved $k_x-k_y$ map at 21.2~eV and at binding energy $E_B = 0.45$~eV. Magenta arrows show in-plane direction of the spin; darkness of arrows shows magnitude of in-plane spin polarization; colour of bands shows magnitude of the out-of-plane $S_z$ polarization. The O-A-B-C plane is the $k_z = -0.2\pi/c$ plane. {\bf l,}  ARPES-measured, $S_x$-resolved Cuts 1 and 2. The $k$ positions of Cuts 1 and 2 are marked in panel d.  Adapted from Ref.~\onlinecite{Zhu2024}.
}
\label{fig:ARPES}
\end{figure}

An analogous decomposition of the local atomic spin densities applies to semiconducting and metallic room-temperature altermagnets MnTe and CrSb\cite{Jaeschke-Ubiergo2025}, which have become workhorse materials in theoretical and experimental research of  altermagnetism \cite{Smejkal2021a,Smejkal2022a,Betancourt2021,Lovesey2023,Mazin2023,Krempasky2024,Lee2024,Osumi2024,Hajlaoui2024,Aoyama2024,Kluczyk2024,Reimers2024,Yang2024,Ding2024,Li2024,Lu2025,Zeng2024,Dale2024}. 
In these crystals, the antiparallel alignment of local atomic dipoles is complemented by the ferroic order of local $g$-wave components of the anisotropic spin density\cite{Jaeschke-Ubiergo2025}, as schematically indicated in Fig.~\ref{fig:ARPES}a. Note that the corresponding  anisotropic exchange interactions were predicted\cite{Smejkal2023} and experimentally demonstrated\cite{Liu2024b} in MnTe to generate alternating chiral splitting of the magnon spectra. 

Recently, the family of experimentally confirmed altermagnets has been  extended by KV$_2$Se$_2$O and RbV$_2$Te$_2$O layered crystals\cite{Jiang2025,Zhang2025a}, shown in Fig.~\ref{fig:ARPES}g. They feature atomic and spin arrangements corresponding to the 2D Lieb-lattice model of $d$-wave  altermagnetism\cite{Mazin2023a,Brekke2023,Antonenko2025,Kaushal2024}, introduced in Fig.~\ref{fig:AM-FM}b. Lieb-lattice altermagnetism was also shown to be realized in the Mott insulating compound La$_2$Mn$_2$Se$_2$O$_3$ \cite{Wei2025}. 

Remarkably, the altermagnetic symmetries can be realized by  the ferroic order of the local $d$-wave (or higher even-parity-wave) spin densities on magnetic atoms without any local atomic dipole components\cite{Jaeschke-Ubiergo2025}. Microscopic calculations identified such a pure form of atomic $d$-wave altermagnetism\cite{Jaeschke-Ubiergo2025} in a Mott insulator Ba$_2$CaOsO$_6$\cite{Maharaj2020,Jaeschke-Ubiergo2025}. This underlines the microscopic distinction of altermagnetism from  conventional N\'eel antiferromagnetism, where the latter has a characteristic antiferroic order ($tT$-symmetry) of atomic dipole moments on an even number of crystal sublattices (Fig.~\ref{fig:split_AHE}f)\cite{Neel1971}.

Other microscopic mechanisms include the realization of the altermagnetic symmetries by crystal lattice deformations\cite{Mazin2023a,Chakraborty2024,Liu2024}. While in the unperturbed crystal the opposite spin densities are related by one of the symmetries excluding the altermagnetic order (translation or inversion), the deformation such as a twist of crystal planes breaks these symmetries while obeying the altermagnetic spin-group symmetry.

There can also be cases where the role of symmetries of the single-particle potential of the ionic crystal lattice are complemented by electronic correlations in the formation of the altermagnetic phase. Here the altermagnetic phase is enabled by correlation-induced orbital ordering  which lowers the symmetry of the crystallographic group, as compared to the symmetry of the ionic lattice. Together with the spin ordering by the exchange interaction, the phase then falls into the altermagnetic spin-group class\cite{Leeb2023,Fernandes2023}. 

Apart from this microscopic mechanism, the role of electronic correlations in promoting or affecting altermagnetism has been studied in various contexts \cite{Das2023,Maier2023,Leeb2023,Sato2024,Roig2024}, including the strongly-coupled regime where the interaction strength is comparable to the bandwidth \cite{Bose2024,Ferrari2024}. Since the strongly-correlated Mott-insulator crystals commonly display a compensated antiparallel magnetic order, several Mott insulating materials with appropriate spin symmetries have been put forward as altermagnetic candidates. Specifically, a wide class of Mott insulating perovskites, such as the manganite CaMnO$_3$ and the titanate LaTiO$_3$ \cite{Smejkal2020,Smejkal2021a,Smejkal2022a,Fernandes2023,Bernardini2025,Rooj2025}, have been identified as platforms to realize and investigate altermagnetism in the regime of strong correlations. In these compounds, the oxygen octahedra rotate to accommodate the cation, lowering the ideal cubic symmetry of the perovskite down to orthorhombic, which essentially changes the spin group to altermagnetic. In different contexts, earlier studies found that the octahedra rotation in the perovskites can directly impact orbital degrees of freedom, which in turn affect the magnetic interactions\cite{Topfer1997,Noda2016}. Subsequent studies made the observation of the $T$-symmetry breaking electronic structure and responses, and the identification of altermagnetism in these systems\cite{Smejkal2020,Smejkal2021a,Smejkal2022a,Fernandes2023,Bernardini2025,Rooj2025}.

Similarly, Mott insulating states realized in oxides with other Ruddlesden-Popper phases, such as the cuprate La$_2$CuO$_4$ \cite{Smejkal2021a} and the nickelate La$_3$Ni$_2$O$_7$ \cite{Bernardini2025}, possess the spin-symmetry requirements to display altermagnetism. Besides 3D oxides, 2D organic charge-transfer salts such as $\alpha$- and $\kappa$-(BEDT-TTF)$_2$X can also realize a Mott insulating altermagnetic phase due to the fact that their organic molecules can form an arrangement similar to the Shastry-Sutherland lattice\cite{Ferrari2024}.

\bigskip
\noindent{\large\bf Spectroscopy signatures}
\bigskip

\noindent An unconventional spin splitting that breaks the $T$-symmetry in theoretical band structures of compensated magnets was initially reported in several  coplanar\cite{Chen2014}, non-coplanar\cite{Zhou2016} and collinear\cite{Smejkal2020} magnets. Employing the spin-group formalism
\cite{Smejkal2021a,Feng2020,Mazin2021,Liu2021,Reichlova2024,Smejkal2022GMR,Betancourt2021,Smejkal2023,Chen2025,Hariki2023,Hellenes2023,Hellenes2023a,McClarty2024,Smolyanyuk2024,Shinohara2024,Watanabe2024,Jiang2024,Zhu2025,Chen2024,Schiff2025a,Xiao2024} then enabled to disentangle by symmetry the spin-ordering and the spin-orbit coupling contributions to this unconventional spin splitting; the same applies  to other prior theoretical reports of non-relativistic electronic structures with momentum-dependent spin splitting in compensated magnets\cite{Lopez-Moreno2016,Noda2016,Ahn2019,Hayami2019,Naka2019,Yuan2020}. 

Within the collinear magnets, the spin-groups allowed to delineate altermagnetism, as discussed in the previous sections. 
Numerous candidate materials with altermagnetic band structures have been identified based on the spin groups when analyzing, e.g., the published magnetic structures in the MAGNDATA database on the Bilbao Crystallographic Server \cite{Smejkal2021a,Smejkal2022a,Guo2023b,Xiao2024,Chen2025,Bai2024}. Besides the numerous 3D inorganic crystals, the spin groups also enabled to establish the altermagnetic symmetry in the spectra of earlier studied or newly identified 2D\cite{Smejkal2022a,Smejkal2022GMR,Ma2021,Egorov2021,Brekke2023,Cui2023,Chen2023b,Mazin2023a,Sodequist2024} or organic\cite{Naka2019,Ferrari2024} materials.

Recently, the unconventional $T$-symmetry breaking spin-splitting in the compensated magnets was reported by angle-resolved photoemission spectroscopy (ARPES) measurements in the $g$-wave altermagnets MnTe, CrSb or CoNb$_4$Se$_8$ (Fig.~\ref{fig:ARPES}a-f)\cite{Krempasky2024,Lee2024,Osumi2024,Hajlaoui2024,Reimers2024,Yang2024,Ding2024,Li2024,Lu2025,Zeng2024,Dale2024}, and the $d$-wave  altermagnets KV$_2$Se$_2$O or RbV$_2$Te$_2$O (Fig.~\ref{fig:ARPES}g-i)\cite{Jiang2025,Zhang2025a}. Unconventional spin splitting was also experimentally observed by ARPES measurements in a non-collinear non-coplanar compensated magnet MnTe$_2$ (Fig.~\ref{fig:ARPES}j-l)\cite{Zhu2024}. In contrast to the collinear altermagnetic ordering, here the non-collinear ordering generates a corresponding non-collinear spin texture in the band structure of MnTe$_2$.

Below we start with a discussion of the electronic structure of altermagnets in the limit of zero spin-orbit coupling. This is followed  by a comparison to non-relativistic electronic structures of several representative magnets with compensated non-collinear spin orders. Finally, we include a discussion of relativistic and topological effects in the altermagnetic band structures.

\smallskip
\noindent{\bf Non-relativistic spectroscopy signatures}
\smallskip

\noindent The symmetry and microscopy signatures of altermagnetism, described in previous sections, are directly reflected in the spin-dependent band structure in the momentum space in the limit of zero relativistic spin-orbit coupling. The symmetries of the collinear spin-only group  ${\bf Z}_2^{C_2T}\ltimes{\rm SO(2)}$ effectively act in the momentum space as the $P$-symmetry. The spin-dependent electronic structure in the momentum space for the altermagnetic order  thus has the  $P$-symmetry, regardless of whether  the magnetic crystal has or has not the $P$-symmetry in the position space\cite{Smejkal2021a}. 

In total, there are 10 non-trivial $P$-symmetric spin point groups (10 non-trivial spin Laue groups) of the form introduced above, $([E\parallel {\bf H}] +[C_2\parallel {\bf G}-{\bf H}])$, describing the spin and momentum dependent band structures in the non-relativistic limit for all altermagnetic spin orders on crystals\cite{Smejkal2021a}. Apart from the spin degeneracy at the $\boldsymbol\Gamma$-point, the symmetries of a first class of altermagnets, described by 4 out of the 10 non-trivial spin Laue groups,  protect spin degeneracy at 2 nodal surfaces in the 3D Brillouin zone crossing the $\boldsymbol\Gamma$-point. In the second class, corresponding to another 4 non-trivial spin Laue groups,  there are 4 symmetry-protected spin-degenerate nodal surfaces. A third altermagnetic class, containing the remaining 2 non-trivial spin Laue groups, has 6 spin-degenerate nodal surfaces protected by symmetry. Outside these nodal surfaces, the spin symmetries allow for lifting the spin degeneracy, resulting in even-parity $T$-symmetry breaking electronic spectra in the three altermagnetic classes of the $d$, $g$ or $i$-wave form, respectively\cite{Smejkal2021a}. The  $d$, $g$ or $i$-wave spectra in the momentum space correspond to the ferroicly ordered $d$, $g$ or $i$-wave spin-density components in the position space of the altermagnetic crystals, discussed in the previous section\cite{Jaeschke-Ubiergo2025}. We note that in comparison to the 10 spin Laue\cite{Smejkal2021a} (37/422 point\cite{Smejkal2021a}/space\cite{Chen2025}$^,$\cite{Jungwirth2025}) groups of altermagnets, there are 11 spin Laue (32/230 point/space)  groups corresponding to the collinear ferromagnetic ordering with split majority-spin and minority-spin bands, and 11 spin Laue (53/769 point/space) groups of the collinear antiferromagnetic ordering with spin-degenerate bands\cite{Smejkal2021a,Jungwirth2025,Chen2025}.

\begin{figure}[h!]
	\centering
	\includegraphics[width=.6\linewidth]{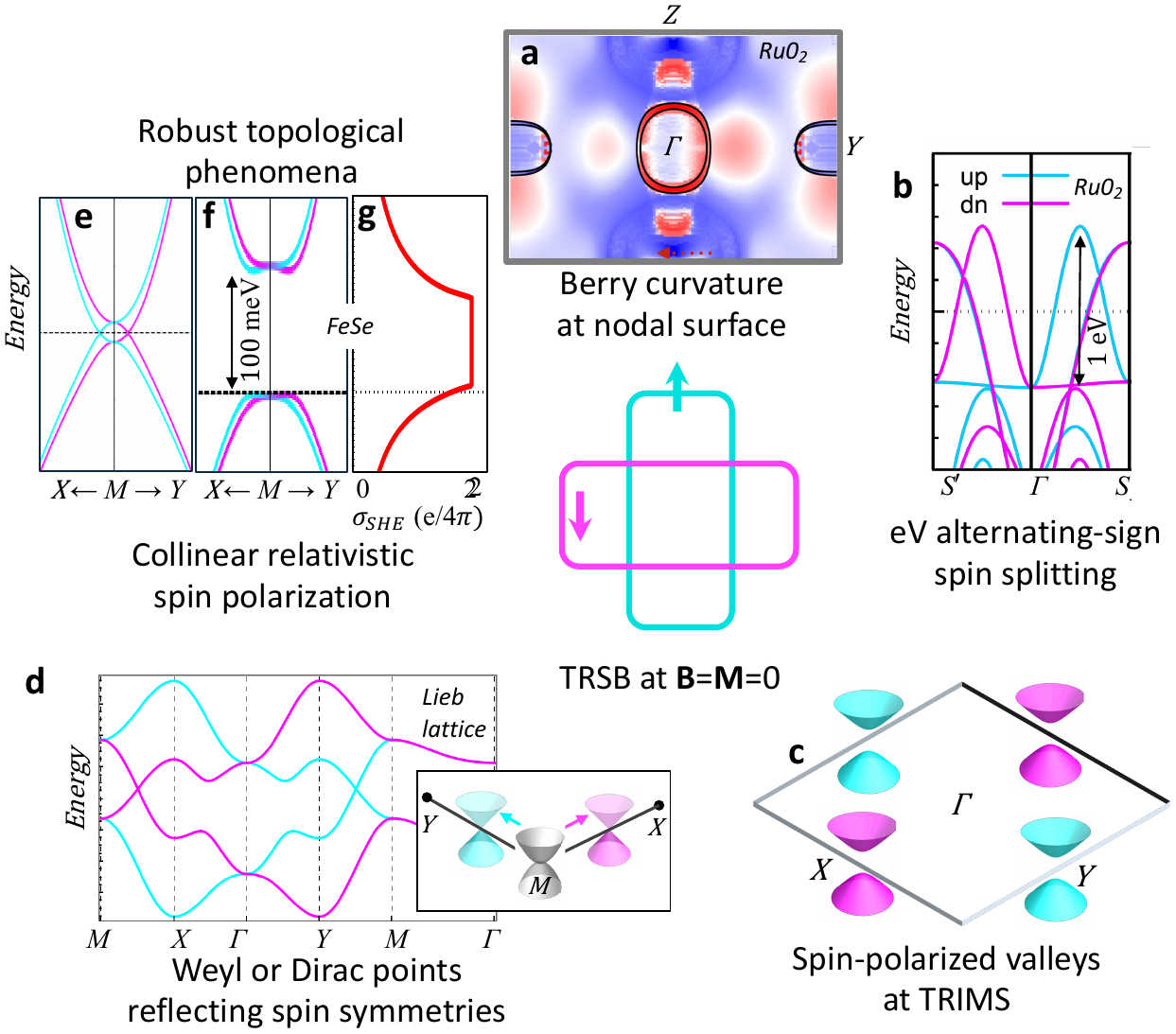}
	\vspace{-.0cm}
	\caption{
	\textbf{Salient electronic structure features of altermagnets.}
Central panel highlight $T$-symmetry breaking (TRSB) in the even-parity electronic structure of altermagnets in the limit of zero spin-orbit coupling and at zero external magnetic field and internal magnetization. {\bf a,} Berry curvature projected on the nodal surface with hot spots around the bands intersecting the nodal surface spin split by the spin-orbit coupling. {\bf b,} Spin splitting without spin-orbit coupling with opposite sign along two perpendicular directions and magnitude reaching eV scale. DFT calculations are for the altermagnetic phase of RuO$_2$, and the panels {\bf a} and {\bf b} are adapted from Ref.~\onlinecite{Smejkal2020}. {\bf c,} Cartoon of spin-split valleys without spin-orbit coupling around $T$-invariant momenta (TRIMS)\cite{Reichlova2024,Smejkal2022GMR,Ma2021,Cui2023,Zhu2024a}.  {\bf d,} Band structure of the altermagnetic 2D Lieb lattice model; the inset shows the emergence of symmetry-related Dirac points from the splitting of the quadratic band crossing. Adapted from Ref.~\onlinecite{Antonenko2025}. {\bf e,}  DFT calculation of bands without spin-orbit coupling in a d-wave altermagnetic phase of electrically-gated 2D FeSe. {\bf f,} Corresponding DFT calculation with spin-orbit coupling showing collinear  spin polarization in a topological insulating phase. {\bf g,} Corresponding quantum spin Hall effect with precise quantization of the spin-Hall conductivity. Panels {\bf e-g}  are adapted from Ref.~\onlinecite{Mazin2023a}.
}
\label{fig:Topo}
\end{figure}

The possibility of a strong $T$-symmetry breaking spin polarization in the non-relativistic electronic structure  away from the nodal surfaces, with spin splitting of the bands on an eV scale,  was pointed out in the initial theoretical studies of the altermagnetic phase of RuO$_2$\cite{Smejkal2020,Ahn2019} (Fig.~\ref{fig:Topo}b). Another distinctive band-structure feature in the limit of zero spin-orbit coupling, introduced in theoretical studies of several altermagnetic candidates, is the presence of spin-polarized valleys\cite{Reichlova2024,Smejkal2022GMR,Ma2021,Cui2023,Zhu2024a}  (Fig.~\ref{fig:Topo}c). Unlike in the relativistic spectra of non-magnetic systems, the non-relativistic band structures of altermagnets can host these spin-polarized valleys at $T$-invariant momenta.

\smallskip
\noindent{\bf Comparison to non-collinear compensated magnets}
\smallskip

\noindent Using the spin symmetries, we will now discuss the salient features of the altermagnetic electronic structure in the limit of zero spin-orbit coupling, highlighted in  Fig.~\ref{fig:split_AHE}a, in comparison to different representative non-collinear magnets with a compensated spin order protected by spin symmetry, illustrated in Figs.~\ref{fig:split_AHE}b-e. All compensated magnetic structures discussed in Fig.~\ref{fig:split_AHE}a-e break the $PT$-symmetry. Recall that the $PT$-symmetry protects Kramers spin degeneracy of the electronic structure across the whole Brillouin zone, both without and with the spin-orbit coupling\cite{Wigner1932,Tang2016,Smejkal2017c,Smejkal2018}. 

We note that  non-collinear spin arrangements on crystals can originate from, e.g., frustrated exchange interactions  in the absence of spin-orbit coupling, or from Dzyaloshinskii-Moriya interaction. (Even in the latter case where spin-orbit coupling contributes to the stabilization of the non-collinear spin ordering, the spin-ordering symmetry can be described by spin-group symmetries.)

In Fig.~\ref{fig:split_AHE}b we show the Mn$_3$Sn member of the intensely studied  Mn$_3$X family of non-collinear compensated magnets\cite{Chen2014,Kubler2014,Nakatsuji2015,Kiyohara2015,Nayak2016,Takagi2023,Ikhlas2017,Li2017a,Higo2018c,Matsuda2020,Kimata2021,Sakamoto2021,Zelezny2017a,Zhang2018h,Kimata2019a,Hu2022,Tsai2020,Takeuchi2021,Higo2022,Pal2022,Chen2023,Qin2023,Smejkal2022AHEReview,Nakatsuji2022,Rimmler2024,Chen2024,Han2025}.
%Its spin group\cite{Liu2021}, containing a symmetry combining spin-space and real-space rotations (specifically the $[C_3||C_6]$ symmetry), was given in the right panel of Fig.~\ref{fig:groups}. 
The right panel of Fig.~\ref{fig:split_AHE}b shows a cartoon of the even-parity $T$-symmetry breaking electronic structure of  Mn$_3$Sn whose  compensated magnetic crystal breaks, besides the  $PT$ symmetry, also the $tT$-symmetry (left panel of Fig.~\ref{fig:split_AHE}b). These momentum-space and position-space symmetry characteristics are analogous to the altermagnetic order (Fig.~\ref{fig:split_AHE}a). Unlike altermagnetism, however, the spin arrangement on the crystal in Fig.~\ref{fig:split_AHE}b is non-collinear (coplanar), resulting in a spin-dependent electronic structure with a non-collinear (coplanar) spin texture in the momentum space which winds twice along the Fermi surface\cite{Zelezny2017a}.  Also in contrast to the anisotropic nodal altermagnetic order, and reminiscent of the nodeless ferromagnetic order, the shape of the energy iso-surfaces and the spin-splitting magnitude do not show a spontaneous anisotropic deformation but preserve the symmetry of the underlying crystal lattice. 

In Figs.~\ref{fig:split_AHE}d,e we show cartoons of non-relativistic spin-dependent electronic structures having odd parity and $T$-symmetry\cite{Hellenes2023,Hellenes2023a,Chen2024,Brekke2024,Ezawa2024d,Sivianes2025a,Chakraborty2024c,Yu2025}. They contrast with the so far discussed even-parity $T$-symmetry-breaking spectra of collinear altermagnets and ferromagnets, and the non-collinear $P$-symmetric magnets represented by Mn$_3$Sn. The odd-parity $T$-symmetric type of non-relativistic spin-split spectra in the momentum space is due to the broken $P$-symmetry in a non-collinear magnetic crystal structure. In addition, it is enabled by the presence of the $tT$-symmetry (Figs.~\ref{fig:split_AHE}d,e). This antiferroic order, in which opposite atomic magnetic dipoles are related by translation in the crystal lattice, contrasts with the ferroic order of the local dipoles in ferromagnets, or the ferroic order of the higher even-parity-wave components of the local spin density in altermagnets. The momentum-space electronic spectra generated by the antiferroic order of the $tT$-symmetric non-collinear magnets in Figs.~\ref{fig:split_AHE}d,e also contrast with the collinear N\'eel antiferromagnets. In the latter case, the $tT$-symmetry together with the collinearity in the position space protects the spin-degeneracy of the even-parity non-relativistic spectra in the momentum space (Figs.~\ref{fig:split_AHE}f). 

Fig.~\ref{fig:split_AHE}d illustrates the odd-parity magnetism on a non-collinear coplanar spin structure in the position space of the crystal. Here the $tT$-symmetry together with the coplanar spin-only group ${\bf Z}_2^{C_2T}$ results in a symmetry $[C_2|| t]$, combining a real-space translation with a two-fold spin-space rotation around the axis orthogonal to the coplanar spins. This allows for a compensated magnetic phase with an odd-parity electronic energy spectrum, $E_{\uparrow}({\bf k})=E_{\downarrow}(-{\bf k})$, and a collinear spin polarization in the momentum space with the polarization aligned along the above $C_2$ spin-rotation axis\cite{Hellenes2023}. To highlight the combination of the $tT$-symmetric antiferroic order in the position space, reminiscent of the N\'eel antiferromagnetism, with the collinear alternating spin polarization in the momentum space, akin to altermagnetism, this type of odd-parity magnetism was recently dubbed antialtermagnetism by some of us\cite{Jungwirth2025}.

Note that while the spins are collinear in the momentum space, the magnitude of their expectation value can vary with momentum. Remarkably, the spin-polarization axis of the collinear spins in the momentum space is perpendicular to the plane of the non-collinear coplanar spins in the position space of the crystal. This is to be contrasted with the case of the collinear ferromagnetic and altermagnetic ordering, for which the momentum-independent spin-quantization axis in the non-relativistic band structure is oriented along the spin-axis in the crystal lattice, and where spin is a good quantum number of the electronic states. 

Fig.~\ref{fig:split_AHE}d illustrates a $p$-wave order featuring nodal spin-polarized energy iso-surfaces that, compared to the normal phase, shift in opposite directions in the momentum space for opposite spin directions. This $p$-wave antialtermagnetism has been predicted\cite{Hellenes2023}  in the $P$-symmetry-breaking $tT$-symmetric non-collinear coplanar magnet CeNiAsO\cite{Hellenes2023}, and is expected to allow for highly efficient charge-to-spin conversion via, e.g., a non-relativistic Edelstein effect\cite{Chakraborty2024c}. 

Fig.~\ref{fig:split_AHE}e illustrates an odd-parity electronic structure generated by a $P$-symmetry-breaking $tT$-symmetric non-coplanar spin arrangement on the crystal of Ce$_3$InN. Because of the absence of the coplanar symmetry ${\bf Z}_2^{C_2T}$, the bands have spin textures with momentum-dependent magnitude and direction of spins\cite{Hellenes2023a} (A.~B.~Hellenes, T.J., J.S. \& L.S. Unconventional B-type magnetism, {\em in preparation}). These odd-parity  $T$-symmetric spin-textures in the momentum space of the magnet represent a non-relativistic spin-ordering counterpart of the  relativistic spin-orbit coupled textures in electronic spectra of non-magnetic crystals with broken $P$-symmetry.

Finally, in Fig.~\ref{fig:split_AHE}c, we highlight an example of a  non-coplanar compensated magnetic crystal CoM$_3$S$_6$ (M = Ta or Nb) breaking both the $tT$ and the $PT$ symmetries. Remarkably, despite the broken $tT$ and the $PT$ symmetries, the non-relativistic band structure has a symmetry-protected spin-degeneracy across the whole Brillouin zone\cite{Feng2020,Takagi2023,Watanabe2024,Chen2024}. This contrasts with the spin-split non-relativistic band structures of altermagnets (Fig.~\ref{fig:split_AHE}a), the class of non-collinear compensated magnets represented in Fig.~\ref{fig:split_AHE}b  by Mn$_3$Sn, or the conventional uncompensated ferromagnets, all also sharing the broken $tT$ and $PT$ symmetries. The spin degeneracy of the electronic spectrum of CoM$_3$S$_6$, illustrated in Fig.~\ref{fig:split_AHE}c, is protected by multiple symmetry elements in the spin space group combing a real-space translation with a spin-space rotation\cite{Feng2020,Watanabe2024,Chen2024}. 

We recall that the $tT$ and $PT$ symmetry-breaking compensated magnetic orders with spin-split non-relativistic electronic spectra, represented in Figs.~\ref{fig:split_AHE}a,b,  can generate (in the presence of the spin-orbit coupling) the anomalous Hall effect\cite{Chen2014,Kubler2014,Nakatsuji2015,Kiyohara2015,Nayak2016,Smejkal2020,Feng2022,Reichlova2024,Betancourt2021,Wang2023a,Han2024,Kluczyk2024,Takagi2025,Ray2025}. The spin ordering in  the non-coplanar compensated magnets CoM$_3$S$_6$, illustrated in Fig.~\ref{fig:split_AHE}c, also  breaks the $tT$ and $PT$ symmetry and allows for the anomalous Hall effect (even in the absence of the spin-orbit coupling), while the non-relativistic electronic spectrum is spin-degenerate\cite{Takagi2023,Chen2024}. {\em Vice versa}, Figs.~\ref{fig:split_AHE}d,e show examples of non-collinear compensated spin orders with the $tT$-symmetry and broken $PT$-symmetry whose non-relativistic band structure is spin split while the anomalous Hall effect is excluded by symmetry (without or with the spin-orbit coupling)\cite{Hellenes2023}. Finally, in the collinear N\'eel antiferromagnet with the $tT$-symmetry and $PT$-symmetry\cite{Neel1971}, illustrated in Fig.~\ref{fig:split_AHE}f, both the spin splitting  and the anomalous Hall effect (without or with the spin-orbit coupling) are excluded by symmetry.

\smallskip
\noindent{\bf Spin-orbit coupling and topological phenomena}
\smallskip

\noindent We now focus on  spin-orbit coupling and topological phenomena in the electronic structure of altermagnets\cite{Krempasky2024,Fakhredine2023,Mazin2023a,Fang2024,Li2024b,Fernandes2023,Zhou2023,Zhan2023,Antonenko2025,Nag2024,Roig2024,Parshukov2024,Rao2024,Li2024,Lu2025,Zhu2023b,Ghorashi2024,Zhao2024}. The spin degeneracy at the nodal surfaces in the non-relativistic band structure of altermagnets can be lifted by the relativistic spin-orbit coupling. Among a range of phenomena further discussed in this section,  this can generate Berry-curvature hot spots (Fig.~\ref{fig:Topo}a) and, correspondingly,  large values of the anomalous Hall effect. This was initially theoretically predicted in  altermagnetic candidates RuO$_2$  \cite{Smejkal2020} or FeSb$_2$\cite{Mazin2021}, and reviewed in detail in Ref.~\onlinecite{Smejkal2022AHEReview}.

In the following paragraphs, we first give an example of the interplay of the spin-orbit coupling with the altermagnetic ordering on electronic energy bands of the 3D altermagnet MnTe and a 2D altermagnet FeSe\cite{Krempasky2024,Mazin2023a}. The feature that we highlight  is a possibility to realize  in altermagnets spin-orbit coupled band structures in which the spin-polarization axis is momentum-independent across high-symmetry directions, planes or the entire Brillouin zone. This is extraordinary given the general form, $\sim {\bf s}\cdot({\bf k}\times{\bf E})$, of the spin-orbit coupling term in the Dirac equation, where {\bf s} denotes spin, {\bf k} momentum and {\bf E} electric field. The coupling between spin and momentum vectors implies the tendency to form spin textures in the electronic structure where the magnitude and direction of spin varies  with momentum, as commonly observed in ferromagnets or ($P$-symmetry-breaking) non-magnetic materials. 

We contrast these spin textures with the effect of the spin-orbit coupling on  the  $k_z=0$ nodal plane in altermagnetic MnTe. This material has a spin-polarized band structure without spin-orbit coupling of the nodal $g$-wave type, thus displaying 4 nodal planes crossing the $\boldsymbol\Gamma$-point\cite{Smejkal2021a}. The spin degeneracy, protected at the nodal planes by the  symmetry of the spin point group $^26/^2m^2m^1m$, can be lifted by the spin-orbit coupling \cite{Krempasky2024}. Remarkably, the spin-orbit coupling can spin-split the bands despite the $P$-symmetry of altermagnetic MnTe. This already indicates the distinct phenomenology of spin-orbit coupling effects in altermagnets, compared to conventional spin splitting by the spin-orbit coupling which requires broken $P$-symmetry. 

When the N\'eel vector in MnTe is in the magnetic easy-plane ($c$-plane of the MnTe crystal shown in Fig.~\ref{fig:ARPES}a), the strong  $g$-wave order without spin-orbit coupling generates a corresponding in-plane spin-polarization component along the N\'eel vector away from the nodal planes. This is complemented on the $k_z=0$ nodal plane by an out-of-plane spin-polarization of the spin-orbit coupled bands, whose sign alternates and the polarization axis is independent of the in-plane momentum (Fig.~\ref{fig:ARPES}b-d) \cite{Krempasky2024}. We point out that the  absence of a momentum-dependent non-collinear spin texture on the $k_z=0$ plane is realized at energies in the band structure with a strong admixture of orbitals from the heavy element Te. Indeed the spin splitting in this part of the spin-orbit coupled band structure reaches large magnitudes $\sim 100$~meV (Fig.~\ref{fig:ARPES}c,f). 

The realization of such a common momentum-independent spin-polarization axis has been a long-sought goal in the research of spin-orbit coupling band-structure effects. Besides altermagnets, this was only considered in odd-parity spectra of non-magnetic non-centrosymmetric systems. Here in 3D, the effect was attributed  to (approximate) symmetry protection  in a vicinity of high-symmetry points of the Brillouin zone\cite{Tao2018}. In 2D, it was predicted to occur due to the presence of a mirror symmetry\cite{Ji2022}, or was observed for  fine-tuned strengths of microscopic Rashba and Dresselhaus spin-orbit coupling \cite{Bernevig2006,Koralek2009}.  

In altermagnetic MnTe,  the collinear  spin-polarization on the $k_z=0$ nodal plane, with all spins pointing along the  normal to the plane, is observed in even-parity spectra and is symmetry protected\cite{Krempasky2024}. Specifically, the absence of any in-plane spin-polarization component on the $k_z=0$ nodal plane is enforced by the mirror symmetry of the magnetic point group $m^\prime m^\prime m$, where the mirror plane is parallel to the crystal $c$-plane. An analogous effect was predicted in a supercell structure based on altermagnetic LaMnO$_3$\cite{Ji2022}.

We emphasize that the $m^\prime m^\prime m$ magnetic-group symmetry is only present in MnTe for the N\'eel vector oriented along the c-plane in the crystal\cite{Krempasky2024}. In contrast, in the $g$-wave altermagnet CrSb, which has identical crystallographic and  spin groups as MnTe\cite{Smejkal2021a}, the spin-orbit coupling leads to the N\'eel vector easy-axis pointing along the crystal $c$-axis. It changes the magnetic point group to $6'/m'mm'$ with a symmetry combining the  $c$-plane mirror with $T$ in CrSb. This CrSb symmetry does not enforce the absence of the in-plane  spin-polarization component in the $k_z=0$ plane; instead it enforces the absence of the out-of-plane spin-polarization component for states with momenta along the $k_x=k_y=0$ line (which turns out to be a completely spin-degenerate nodal line because of the interplay with another magnetic-group mirror symmetry)\cite{Fernandes2023}.

An additional extraordinary feature of the  spin-orbit coupling in the MnTe altermagnet with in-plane N\'eel vector is a quadratic band dispersion and spin splitting near the $\boldsymbol\Gamma$-point (Fig.~\ref{fig:ARPES}c,f). The absence of the constant and linear spin-splitting terms highlights the principal distinction  from the exchange spin splitting present in ferromagnets and the spin-orbit coupling induced spin splitting present in the non-magnetic $P$-symmetry-breaking crystals. 

The collinear spin-polarization is not a feature seen exclusively in the spin-orbit coupled band structure of MnTe. In Fig.~\ref{fig:Topo}e,f we show the electronic structure of a candidate 2D altermagnet FeSe \cite{Mazin2023a}. Without spin-orbit coupling (Fig.~\ref{fig:Topo}e), the spin-polarized 2D bands have a $d$-wave ordering. When spin-orbit coupling is included and the N\'eel vector is oriented in the direction normal to the 2D plane (Fig.~\ref{fig:Topo}f), spin-polarized states in valleys around {\bf M}-points in the Brillouin zone acquire a common spin axis, this time parallel to the N\'eel vector. For energy ranges where the spectrum contains only the {\bf M}-point valleys, the entire energy iso-surface in the 2D momentum space has a common momentum-independent spin-polarization axis. This is again despite the large spin-orbit-coupling strength introduced by Se, which generates a splitting in the {\bf M}-point valleys  on the $\sim 100$~meV scale (Fig.~\ref{fig:Topo}f).

We now move on to topological phenomena. In the non-collinear compensated magnets, the observation of the large anomalous Hall effect prompted the research of Weyl fermions and topological signatures in the magnetotransport\cite{Kuroda2017,Liu2018b,Xu2020b,Feng2020,Chen2021a,Takagi2023}. Below we highlight  topological band-structure phenomena which are characteristic of altermagnets\cite{Mazin2023a,Ghorashi2024,Fang2024,Li2024b,Fernandes2023,Zhou2023,Zhan2023,Antonenko2025,Nag2024,Roig2024,Parshukov2024,Rao2024,Li2024,Lu2025,Li2023,Zhu2023b,Zhao2024,Ghosh2025}. Several of these phenomena are related to the spin-degenerate nodal lines in the Brillouin zone, which are the remnants of the  nodal planes of the altermagnet when the spin-orbit coupling is included. They are topologically trivial with respect to non-spatial symmetries. When present, however, these Brillouin-zone nodal lines, and the corresponding Weyl nodes in the band structure, can be protected by mirror symmetries of the crystal\cite{Fernandes2023,Zhou2023,Zhan2023}. Therefore, they remain stable against small perturbations (e.g. by magnetic field or strain) that preserve the mirror symmetries \cite{Fernandes2023}. 

We again illustrate these and other topological phenomena  on representative altermagnetic materials. In CrSb, the splitting of the $k_z = 0$ nodal plane by the spin-orbit coupling leads to the emergence of pairs of Weyl points and thus Fermi surface arcs\cite{Li2024}. Here the Weyl points result from the crossing of  bands with opposite spin. Conversely, along the spin-split parts of the Brillouin zone in the  band structure without spin-orbit coupling, crossings of  bands with the same spin give rise to spin-polarized Weyl points. Fermi arcs connecting the surface projections of the Weyl points with the same total spin are also spin polarized\cite{Li2024,Lu2025}. In altermagnetic VNb$_3$S$_6$, signatures of linearly dispersing Weyl nodes were experimentally detected by Raman scattering spectroscopy\cite{Ghosh2025}.

An extraordinary interplay of topology and  spin-orbit coupling can be illustrated on the 2D altermagnetic candidate FeSe (Fig.~\ref{fig:Topo}e-g)  \cite{Mazin2023a}. Without spin-orbit coupling, the bands feature spin-degeneracy above and below the Fermi level at the {\bf M}-points which are located at the two nodal lines in the 2D Brillouin zone of this $d$-wave altermagnet (Fig.~\ref{fig:Topo}e). In addition, there are two crossings of bands with the same spin  at the Fermi level. They are located on {\bf M}-{\bf X} and {\bf M}-{\bf Y} lines (Fig.~\ref{fig:Topo}e), respectively, and are connected by the altermagnetic spin symmetries.  Spin-orbit coupling splits these same-spin band crossings and shifts further apart in energy the two spin-degenerate bands at the {\bf M}-point by $\sim 100$~meV, resulting in the formation of a 2D topological spin-Chern insulator  (Fig.~\ref{fig:Topo}f). As already highlighted above in the discussion of the  spin-orbit coupling effects in FeSe,  the spin polarization maintains a momentum-independent axis in the valleys around the {\bf M}-points, even for this relatively large strength of the spin-orbit coupling in FeSe.  This leads to an exceptionally precise quantization of the quantum spin Hall effect over a broad range of energies (Fig.~\ref{fig:Topo}g)\cite{Mazin2023a}.

Crossings of bands with the same spin, and their connection to non-trivial topology, are widespread in altermagnetic materials \cite{Antonenko2025,Parshukov2024,Roig2024}. Further insights about the topological nature of such crossings follow from analyzing the 2D Lieb lattice model (Fig.~\ref{fig:AM-FM}) \cite{Mazin2023a,Brekke2023,Antonenko2025,Kaushal2024}, of which monolayer FeSe is a particular material realization. A characteristic feature of this model, illustrated in Fig.~\ref{fig:Topo}d, is the existence of a quadratic band crossing  at the {\bf M}-point \cite{Sun2009}. Altermagnetic order splits the quadratic band crossing into pairs of crossings of same-spin bands (Dirac points in 2D) located at the {\bf M}-{\bf X} and {\bf M}-{\bf Y} zone edges and related by the $[C_2||C_4]$ spin symmetry (see inset of Fig.~\ref{fig:Topo}d). While these Dirac points are guaranteed to exist for an infinitesimally small sublattice magnetic moment, a large enough moment will bring the Dirac crossings to the {\bf X} and {\bf Y} points and remove them. The impact of the spin-orbit coupling depends on the direction of the N\'eel vector\cite{Antonenko2025}. For the out-of-plane direction, like in the case of the above discussed FeSe, the Dirac points are gapped out, resulting in the topological quantum spin Hall state and mirror spin-Chern bands.\cite{Mazin2023a,Antonenko2025}. For the in-plane N\'eel vector, however, even if the Dirac points are gapped, the bands are topologically trivial. This illustrates that the direction of the N\'eel vector has a substantial impact on the opposite-spin and same-spin band crossings. In 3D lattices, same-spin band crossings can also occur, where they can give rise to Weyl nodal loops in the presence of spin-orbit coupling \cite{Antonenko2025}.

Another promising route to imprint unusual electronic properties in altermagnetic materials is to combine their characteristic anisotropic nodal spin ordering with other phases, such as quantum anomalous Hall Chern insulators, axion insulators, multiferroics or superconductors\cite{Smejkal2021a,Smejkal2022a,Mazin2025,Mazin2023a,Beenakker2023,Brekke2023,Li2023,Zhu2023b,Sumita2023,Ghorashi2024,Papaj2023,Wei2024,Zhang2023,Cheng2024,Zhao2024,Maeland2024,Chakraborty2024b,Banerjee2024a,Jeschke2024,Verbeek2023,Bernardini2025,Zyuzin2024,Sim2025,Hu2025,Smejkal2024,Duan2025,Gu2025,Parthenios2025}.
The interplay between superconductivity and altermagnetism or altermagnetic fluctuations has been investigated as intrinsic phenomena in a single material or in the context of heterostructures, where pairing is induced by the proximity effect, and bulk systems, where pairing and altermagnetic ordering coexist. These studies have revealed the emergence of intriguing phenomena, such as unconventional Andreev reflection, pair density waves, non-trivial topological modes, and non-reciprocal supercurrents.

\bigskip
\noindent{\large\bf Outlook}
\bigskip

\noindent The salient band structure features of altermagnets can have unparalleled practical utility in spintronics, sharing the merits of the vanishing net magnetization with the earlier studied collinear and non-collinear compensated magnets, and sharing the merits of the well separated and conserved spin-up and spin-down channels with the conventional ferromagnets. 

Altermagnetism can lead to unique spin-orbit coupling and topological phenomena, such as spin polarizations that are collinear despite their spin-orbit coupling origin, or topological Weyl points that reflect the specific altermagnetic spin-group symmetries. In addition, altermagnets represent a unique platform to search for quantized topological responses at zero magnetic field and high temperatures, or to  combine this anisotropic compensated $T$-symmetry-breaking magnetic ordering with superconductivity.

Notwithstanding the remarkable and rapid progress in the field, several open questions remain and new research directions emerge on the horizon. For instance, it is an intriguing question whether an altermagnetic quantum critical point can promote new phenomena\cite{Steward2023,Ray2025}, perhaps related to non-Fermi liquid behavior and unconventional superconductivity, that is not encountered in other widely studied quantum-critical ferroic orders, like ferromagnetism, ferroelectricity, and nematicity. A recent report on the coexistence of the strange metallic behavior and the anomalous Hall effect in an altermagnetic phase of a Kondo  lattice-like flat-band system\cite{Ray2025} illustrates the emergence of a new research path exploring the impact of correlations and bad metal behavior on altermagnetic properties of strongly-correlated materials. 

Finally, an extensive experimental effort, complementing the parallel fruitful research of other compensated magnetic phases,  is needed to establish all the predicted specific symmetry, microscopy and spectroscopy signatures of altermagnetism, to further explore the corresponding unconventional electronic and optical responses, and to exploit them in novel devices.

\bigskip

\noindent {\bf Acknowledgments}
We acknowledge Rodrigo Jaeschke-Ubiergo and Anna Birk Hellenes for helpful discussions and sharing unpublished results. TJ acknowledges support by the Ministry of Education of the Czech Republic CZ.02.01.01/00/22008/0004594 and ERC Advanced Grant no. 101095925. RMF acknowledges support by the Air Force Office of Scientific Research under Award No. FA9550-21-1-0423. JS and LŠ acknowledge support by Deutsche Forschungsgemeinschaft (DFG, German Research Foundation) - DFG (Project 452301518) and TRR 288 – 422213477 (project A09). LŠ acknowledges support by the ERC Starting Grant No. 101165122. SN acknowledges support  by JST-MIRAI Program (JPMJMI20A1), JST-ASPIRE Program (JPMJAP2317) and by the fund made by Canadian Institute for Advanced Research; the work at the Institute for Quantum Matter was funded by DOE, Office of Science, Basic Energy Sciences under Award \# DE-SC0024469. SM acknowledges support  by Japan Society for the Promotion of Science (JSPS), KAKENHI Grant No.JP22H00108, and No.JP24H02231.

%\bibliographystyle{naturemag}  % ama, nar, alpha, plain, chicago, abbrv, siam
%\bibliography{Refs}

\begin{thebibliography}{100}
\expandafter\ifx\csname url\endcsname\relax
  \def\url#1{\texttt{#1}}\fi
\expandafter\ifx\csname urlprefix\endcsname\relax\def\urlprefix{URL }\fi
\providecommand{\bibinfo}[2]{#2}
\providecommand{\eprint}[2][]{\url{#2}}

\bibitem{Smejkal2021a}
\bibinfo{author}{{\v{S}}mejkal, L.}, \bibinfo{author}{Sinova, J.} \&
  \bibinfo{author}{Jungwirth, T.}
\newblock \bibinfo{title}{{Beyond Conventional Ferromagnetism and
  Antiferromagnetism: A Phase with Nonrelativistic Spin and Crystal Rotation
  Symmetry}}.
\newblock \emph{\bibinfo{journal}{Physical Review X}}
  \textbf{\bibinfo{volume}{12}}, \bibinfo{pages}{031042}
  (\bibinfo{year}{2022}).

\newblock {\bf This paper delineated altermagnetism as an exclusively distinct spin-group symmetry class of d, g or i-wave spin-ordered phases.}  
%\newblock \urlprefix\url{https://link.aps.org/doi/10.1103/PhysRevX.12.031042}.
%\newblock \eprint{2105.05820}.

\bibitem{Smejkal2022a}
\bibinfo{author}{{\v{S}}mejkal, L.}, \bibinfo{author}{Sinova, J.} \&
  \bibinfo{author}{Jungwirth, T.}
\newblock \bibinfo{title}{{Emerging Research Landscape of Altermagnetism}}.
\newblock \emph{\bibinfo{journal}{Physical Review X}}
  \textbf{\bibinfo{volume}{12}}, \bibinfo{pages}{040501}
  (\bibinfo{year}{2022}).
%\newblock \urlprefix\url{http://arxiv.org/abs/2204.10844
%  https://link.aps.org/doi/10.1103/PhysRevX.12.040501}.
%\newblock \eprint{2204.10844}.

\newblock {\bf This perspective outlined  envisaged research directions of altermagnetism.}  

\bibitem{Smejkal2020}
\bibinfo{author}{{\v{S}}mejkal, L.},
  \bibinfo{author}{Gonz{\'{a}}lez-Hern{\'{a}}ndez, R.},
  \bibinfo{author}{Jungwirth, T.} \& \bibinfo{author}{Sinova, J.}
\newblock \bibinfo{title}{{Crystal time-reversal symmetry breaking and
  spontaneous Hall effect in collinear antiferromagnets}}.
\newblock \emph{\bibinfo{journal}{Science Advances}}
  \textbf{\bibinfo{volume}{6}}, \bibinfo{pages}{eaaz8809}
  (\bibinfo{year}{2020}).
%\newblock \urlprefix\url{https://www.science.org/doi/10.1126/sciadv.aaz8809}.
%\newblock \eprint{1901.00445}.

\newblock {\bf This paper predicted that an interplay of crystal and collinear spin orders in a class of materials can lead to a time-reversal symmetry breaking spin-split electronic structure and, when spin-orbit coupling is included, to the anomalous Hall effect.}  

\bibitem{Jungwirth2025}
\bibinfo{author}{Jungwirth, T.} \emph{et~al.}
\newblock \bibinfo{title}{{Altermagnetism: An unconventional spin-ordered phase
  of matter}}.
\newblock \emph{\bibinfo{journal}{Newton}} \textbf{\bibinfo{volume}{1}},
  \bibinfo{pages}{100162} (\bibinfo{year}{2025}).
%\newblock
%  \urlprefix\url{https://linkinghub.elsevier.com/retrieve/pii/S2950636025001549}.

\bibitem{Mazin2025}
\bibinfo{author}{Mazin, I.~I.}
\newblock \bibinfo{title}{{Notes on altermagnetism and superconductivity}}.
\newblock \emph{\bibinfo{journal}{AAPPS Bulletin}}
  \textbf{\bibinfo{volume}{35}}, \bibinfo{pages}{18} (\bibinfo{year}{2025}).
%\newblock \urlprefix\url{https://link.springer.com/10.1007/s43673-025-00158-6}.

\newblock {\bf These informal notes, originally posted on arXiv:2203.05000, aimed to stimulate the exploration of the interplay of altermagnetism with superconductivity.} 
\bibitem{Beenakker2023}
\bibinfo{author}{Beenakker, C. W.~J.} \& \bibinfo{author}{Vakhtel, T.}
\newblock \bibinfo{title}{{Phase-shifted Andreev levels in an altermagnet
  Josephson junction}}.
\newblock \emph{\bibinfo{journal}{Physical Review B}}
  \textbf{\bibinfo{volume}{108}}, \bibinfo{pages}{075425}
  (\bibinfo{year}{2023}).
%\newblock \urlprefix\url{http://arxiv.org/abs/2306.16300
%  http://dx.doi.org/10.1103/PhysRevB.108.075425
%  https://link.aps.org/doi/10.1103/PhysRevB.108.075425}.
%\newblock \eprint{2306.16300}.

\bibitem{Mazin2023a}
\bibinfo{author}{Mazin, I.}, \bibinfo{author}{Gonz{\'{a}}lez-Hern{\'{a}}ndez,
  R.} \& \bibinfo{author}{{\v{S}}mejkal, L.}
\newblock \bibinfo{title}{{Induced Monolayer Altermagnetism in MnP(S,Se)$_3$
  and FeSe}}.
\newblock \emph{\bibinfo{journal}{Arxiv Preprint}} \bibinfo{pages}{1--11}
  (\bibinfo{year}{2023}).
\newblock \urlprefix\url{http://arxiv.org/abs/2309.02355}.
%\newblock \eprint{2309.02355}.

\bibitem{Brekke2023}
\bibinfo{author}{Brekke, B.}, \bibinfo{author}{Brataas, A.} \&
  \bibinfo{author}{Sudb{\o}, A.}
\newblock \bibinfo{title}{{Two-dimensional altermagnets: Superconductivity in a
  minimal microscopic model}}.
\newblock \emph{\bibinfo{journal}{Physical Review B}}
  \textbf{\bibinfo{volume}{108}}, \bibinfo{pages}{224421}
  (\bibinfo{year}{2023}).
%\newblock \urlprefix\url{http://arxiv.org/abs/2308.08606
%  https://link.aps.org/doi/10.1103/PhysRevB.108.224421}.
%\newblock \eprint{2308.08606}.

\bibitem{Li2023}
\bibinfo{author}{Li, Y.-X.} \& \bibinfo{author}{Liu, C.-C.}
\newblock \bibinfo{title}{{Majorana corner modes and tunable patterns in an
  altermagnet heterostructure}}.
\newblock \emph{\bibinfo{journal}{Physical Review B}}
  \textbf{\bibinfo{volume}{108}}, \bibinfo{pages}{205410}
  (\bibinfo{year}{2023}).
%\newblock \urlprefix\url{https://link.aps.org/doi/10.1103/PhysRevB.108.205410}.

\bibitem{Zhu2023b}
\bibinfo{author}{Zhu, D.}, \bibinfo{author}{Zhuang, Z.-Y.},
  \bibinfo{author}{Wu, Z.} \& \bibinfo{author}{Yan, Z.}
\newblock \bibinfo{title}{{Topological superconductivity in two-dimensional
  altermagnetic metals}}.
\newblock \emph{\bibinfo{journal}{Physical Review B}}
  \textbf{\bibinfo{volume}{108}}, \bibinfo{pages}{184505}
  (\bibinfo{year}{2023}).
%\newblock \urlprefix\url{http://arxiv.org/abs/2305.10479
%  https://link.aps.org/doi/10.1103/PhysRevB.108.184505}.
%\newblock \eprint{2305.10479}.

\bibitem{Sumita2023}
\bibinfo{author}{Sumita, S.}, \bibinfo{author}{Naka, M.} \&
  \bibinfo{author}{Seo, H.}
\newblock \bibinfo{title}{{Fulde-Ferrell-Larkin-Ovchinnikov state induced by
  antiferromagnetic order in k-type organic conductors}}.
\newblock \emph{\bibinfo{journal}{Physical Review Research}}
  \textbf{\bibinfo{volume}{5}}, \bibinfo{pages}{043171} (\bibinfo{year}{2023}).
%\newblock
%  \urlprefix\url{https://link.aps.org/doi/10.1103/PhysRevResearch.5.043171}.
%\newblock \eprint{2308.14227}.

\bibitem{Ghorashi2024}
\bibinfo{author}{Ghorashi, S. A.~A.}, \bibinfo{author}{Hughes, T.~L.} \&
  \bibinfo{author}{Cano, J.}
\newblock \bibinfo{title}{{Altermagnetic Routes to Majorana Modes in Zero Net
  Magnetization}}.
\newblock \emph{\bibinfo{journal}{Physical Review Letters}}
  \textbf{\bibinfo{volume}{133}}, \bibinfo{pages}{106601}
  (\bibinfo{year}{2024}).
%\newblock
%  \urlprefix\url{https://link.aps.org/doi/10.1103/PhysRevLett.133.106601}.

\bibitem{Papaj2023}
\bibinfo{author}{Papaj, M.}
\newblock \bibinfo{title}{{Andreev reflection at the altermagnet-superconductor
  interface}}.
\newblock \emph{\bibinfo{journal}{Physical Review B}}
  \textbf{\bibinfo{volume}{108}}, \bibinfo{pages}{1--7} (\bibinfo{year}{2023}).
%\newblock \urlprefix\url{http://arxiv.org/abs/2305.03856
%  https://link.aps.org/doi/10.1103/PhysRevB.108.L060508}.
%\newblock \eprint{2305.03856}.

\bibitem{Wei2024}
\bibinfo{author}{Wei, M.} \emph{et~al.}
\newblock \bibinfo{title}{{Gapless superconducting state and mirage gap in
  altermagnets}}.
\newblock \emph{\bibinfo{journal}{Physical Review B}}
  \textbf{\bibinfo{volume}{109}}, \bibinfo{pages}{L201404}
  (\bibinfo{year}{2023}).
%\newblock \urlprefix\url{http://arxiv.org/abs/2308.00248
%  https://link.aps.org/doi/10.1103/PhysRevB.109.L201404}.
%\newblock \eprint{2308.00248}.

\bibitem{Zhang2023}
\bibinfo{author}{Zhang, S.-B.}, \bibinfo{author}{Hu, L.-H.} \&
  \bibinfo{author}{Neupert, T.}
\newblock \bibinfo{title}{{Finite-momentum Cooper pairing in proximitized
  altermagnets}}.
\newblock \emph{\bibinfo{journal}{Nature Communications}}
  \textbf{\bibinfo{volume}{15}}, \bibinfo{pages}{1801} (\bibinfo{year}{2024}).
%\newblock \urlprefix\url{http://arxiv.org/abs/2302.13185
%  https://www.nature.com/articles/s41467-024-45951-3}.
%\newblock \eprint{2302.13185}.

\bibitem{Cheng2024}
\bibinfo{author}{Cheng, Q.} \& \bibinfo{author}{Sun, Q.-F.}
\newblock \bibinfo{title}{{Orientation-dependent Josephson effect in
  spin-singlet superconductor/altermagnet/spin-triplet superconductor
  junctions}}.
\newblock \emph{\bibinfo{journal}{Physical Review B}}
  \textbf{\bibinfo{volume}{109}}, \bibinfo{pages}{024517}
  (\bibinfo{year}{2024}).
%\newblock
%  \urlprefix\url{http://arxiv.org/abs/2402.02810%0Ahttp://dx.doi.org/10.1103/PhysRevB.109.024517
%  https://link.aps.org/doi/10.1103/PhysRevB.109.024517}.
%\newblock \eprint{2402.02810}.

\bibitem{Zhao2024}
\bibinfo{author}{Zhao, Y.} \emph{et~al.}
\newblock \bibinfo{title}{{Hybrid-order topology in unconventional magnets of
  Eu-based Zintl compounds with surface-dependent quantum geometry}}.
\newblock \emph{\bibinfo{journal}{Physical Review B}}
  \textbf{\bibinfo{volume}{110}}, \bibinfo{pages}{205111}
  (\bibinfo{year}{2024}).
%\newblock \urlprefix\url{http://arxiv.org/abs/2403.06304
%  https://link.aps.org/doi/10.1103/PhysRevB.110.205111}.
%\newblock \eprint{2403.06304}.

\bibitem{Maeland2024}
\bibinfo{author}{M{\ae}land, K.}, \bibinfo{author}{Brekke, B.} \&
  \bibinfo{author}{Sudb{\o}, A.}
\newblock \bibinfo{title}{{Many-body effects on superconductivity mediated by
  double-magnon processes in altermagnets}}.
\newblock \emph{\bibinfo{journal}{Physical Review B}}
  \textbf{\bibinfo{volume}{109}}, \bibinfo{pages}{134515}
  (\bibinfo{year}{2024}).
%\newblock \urlprefix\url{http://arxiv.org/abs/2402.14061
%  http://dx.doi.org/10.1103/PhysRevB.109.134515
%  https://link.aps.org/doi/10.1103/PhysRevB.109.134515}.
%\newblock \eprint{2402.14061}.

\bibitem{Chakraborty2024b}
\bibinfo{author}{Chakraborty, D.} \& \bibinfo{author}{Black-Schaffer, A.~M.}
\newblock \bibinfo{title}{{Zero-field finite-momentum and field-induced
  superconductivity in altermagnets}}.
\newblock \emph{\bibinfo{journal}{Physical Review B}}
  \textbf{\bibinfo{volume}{110}}, \bibinfo{pages}{L060508}
  (\bibinfo{year}{2024}).
%\newblock
%  \urlprefix\url{https://link.aps.org/doi/10.1103/PhysRevB.110.L060508}.

\bibitem{Banerjee2024a}
\bibinfo{author}{Banerjee, S.} \& \bibinfo{author}{Scheurer, M.~S.}
\newblock \bibinfo{title}{{Altermagnetic superconducting diode effect}}.
\newblock \emph{\bibinfo{journal}{Physical Review B}}
  \textbf{\bibinfo{volume}{110}}, \bibinfo{pages}{024503}
  (\bibinfo{year}{2024}).
%\newblock \urlprefix\url{http://arxiv.org/abs/2402.14071
 % http://dx.doi.org/10.1103/PhysRevB.110.024503
%  https://link.aps.org/doi/10.1103/PhysRevB.110.024503}.
%\newblock \eprint{2402.14071}.

\bibitem{Jeschke2024}
\bibinfo{author}{Jeschke, H.~O.}, \bibinfo{author}{Shimizu, M.} \&
  \bibinfo{author}{Mazin, I.~I.}
\newblock \bibinfo{title}{{CuAg(SO4)2: A doubly strongly correlated
  altermagnetic three-dimensional analog of the parent compounds of high-$T_c$ cuprates}}.
\newblock \emph{\bibinfo{journal}{Physical Review B}}
  \textbf{\bibinfo{volume}{109}}, \bibinfo{pages}{L220412}
  (\bibinfo{year}{2024}).
%\newblock \urlprefix\url{http://arxiv.org/abs/2403.02201
%  https://link.aps.org/doi/10.1103/PhysRevB.109.L220412}.
%\newblock \eprint{2403.02201}.

\bibitem{Verbeek2023}
\bibinfo{author}{Verbeek, X.~H.}, \bibinfo{author}{Urru, A.} \&
  \bibinfo{author}{Spaldin, N.~A.}
\newblock \bibinfo{title}{{Hidden orders and (anti-)magnetoelectric effects in
  Cr$_2$O$_3$ and Fe$_2$O$_3$}}.
\newblock \emph{\bibinfo{journal}{Physical Review Research}}
  \textbf{\bibinfo{volume}{5}}, \bibinfo{pages}{L042018}
  (\bibinfo{year}{2023}).
%\newblock
%  \urlprefix\url{https://link.aps.org/doi/10.1103/PhysRevResearch.5.L042018}.

\bibitem{Bernardini2025}
\bibinfo{author}{Bernardini, F.}, \bibinfo{author}{Fiebig, M.} \&
  \bibinfo{author}{Cano, A.}
\newblock \bibinfo{title}{{Ruddlesden–Popper and perovskite phases as a
  material platform for altermagnetism}}.
\newblock \emph{\bibinfo{journal}{Journal of Applied Physics}}
  \textbf{\bibinfo{volume}{137}} (\bibinfo{year}{2025}).
%\newblock
%  \urlprefix\url{https://pubs.aip.org/jap/article/137/10/103903/3339285/Ruddlesden-Popper-and-perovskite-phases-as-a}.

\bibitem{Zyuzin2024}
\bibinfo{author}{Zyuzin, A.~A.}
\newblock \bibinfo{title}{{Magnetoelectric effect in superconductors with
  d-wave magnetization}}.
\newblock \emph{\bibinfo{journal}{Physical Review B}}
  \textbf{\bibinfo{volume}{109}}, \bibinfo{pages}{L220505}
  (\bibinfo{year}{2024}).
%\newblock
%  \urlprefix\url{https://link.aps.org/doi/10.1103/PhysRevB.109.L220505}.

\bibitem{Sim2025}
\bibinfo{author}{Sim, G.} \& \bibinfo{author}{Knolle, J.}
\newblock \bibinfo{title}{{Pair density waves and supercurrent diode effect in
  altermagnets}}.
\newblock \emph{\bibinfo{journal}{Physical Review B}}
  \textbf{\bibinfo{volume}{112}}, \bibinfo{pages}{L020502}
  (\bibinfo{year}{2025}).
%\newblock \urlprefix\url{https://link.aps.org/doi/10.1103/b7rh-v7nq}.

\bibitem{Hu2025}
\bibinfo{author}{Hu, J.-X.}, \bibinfo{author}{Matsyshyn, O.} \&
  \bibinfo{author}{Song, J. C.~W.}
\newblock \bibinfo{title}{{Nonlinear Superconducting Magnetoelectric Effect}}.
\newblock \emph{\bibinfo{journal}{Physical Review Letters}}
  \textbf{\bibinfo{volume}{134}}, \bibinfo{pages}{026001}
  (\bibinfo{year}{2025}).
%\newblock
%  \urlprefix\url{https://link.aps.org/doi/10.1103/PhysRevLett.134.026001}.

\bibitem{Smejkal2024}
\bibinfo{author}{{\v{S}}mejkal, L.}
\newblock \bibinfo{title}{{Altermagnetic multiferroics and altermagnetoelectric
  effect}}  (\bibinfo{year}{2024}).
\newblock \urlprefix\url{http://arxiv.org/abs/2411.19928}.
%\newblock \eprint{2411.19928}.

\newblock {\bf This paper and papers in Refs. 28 and 29 theoretically predicted an interplay of altermagnetism with ferroelectricity.}  

\bibitem{Duan2025}
\bibinfo{author}{Duan, X.} \emph{et~al.}
\newblock \bibinfo{title}{{Antiferroelectric Altermagnets: Antiferroelectricity
  Alters Magnets}}.
\newblock \emph{\bibinfo{journal}{Physical Review Letters}}
  \textbf{\bibinfo{volume}{134}}, \bibinfo{pages}{106801}
  (\bibinfo{year}{2025}).
%\newblock
%  \urlprefix\url{https://link.aps.org/doi/10.1103/PhysRevLett.134.106801}.

\bibitem{Gu2025}
\bibinfo{author}{Gu, M.} \emph{et~al.}
\newblock \bibinfo{title}{{Ferroelectric switchable altermagnetism}}.
\newblock \emph{\bibinfo{journal}{Physical Review Letters}}
  \textbf{\bibinfo{volume}{134}}, \bibinfo{pages}{106802}
  (\bibinfo{year}{2024}).
%\newblock \urlprefix\url{http://arxiv.org/abs/2411.14216
%  https://link.aps.org/doi/10.1103/PhysRevLett.134.106802}.
%\newblock \eprint{2411.14216}.

\bibitem{Parthenios2025}
\bibinfo{author}{Parthenios, N.} \emph{et~al.}
\newblock \bibinfo{title}{{Spin and pair density waves in 2D altermagnetic
  metals}}  (\bibinfo{year}{2025}).
\newblock \urlprefix\url{http://arxiv.org/abs/2502.19270}.
%\newblock \eprint{2502.19270}.

\bibitem{Smejkal2022AHEReview}
\bibinfo{author}{{\v{S}}mejkal, L.}, \bibinfo{author}{MacDonald, A.~H.},
  \bibinfo{author}{Sinova, J.}, \bibinfo{author}{Nakatsuji, S.} \&
  \bibinfo{author}{Jungwirth, T.}
\newblock \bibinfo{title}{{Anomalous Hall antiferromagnets}}.
\newblock \emph{\bibinfo{journal}{Nature Reviews Materials}}
  \textbf{\bibinfo{volume}{7}}, \bibinfo{pages}{482--496}
  (\bibinfo{year}{2022}).
%\newblock \urlprefix\url{http://arxiv.org/abs/2107.03321
%  https://www.nature.com/articles/s41578-022-00430-3}.
%\newblock \eprint{2107.03321}.

\newblock {\bf This paper reviewed  the anomalous Hall effect in altermagnets and non-collinear compensated magnets.}  

\bibitem{Bai2024}
\bibinfo{author}{Bai, L.} \emph{et~al.}
\newblock \bibinfo{title}{{Altermagnetism: Exploring New Frontiers in Magnetism
  and Spintronics}}.
\newblock \emph{\bibinfo{journal}{Advanced Functional Materials}}
  \bibinfo{pages}{1--49} (\bibinfo{year}{2024}).
%\newblock
%  \urlprefix\url{https://onlinelibrary.wiley.com/doi/10.1002/adfm.202409327
%  http://arxiv.org/abs/2406.02123}.
%\newblock \eprint{2406.02123}.

\bibitem{Liu2025}
\bibinfo{author}{Liu, Q.}, \bibinfo{author}{Dai, X.} \&
  \bibinfo{author}{Bl{\"{u}}gel, S.}
\newblock \bibinfo{title}{{Different facets of unconventional magnetism}}.
\newblock \emph{\bibinfo{journal}{Nature Physics}}
  \textbf{\bibinfo{volume}{21}}, \bibinfo{pages}{329--331}
  (\bibinfo{year}{2025}).
%\newblock \urlprefix\url{https://www.nature.com/articles/s41567-024-02750-3}.

\bibitem{Song2025}
\bibinfo{author}{Song, C.} \emph{et~al.}
\newblock \bibinfo{title}{{Altermagnets as a new class of functional
  materials}}.
\newblock \emph{\bibinfo{journal}{Nature Reviews Materials}}
  (\bibinfo{year}{2025}).
%\newblock \urlprefix\url{https://www.nature.com/articles/s41578-025-00779-1}.

\bibitem{Jungwirth2025b}
\bibinfo{author}{Jungwirth, T.} \emph{et~al.}
\newblock \bibinfo{title}{{Altermagnetic spintronics}}  (\bibinfo{year}{2025}).
\newblock \urlprefix\url{http://arxiv.org/abs/2508.09748}.
%\newblock \eprint{2508.09748}.

\bibitem{Liu2021}
\bibinfo{author}{Liu, P.}, \bibinfo{author}{Li, J.}, \bibinfo{author}{Han, J.},
  \bibinfo{author}{Wan, X.} \& \bibinfo{author}{Liu, Q.}
\newblock \bibinfo{title}{{Spin-Group Symmetry in Magnetic Materials with
  Negligible Spin-Orbit Coupling}}.
\newblock \emph{\bibinfo{journal}{Physical Review X}}
  \textbf{\bibinfo{volume}{12}}, \bibinfo{pages}{21016} (\bibinfo{year}{2022}).
%\newblock \urlprefix\url{http://arxiv.org/abs/2103.15723
%  https://link.aps.org/doi/10.1103/PhysRevX.12.021016}.
%\newblock \eprint{2103.15723}.

\newblock {\bf This paper discussed spin space group symmetries of magnetic materials and corresponding band degeneracies and emergent topological phases.}  

\bibitem{Litvin1974}
\bibinfo{author}{Litvin, D.~B.} \& \bibinfo{author}{Opechowski, W.}
\newblock \bibinfo{title}{{Spin groups}}.
\newblock \emph{\bibinfo{journal}{Physica}} \textbf{\bibinfo{volume}{76}},
  \bibinfo{pages}{538--554} (\bibinfo{year}{1974}).
%\newblock
%  \urlprefix\url{https://linkinghub.elsevier.com/retrieve/pii/0031891474901578
%  https://www.sciencedirect.com/science/article/abs/pii/0031891474901578?via%3Dihub}.

\newblock {\bf This paper described the mathematical spin-group formalism.} 

\bibitem{Andreev1984}
\bibinfo{author}{Andreev, A.} \& \bibinfo{author}{Grishchuk, I.}
\newblock \bibinfo{title}{{Spin nematics}}.
\newblock \emph{\bibinfo{journal}{Sov. Phys. JETP}}
  \textbf{\bibinfo{volume}{60}}, \bibinfo{pages}{267} (\bibinfo{year}{1984}).

\bibitem{Gorkov1992}
\bibinfo{author}{Gor'kov, L.~P.} \& \bibinfo{author}{Sokol, A.}
\newblock \bibinfo{title}{{Nontrivial magnetic order: Localized versus
  itinerant systems}}.
\newblock \emph{\bibinfo{journal}{Physical Review Letters}}
  \textbf{\bibinfo{volume}{69}}, \bibinfo{pages}{2586--2589}
  (\bibinfo{year}{1992}).
%\newblock \urlprefix\url{https://link.aps.org/doi/10.1103/PhysRevLett.69.2586}.

\bibitem{Leggett2003}
\bibinfo{author}{Leggett, A.~J.}
\newblock \bibinfo{title}{{Nobel Lecture: Superfluid 3-He: The early days as
  seen by a theorist}}.
\newblock \emph{\bibinfo{journal}{Reviews of Modern Physics}}
  \textbf{\bibinfo{volume}{76}}, \bibinfo{pages}{999} (\bibinfo{year}{2004}).
%\newblock
%  \urlprefix\url{https://www.nobelprize.org/uploads/2018/06/leggett-lecture.pdf
%  https://link.aps.org/doi/10.1103/RevModPhys.76.999}.

\newblock {\bf This paper reviewed the superfluid phase of $^3$He, including the spontaneous breaking of real-space and spin-space rotation symmetries.} 

\bibitem{Moessner2021}
\bibinfo{author}{Moessner, R.} \& \bibinfo{author}{Moore, J.~E.}
\newblock \emph{\bibinfo{title}{{Topological Phases of Matter}}}
  (\bibinfo{publisher}{Cambridge University Press}, \bibinfo{year}{2021}).
%\newblock
%  \urlprefix\url{https://www.cambridge.org/core/product/identifier/9781316226308/type/book}.

\bibitem{Strange1998}
\bibinfo{author}{Strange, P.}
\newblock \emph{\bibinfo{title}{{Relativistic Quantum Mechanics}}}
  (\bibinfo{publisher}{Cambridge University Press},
  \bibinfo{address}{Cambridge}, \bibinfo{year}{1998}), \bibinfo{edition}{1}
  edn.
%\newblock
%  \urlprefix\url{https://books.google.de/books?id=sdVrBM2w0OwC&printsec=frontcover&dq=Relativistic+quantum+mechanics&hl=en&sa=X&ved=0ahUKEwjvitehpeniAhVusYsKHXTsDFYQ6AEIMzAC#v=onepage&q=Relativistic quantum mechanics&f=false http://ebooks.cambridge.org/ref/id/CBO978051162}.

\bibitem{Winkler2003}
\bibinfo{author}{Winkler, R.}
\newblock \emph{\bibinfo{title}{{Spin—Orbit Coupling Effects in
  Two-Dimensional Electron and Hole Systems}}}, vol. \bibinfo{volume}{191} of
  \emph{\bibinfo{series}{Springer Tracts in Modern Physics}}
  (\bibinfo{publisher}{Springer Berlin Heidelberg}, \bibinfo{address}{Berlin,
  Heidelberg}, \bibinfo{year}{2003}).
%\newblock \urlprefix\url{http://link.springer.com/10.1007/b13586
%  https://link.springer.com/10.1007/b13586}.

\bibitem{Landau1984}
\bibinfo{author}{Landau, L.} \& \bibinfo{author}{Lifshitz, E.}
\newblock \emph{\bibinfo{title}{{Electrodynamics of Continuous Media, vol. 8 of
  Course of Theoretical Physics}}} (\bibinfo{publisher}{Pergamon Press,
  Oxford}, \bibinfo{year}{1965}), \bibinfo{edition}{2nd} edn.

\bibitem{Sinova2015}
\bibinfo{author}{Sinova, J.}, \bibinfo{author}{Valenzuela, S.~O.},
  \bibinfo{author}{Wunderlich, J.}, \bibinfo{author}{Back, C.~H.} \&
  \bibinfo{author}{Jungwirth, T.}
\newblock \bibinfo{title}{{Spin Hall effects}}.
\newblock \emph{\bibinfo{journal}{Reviews of Modern Physics}}
  \textbf{\bibinfo{volume}{87}}, \bibinfo{pages}{1213--1260}
  (\bibinfo{year}{2015}).
%\newblock \urlprefix\url{https://link.aps.org/doi/10.1103/RevModPhys.87.1213}.

\bibitem{Nagaosa2010}
\bibinfo{author}{Nagaosa, N.}, \bibinfo{author}{Sinova, J.},
  \bibinfo{author}{Onoda, S.}, \bibinfo{author}{MacDonald, A.~H.} \&
  \bibinfo{author}{Ong, N.~P.}
\newblock \bibinfo{title}{{Anomalous Hall effect}}.
\newblock \emph{\bibinfo{journal}{Reviews of Modern Physics}}
  \textbf{\bibinfo{volume}{82}}, \bibinfo{pages}{1539--1592}
  (\bibinfo{year}{2010}).
%\newblock \urlprefix\url{https://link.aps.org/doi/10.1103/RevModPhys.82.1539}.
%\newblock \eprint{0904.4154}.

\bibitem{Franz2013}
\bibinfo{editor}{Franz, M.} \& \bibinfo{editor}{Molenkamp, L.} (eds.)
  \emph{\bibinfo{title}{{Contemporary Concepts of Condensed Matter Science,
  vol. 6 - Topological Insulators}}} (\bibinfo{publisher}{Elsevier},
  \bibinfo{year}{2013}).

\bibitem{Murakami2017a}
\bibinfo{author}{Murakami, S.} \& \bibinfo{author}{Yokoyama, T.}
\newblock \emph{\bibinfo{title}{{Quantum spin Hall effect and topological
  insulators}}}, vol.~\bibinfo{volume}{1} (\bibinfo{publisher}{Oxford
  University Press}, \bibinfo{year}{2017}).
%\newblock
%  \urlprefix\url{https://academic.oup.com/book/27906/chapter/203910964}.

\bibitem{Bradlyn2017}
\bibinfo{author}{Bradlyn, B.} \emph{et~al.}
\newblock \bibinfo{title}{{Topological quantum chemistry}}.
\newblock \emph{\bibinfo{journal}{Nature}} \textbf{\bibinfo{volume}{547}},
  \bibinfo{pages}{298--305} (\bibinfo{year}{2017}).
%\newblock \urlprefix\url{http://dx.doi.org/10.1038/nature23268
%  http://www.nature.com/articles/nature23268
%  https://www.nature.com/articles/nature23268}.
%\newblock \eprint{1703.02050}.

\bibitem{Armitage2018}
\bibinfo{author}{Armitage, N.~P.}, \bibinfo{author}{Mele, E.~J.} \&
  \bibinfo{author}{Vishwanath, A.}
\newblock \bibinfo{title}{{Weyl and Dirac semimetals in three-dimensional
  solids}}.
\newblock \emph{\bibinfo{journal}{Reviews of Modern Physics}}
  \textbf{\bibinfo{volume}{90}}, \bibinfo{pages}{015001}
  (\bibinfo{year}{2018}).
%\newblock
%  \urlprefix\url{https://link.aps.org/doi/10.1103/RevModPhys.90.015001}.
%\newblock \eprint{1705.01111}.

\bibitem{Elcoro2021}
\bibinfo{author}{Elcoro, L.} \emph{et~al.}
\newblock \bibinfo{title}{{Magnetic Topological Quantum Chemistry}}.
\newblock \emph{\bibinfo{journal}{Nature Communications}}
  \textbf{\bibinfo{volume}{12}}, \bibinfo{pages}{5965} (\bibinfo{year}{2021}).
%\newblock \urlprefix\url{http://dx.doi.org/10.1038/s41467-021-26241-8
%  https://www.nature.com/articles/s41467-021-26241-8 http://www.cryst.ehu.
%  http://arxiv.org/abs/2010.00598}.
%\newblock \eprint{2010.00598}.

\bibitem{Mazin2021}
\bibinfo{author}{Mazin, I.~I.}, \bibinfo{author}{Koepernik, K.},
  \bibinfo{author}{Johannes, M.~D.},
  \bibinfo{author}{Gonz{\'{a}}lez-Hern{\'{a}}ndez, R.} \&
  \bibinfo{author}{{\v{S}}mejkal, L.}
\newblock \bibinfo{title}{{Prediction of unconventional magnetism in doped FeSb
  2}}.
\newblock \emph{\bibinfo{journal}{Proceedings of the National Academy of
  Sciences}} \textbf{\bibinfo{volume}{118}}, \bibinfo{pages}{e2108924118}
  (\bibinfo{year}{2021}).
%\newblock \urlprefix\url{http://arxiv.org/abs/2105.06356
%  https://pnas.org/doi/full/10.1073/pnas.2108924118}.
%\newblock \eprint{2105.06356}.

\bibitem{Reichlova2024}
\bibinfo{author}{Reichlova, H.} \emph{et~al.}
\newblock \bibinfo{title}{{Observation of a spontaneous anomalous Hall response
  in the Mn$_5$Si$_3$ $d$-wave altermagnet candidate}}.
\newblock \emph{\bibinfo{journal}{Nature Communications}}
  \textbf{\bibinfo{volume}{15}}, \bibinfo{pages}{4961} (\bibinfo{year}{2024}).
%\newblock \urlprefix\url{http://arxiv.org/abs/2012.15651
%  https://www.nature.com/articles/s41467-024-48493-w}.
%\newblock \eprint{2012.15651}.

\bibitem{Smejkal2022GMR}
\bibinfo{author}{{\v{S}}mejkal, L.}, \bibinfo{author}{Hellenes, A.~B.},
  \bibinfo{author}{Gonz{\'{a}}lez-Hern{\'{a}}ndez, R.},
  \bibinfo{author}{Sinova, J.} \& \bibinfo{author}{Jungwirth, T.}
\newblock \bibinfo{title}{{Giant and Tunneling Magnetoresistance in
  Unconventional Collinear Antiferromagnets with Nonrelativistic Spin-Momentum
  Coupling}}.
\newblock \emph{\bibinfo{journal}{Physical Review X}}
  \textbf{\bibinfo{volume}{12}}, \bibinfo{pages}{011028}
  (\bibinfo{year}{2022}).
%\newblock \urlprefix\url{https://link.aps.org/doi/10.1103/PhysRevX.12.011028
%  http://arxiv.org/abs/2103.12664}.
%\newblock \eprint{2103.12664}.

\bibitem{Betancourt2021}
\bibinfo{author}{{Gonzalez Betancourt}, R.~D.} \emph{et~al.}
\newblock \bibinfo{title}{{Spontaneous Anomalous Hall Effect Arising from an
  Unconventional Compensated Magnetic Phase in a Semiconductor}}.
\newblock \emph{\bibinfo{journal}{Physical Review Letters}}
  \textbf{\bibinfo{volume}{130}}, \bibinfo{pages}{036702}
  (\bibinfo{year}{2023}).
%\newblock \urlprefix\url{https://arxiv.org/abs/2112.06805v1
%  https://link.aps.org/doi/10.1103/PhysRevLett.130.036702}.
%\newblock \eprint{2112.06805}.

\bibitem{Smejkal2023}
\bibinfo{author}{{\v{S}}mejkal, L.} \emph{et~al.}
\newblock \bibinfo{title}{{Chiral magnons in altermagnetic RuO$_2$}}.
\newblock \emph{\bibinfo{journal}{Physical Review Letters}}
  \textbf{\bibinfo{volume}{131}}, \bibinfo{pages}{256703}
  (\bibinfo{year}{2023}).
%\newblock
%  \urlprefix\url{https://link.aps.org/doi/10.1103/PhysRevLett.131.256703}.

\bibitem{Chen2025}
\bibinfo{author}{Chen, X.} \emph{et~al.}
\newblock \bibinfo{title}{{Unconventional magnons in collinear magnets dictated
  by spin space groups}}.
\newblock \emph{\bibinfo{journal}{Nature}} \textbf{\bibinfo{volume}{640}},
  \bibinfo{pages}{349--354} (\bibinfo{year}{2025}).
%\newblock \urlprefix\url{https://www.nature.com/articles/s41586-025-08715-7}.

\bibitem{Hariki2023}
\bibinfo{author}{Hariki, A.} \emph{et~al.}
\newblock \bibinfo{title}{{X-Ray Magnetic Circular Dichroism in Altermagnetic
  $\alpha$-MnTe}}.
\newblock \emph{\bibinfo{journal}{Physical Review Letters}}
  \textbf{\bibinfo{volume}{132}}, \bibinfo{pages}{176701}
  (\bibinfo{year}{2024}).
%\newblock \urlprefix\url{http://arxiv.org/abs/2305.03588
%  https://link.aps.org/doi/10.1103/PhysRevLett.132.176701}.
%\newblock \eprint{2305.03588}.

\bibitem{Hellenes2023}
\bibinfo{author}{Hellenes, A.~B.} \emph{et~al.}
\newblock \bibinfo{title}{{P-wave magnets}}.
\newblock \emph{\bibinfo{journal}{ArXiv 2309.01607v3}}  (\bibinfo{year}{2023}).
\newblock \urlprefix\url{http://arxiv.org/abs/2309.01607}.
%\newblock \eprint{2309.01607}.

\bibitem{Hellenes2023a}
\bibinfo{author}{Hellenes, A.~B.}, \bibinfo{author}{Jungwirth, T.},
  \bibinfo{author}{Sinova, J.} \& \bibinfo{author}{{\v{S}}mejkal, L.}
\newblock \bibinfo{title}{{Exchange spin-orbit coupling and unconventional
  p-wave magnetism}}.
\newblock \emph{\bibinfo{journal}{Arxiv Preprint}}  (\bibinfo{year}{2023}).
\newblock \urlprefix\url{http://arxiv.org/abs/2309.01607v1}.
%\newblock \eprint{2309.01607v1}.

\bibitem{McClarty2024}
\bibinfo{author}{McClarty, P.~A.} \& \bibinfo{author}{Rau, J.~G.}
\newblock \bibinfo{title}{{Landau Theory of Altermagnetism}}.
\newblock \emph{\bibinfo{journal}{Physical Review Letters}}
  \textbf{\bibinfo{volume}{132}}, \bibinfo{pages}{176702}
  (\bibinfo{year}{2023}).
%\newblock
%  \urlprefix\url{https://link.aps.org/doi/10.1103/PhysRevLett.132.176702
%  http://arxiv.org/abs/2308.04484}.
%\newblock \eprint{2308.04484}.

\bibitem{Smolyanyuk2024}
\bibinfo{author}{Smolyanyuk, A.}, \bibinfo{author}{{\v{S}}mejkal, L.} \&
  \bibinfo{author}{Mazin, I.~I.}
\newblock \bibinfo{title}{{A tool to check whether a symmetry-compensated
  collinear magnetic material is antiferro- or altermagnetic}}.
\newblock \emph{\bibinfo{journal}{SciPost Physics Codebases}}
  \textbf{\bibinfo{volume}{30}}, \bibinfo{pages}{1--16} (\bibinfo{year}{2024}).
%\newblock \urlprefix\url{https://scipost.org/10.21468/SciPostPhysCodeb.30
%  http://arxiv.org/abs/2401.08784}.
%\newblock \eprint{2401.08784}.

\bibitem{Shinohara2024}
\bibinfo{author}{Shinohara, K.} \emph{et~al.}
\newblock \bibinfo{title}{{Algorithm for spin symmetry operation search}}.
\newblock \emph{\bibinfo{journal}{Acta Crystallographica Section A Foundations
  and Advances}} \textbf{\bibinfo{volume}{80}}, \bibinfo{pages}{94--103}
  (\bibinfo{year}{2024}).
%\newblock
 % \urlprefix\url{https://scripts.iucr.org/cgi-bin/paper?S2053273323009257}.

\bibitem{Watanabe2024}
\bibinfo{author}{Watanabe, H.}, \bibinfo{author}{Shinohara, K.},
  \bibinfo{author}{Nomoto, T.}, \bibinfo{author}{Togo, A.} \&
  \bibinfo{author}{Arita, R.}
\newblock \bibinfo{title}{{Symmetry analysis with spin crystallographic groups:
  Disentangling effects free of spin-orbit coupling in emergent
  electromagnetism}}.
\newblock \emph{\bibinfo{journal}{Physical Review B}}
  \textbf{\bibinfo{volume}{109}}, \bibinfo{pages}{094438}
  (\bibinfo{year}{2024}).
%\newblock \urlprefix\url{https://link.aps.org/doi/10.1103/PhysRevB.109.094438}.

\bibitem{Jiang2024}
\bibinfo{author}{Jiang, Y.} \emph{et~al.}
\newblock \bibinfo{title}{{Enumeration of Spin-Space Groups: Toward a Complete
  Description of Symmetries of Magnetic Orders}}.
\newblock \emph{\bibinfo{journal}{Physical Review X}}
  \textbf{\bibinfo{volume}{14}}, \bibinfo{pages}{031039}
  (\bibinfo{year}{2024}).
%\newblock \urlprefix\url{http://arxiv.org/abs/2307.10371
%  https://link.aps.org/doi/10.1103/PhysRevX.14.031039}.
%\newblock \eprint{2307.10371}.

\bibitem{Zhu2025}
\bibinfo{author}{Zhu, H.}, \bibinfo{author}{Li, J.}, \bibinfo{author}{Chen,
  X.}, \bibinfo{author}{Yu, Y.} \& \bibinfo{author}{Liu, Q.}
\newblock \bibinfo{title}{{Magnetic geometry induced quantum geometry and
  nonlinear transports}}.
\newblock \emph{\bibinfo{journal}{Nature Communications}}
  \textbf{\bibinfo{volume}{16}}, \bibinfo{pages}{4882} (\bibinfo{year}{2025}).
%\newblock \urlprefix\url{https://www.nature.com/articles/s41467-025-60128-2}.

\bibitem{Chen2024}
\bibinfo{author}{Chen, X.} \emph{et~al.}
\newblock \bibinfo{title}{{Enumeration and Representation Theory of Spin Space
  Groups}}.
\newblock \emph{\bibinfo{journal}{Physical Review X}}
  \textbf{\bibinfo{volume}{14}}, \bibinfo{pages}{031038}
  (\bibinfo{year}{2024}).
%\newblock \urlprefix\url{http://arxiv.org/abs/2307.10369
%  https://link.aps.org/doi/10.1103/PhysRevX.14.031038}.
%\newblock \eprint{2307.10369}.

\bibitem{Schiff2025a}
\bibinfo{author}{Schiff, H.}, \bibinfo{author}{Corticelli, A.},
  \bibinfo{author}{Guerreiro, A.}, \bibinfo{author}{Romh{\'{a}}nyi, J.} \&
  \bibinfo{author}{McClarty, P.}
\newblock \bibinfo{title}{{The crystallographic spin point groups and their
  representations}}.
\newblock \emph{\bibinfo{journal}{SciPost Physics}}
  \textbf{\bibinfo{volume}{18}} (\bibinfo{year}{2025}).
%\newblock \eprint{2307.12784}.

\bibitem{Xiao2024}
\bibinfo{author}{Xiao, Z.}, \bibinfo{author}{Zhao, J.}, \bibinfo{author}{Li,
  Y.}, \bibinfo{author}{Shindou, R.} \& \bibinfo{author}{Song, Z.-D.}
\newblock \bibinfo{title}{{Spin Space Groups: Full Classification and
  Applications}}.
\newblock \emph{\bibinfo{journal}{Physical Review X}}
  \textbf{\bibinfo{volume}{14}}, \bibinfo{pages}{031037}
  (\bibinfo{year}{2024}).
%\newblock \urlprefix\url{https://link.aps.org/doi/10.1103/PhysRevX.14.031037}.

\bibitem{Litvin2013}
\bibinfo{author}{Litvin, D.~B.}
\newblock \bibinfo{title}{{Magnetic Group Tables}} (\bibinfo{year}{2013}).
\newblock \urlprefix\url{http://www.iucr.org/publ/978-0-9553602-2-0}.

\bibitem{MAGNDATA}
\bibinfo{title}{{MAGNDATA: A Collection of magnetic structures with portable
  cif-type files}}.
\newblock \urlprefix\url{https://www.cryst.ehu.es/magndata/}.

\bibitem{Berlijn2017a}
\bibinfo{author}{Berlijn, T.} \emph{et~al.}
\newblock \bibinfo{title}{{Itinerant Antiferromagnetism in RuO$_2$}}.
\newblock \emph{\bibinfo{journal}{Physical Review Letters}}
  \textbf{\bibinfo{volume}{118}}, \bibinfo{pages}{077201}
  (\bibinfo{year}{2017}).
%\newblock
 % \urlprefix\url{https://link.aps.org/doi/10.1103/PhysRevLett.118.077201}.
%\newblock \eprint{1612.09589}.

\bibitem{Zhu2018}
\bibinfo{author}{Zhu, Z.~H.} \emph{et~al.}
\newblock \bibinfo{title}{{Anomalous Antiferromagnetism in Metallic RuO$_2$
  Determined by Resonant X-ray Scattering}}.
\newblock \emph{\bibinfo{journal}{Physical Review Letters}}
  \textbf{\bibinfo{volume}{122}}, \bibinfo{pages}{017202}
  (\bibinfo{year}{2019}).
%\newblock \urlprefix\url{http://arxiv.org/abs/1806.02036
%  http://dx.doi.org/10.1103/PhysRevLett.122.017202
%  https://link.aps.org/doi/10.1103/PhysRevLett.122.017202}.
%\newblock \eprint{1806.02036}.

\bibitem{Lovesey2022}
\bibinfo{author}{Lovesey, S.~W.}, \bibinfo{author}{Khalyavin, D.~D.} \&
  \bibinfo{author}{van~der Laan, G.}
\newblock \bibinfo{title}{{Magnetic properties of RuO$_2$ and charge-magnetic
  interference in Bragg diffraction of circularly polarized x-rays}}.
\newblock \emph{\bibinfo{journal}{Physical Review B}}
  \textbf{\bibinfo{volume}{105}}, \bibinfo{pages}{014403}
  (\bibinfo{year}{2022}).
%\newblock \urlprefix\url{https://link.aps.org/doi/10.1103/PhysRevB.105.014403}.

\bibitem{Occhialini2021}
\bibinfo{author}{Occhialini, C.~A.} \emph{et~al.}
\newblock \bibinfo{title}{{Local electronic structure of rutile RuO$_2$}}.
\newblock \emph{\bibinfo{journal}{Physical Review Research}}
  \textbf{\bibinfo{volume}{3}}, \bibinfo{pages}{033214} (\bibinfo{year}{2021}).
%\newblock
 % \urlprefix\url{https://journals.aps.org/prresearch/abstract/10.1103/PhysRevResearch.3.033214}.
%\newblock \eprint{2108.06256}.

\bibitem{Feng2022}
\bibinfo{author}{Feng, Z.} \emph{et~al.}
\newblock \bibinfo{title}{{An anomalous Hall effect in altermagnetic ruthenium
  dioxide}}.
\newblock \emph{\bibinfo{journal}{Nature Electronics}}
  \textbf{\bibinfo{volume}{5}}, \bibinfo{pages}{735--743}
  (\bibinfo{year}{2022}).
%\newblock \urlprefix\url{http://arxiv.org/abs/2002.08712
%  https://www.nature.com/articles/s41928-022-00866-z}.
%\newblock \eprint{2002.08712}.

\bibitem{Bose2022}
\bibinfo{author}{Bose, A.} \emph{et~al.}
\newblock \bibinfo{title}{{Tilted spin current generated by the collinear
  antiferromagnet ruthenium dioxide}}.
\newblock \emph{\bibinfo{journal}{Nature Electronics}}
  \textbf{\bibinfo{volume}{5}}, \bibinfo{pages}{267--274}
  (\bibinfo{year}{2022}).
%\newblock \urlprefix\url{http://arxiv.org/abs/2108.09150
%  https://www.nature.com/articles/s41928-022-00758-2
%  https://www.nature.com/articles/s41928-022-00744-8
%  http://dx.doi.org/10.1038/s41928-022-00744-8}.
%\newblock \eprint{2108.09150}.

\bibitem{Bai2022}
\bibinfo{author}{Bai, H.} \emph{et~al.}
\newblock \bibinfo{title}{{Observation of Spin Splitting Torque in a Collinear
  Antiferromagnet RuO$_2$}}.
\newblock \emph{\bibinfo{journal}{Physical Review Letters}}
  \textbf{\bibinfo{volume}{128}}, \bibinfo{pages}{197202}
  (\bibinfo{year}{2022}).
%\newblock \urlprefix\url{http://arxiv.org/abs/2109.05933
%  https://link.aps.org/doi/10.1103/PhysRevLett.128.197202}.
%\newblock \eprint{2109.05933}.

\bibitem{Karube2022}
\bibinfo{author}{Karube, S.} \emph{et~al.}
\newblock \bibinfo{title}{{Observation of Spin-Splitter Torque in Collinear
  Antiferromagnetic RuO$_2$}}.
\newblock \emph{\bibinfo{journal}{Physical Review Letters}}
  \textbf{\bibinfo{volume}{129}}, \bibinfo{pages}{137201}
  (\bibinfo{year}{2022}).
%\newblock \urlprefix\url{https://doi.org/10.1103/PhysRevLett.129.137201
%  https://link.aps.org/doi/10.1103/PhysRevLett.129.137201}.
%\newblock \eprint{2111.07487}.

\bibitem{Lovesey2023c}
\bibinfo{author}{Lovesey, S.~W.}, \bibinfo{author}{Khalyavin, D.~D.} \&
  \bibinfo{author}{van~der Laan, G.}
\newblock \bibinfo{title}{{Magnetic structure of RuO$_2$ in view of
  altermagnetism}}.
\newblock \emph{\bibinfo{journal}{Physical Review B}}
  \textbf{\bibinfo{volume}{108}}, \bibinfo{pages}{L121103}
  (\bibinfo{year}{2023}).
%\newblock
%  \urlprefix\url{https://link.aps.org/doi/10.1103/PhysRevB.108.L121103}.

\bibitem{Liu2023}
\bibinfo{author}{Liu, Y.} \emph{et~al.}
\newblock \bibinfo{title}{{Inverse Altermagnetic Spin Splitting
  Effect‐Induced Terahertz Emission in RuO$_2$}}.
\newblock \emph{\bibinfo{journal}{Advanced Optical Materials}}
  \textbf{\bibinfo{volume}{11}}, \bibinfo{pages}{1--7} (\bibinfo{year}{2023}).
%\newblock
%  \urlprefix\url{https://onlinelibrary.wiley.com/doi/10.1002/adom.202300177}.

\bibitem{Fedchenko2024}
\bibinfo{author}{Fedchenko, O.} \emph{et~al.}
\newblock \bibinfo{title}{{Observation of time-reversal symmetry breaking in
  the band structure of altermagnetic RuO$_2$}}.
\newblock \emph{\bibinfo{journal}{Science Advances}}
  \textbf{\bibinfo{volume}{10}}, \bibinfo{pages}{31} (\bibinfo{year}{2024}).
%\newblock \urlprefix\url{https://www.science.org
%  http://arxiv.org/abs/2306.02170
%  https://www.science.org/doi/10.1126/sciadv.adj4883}.
%\newblock \eprint{2306.02170}.

\bibitem{Smolyanyuk2023}
\bibinfo{author}{Smolyanyuk, A.}, \bibinfo{author}{Mazin, I.~I.},
  \bibinfo{author}{Garcia-Gassull, L.} \& \bibinfo{author}{Valent{\'{i}}, R.}
\newblock \bibinfo{title}{{Fragility of the magnetic order in the prototypical
  altermagnet RuO$_2$}}.
\newblock \emph{\bibinfo{journal}{Physical Review B}}
  \textbf{\bibinfo{volume}{109}}, \bibinfo{pages}{134424}
  (\bibinfo{year}{2024}).
%\newblock \urlprefix\url{http://arxiv.org/abs/2310.06909
%  https://link.aps.org/doi/10.1103/PhysRevB.109.134424}.
%\newblock \eprint{2310.06909}.

\bibitem{Lin2024}
\bibinfo{author}{Lin, Z.} \emph{et~al.}
\newblock \bibinfo{title}{{Observation of Giant Spin Splitting and d-wave Spin
  Texture in Room Temperature Altermagnet RuO$_2$}}.
\newblock \emph{\bibinfo{journal}{Arxiv Preprint}}  (\bibinfo{year}{2024}).
\newblock \urlprefix\url{http://arxiv.org/abs/2402.04995}.
%\newblock \eprint{2402.04995}.

\bibitem{Kessler2024}
\bibinfo{author}{Ke{\ss}ler, P.} \emph{et~al.}
\newblock \bibinfo{title}{{Absence of magnetic order in RuO$_2$: insights from
  $\mu$SR spectroscopy and neutron diffraction}}.
\newblock \emph{\bibinfo{journal}{Arxiv Preprint}}  (\bibinfo{year}{2024}).
\newblock \urlprefix\url{http://arxiv.org/abs/2405.10820}.
%\newblock \eprint{2405.10820}.

\bibitem{Li2024a}
\bibinfo{author}{Li, Z.}, \bibinfo{author}{Zhang, Z.}, \bibinfo{author}{Lu, X.}
  \& \bibinfo{author}{Xu, Y.}
\newblock \bibinfo{title}{{Spin Splitting in Altermagnetic RuO$_2$ Enables
  Field-free Spin-Orbit Torque Switching via Dominant Out-of-Plane Spin
  Polarization}}.
\newblock \emph{\bibinfo{journal}{Arxiv Preprint}}  (\bibinfo{year}{2024}).
\newblock \urlprefix\url{http://arxiv.org/abs/2407.07447}.
%\newblock \eprint{2407.07447}.

\bibitem{Wenzel2025}
\bibinfo{author}{Wenzel, M.} \emph{et~al.}
\newblock \bibinfo{title}{{Fermi-liquid behavior of nonaltermagnetic RuO$_2$}}.
\newblock \emph{\bibinfo{journal}{Physical Review B}}
  \textbf{\bibinfo{volume}{111}}, \bibinfo{pages}{L041115}
  (\bibinfo{year}{2025}).
%\newblock
%  \urlprefix\url{https://link.aps.org/doi/10.1103/PhysRevB.111.L041115}.

\bibitem{Jeong2024}
\bibinfo{author}{Jeong, S.~G.} \emph{et~al.}
\newblock \bibinfo{title}{{Altermagnetic Polar Metallic phase in Ultra-Thin
  Epitaxially-Strained RuO$_2$ Films}}.
\newblock \emph{\bibinfo{journal}{Arxiv Preprint}} \bibinfo{pages}{1--15}
  (\bibinfo{year}{2024}).
\newblock \urlprefix\url{http://arxiv.org/abs/2405.05838}.
%\newblock \eprint{2405.05838}.

\bibitem{Hiraishi2024}
\bibinfo{author}{Hiraishi, M.} \emph{et~al.}
\newblock \bibinfo{title}{{Nonmagnetic Ground State in RuO$_2$ Revealed by Muon
  Spin Rotation}}.
\newblock \emph{\bibinfo{journal}{Physical Review Letters}}
  \textbf{\bibinfo{volume}{132}}, \bibinfo{pages}{166702}
  (\bibinfo{year}{2024}).
%\newblock \urlprefix\url{http://arxiv.org/abs/2403.10028
%  https://link.aps.org/doi/10.1103/PhysRevLett.132.166702}.
%\newblock \eprint{2403.10028}.


%\newblock \urlprefix\url{http://xlink.rsc.org/?DOI=C5CP07806G}.

\bibitem{Lopez-Moreno2016}
\bibinfo{author}{L{\'{o}}pez-Moreno, S.}, \bibinfo{author}{Romero, A.~H.},
  \bibinfo{author}{Mej{\'{i}}a-L{\'{o}}pez, J.} \&
  \bibinfo{author}{Mu{\~{n}}oz, A.}
\newblock \bibinfo{title}{{First-principles study of pressure-induced
  structural phase transitions in MnF$_2$}}.
\newblock \emph{\bibinfo{journal}{Physical Chemistry Chemical Physics}}
  \textbf{\bibinfo{volume}{18}}, \bibinfo{pages}{33250--33263}
  (\bibinfo{year}{2016}).
%\newblock \urlprefix\url{http://xlink.rsc.org/?DOI=C6CP05467F}.

\newblock {\bf This paper reported density-functional theory calculations of non-relativistic band structure of MnF$_2$ showing spin-split bands.} 

\bibitem{Noda2016}
\bibinfo{author}{Noda, Y.}, \bibinfo{author}{Ohno, K.} \&
  \bibinfo{author}{Nakamura, S.}
\newblock \bibinfo{title}{{Momentum-dependent band spin splitting in
  semiconducting MnO$_2$ : a density functional calculation}}.
\newblock \emph{\bibinfo{journal}{Physical Chemistry Chemical Physics}}
  \textbf{\bibinfo{volume}{18}}, \bibinfo{pages}{13294--13303}
  (\bibinfo{year}{2016}).
  
\bibitem{Hayami2019}
\bibinfo{author}{Hayami, S.}, \bibinfo{author}{Yanagi, Y.} \&
  \bibinfo{author}{Kusunose, H.}
\newblock \bibinfo{title}{{Momentum-Dependent Spin Splitting by Collinear
  Antiferromagnetic Ordering}}.
\newblock \emph{\bibinfo{journal}{Journal of the Physical Society of Japan}}
  \textbf{\bibinfo{volume}{88}}, \bibinfo{pages}{123702}
  (\bibinfo{year}{2019}).
%\newblock \urlprefix\url{https://journals.jps.jp/doi/10.7566/JPSJ.88.123702}.
%\newblock \eprint{1908.08680}.

\bibitem{Yuan2020}
\bibinfo{author}{Yuan, L.-D.}, \bibinfo{author}{Wang, Z.},
  \bibinfo{author}{Luo, J.-W.}, \bibinfo{author}{Rashba, E.~I.} \&
  \bibinfo{author}{Zunger, A.}
\newblock \bibinfo{title}{{Giant momentum-dependent spin splitting in
  centrosymmetric low-Z antiferromagnets}}.
\newblock \emph{\bibinfo{journal}{Physical Review B}}
  \textbf{\bibinfo{volume}{102}}, \bibinfo{pages}{014422}
  (\bibinfo{year}{2020}).
%\newblock \urlprefix\url{https://arxiv.org/pdf/1912.12689.pdf
%  https://link.aps.org/doi/10.1103/PhysRevB.102.014422}.
%\newblock \eprint{1912.12689}.

\bibitem{Bhowal2024}
\bibinfo{author}{Bhowal, S.} \& \bibinfo{author}{Spaldin, N.~A.}
\newblock \bibinfo{title}{{Ferroically Ordered Magnetic Octupoles in d-Wave
  Altermagnets}}.
\newblock \emph{\bibinfo{journal}{Physical Review X}}
  \textbf{\bibinfo{volume}{14}}, \bibinfo{pages}{011019}
  (\bibinfo{year}{2024}).
%\newblock \urlprefix\url{https://link.aps.org/doi/10.1103/PhysRevX.14.011019}.

\bibitem{Jaeschke-Ubiergo2025}
\bibinfo{author}{Jaeschke-Ubiergo, R.} \emph{et~al.}
\newblock \bibinfo{title}{{Atomic Altermagnetism}}.
\newblock \emph{\bibinfo{journal}{Arxiv Preprint}}  (\bibinfo{year}{2025}).
\newblock \urlprefix\url{https://arxiv.org/pdf/2503.10797}.
%\newblock \eprint{2503.10797}.

\bibitem{Lovesey2023}
\bibinfo{author}{Lovesey, S.~W.}, \bibinfo{author}{Khalyavin, D.~D.} \&
  \bibinfo{author}{van~der Laan, G.}
\newblock \bibinfo{title}{{Templates for magnetic symmetry and altermagnetism
  in hexagonal MnTe}}.
\newblock \emph{\bibinfo{journal}{Physical Review B}}
  \textbf{\bibinfo{volume}{108}}, \bibinfo{pages}{174437}
  (\bibinfo{year}{2023}).
%\newblock \urlprefix\url{http://arxiv.org/abs/2309.04282
%  https://link.aps.org/doi/10.1103/PhysRevB.108.174437}.
%\newblock \eprint{2309.04282}.

\bibitem{Mazin2023}
\bibinfo{author}{Mazin, I.~I.}
\newblock \bibinfo{title}{{Altermagnetism in MnTe: Origin, predicted
  manifestations, and routes to detwinning}}.
\newblock \emph{\bibinfo{journal}{Physical Review B}}
  \textbf{\bibinfo{volume}{107}}, \bibinfo{pages}{L100418}
  (\bibinfo{year}{2023}).
%\newblock \urlprefix\url{http://arxiv.org/abs/2301.08573
%  http://dx.doi.org/10.1103/PhysRevB.107.L100418
%  https://link.aps.org/doi/10.1103/PhysRevB.107.L100418}.
%\newblock \eprint{2301.08573}.

\bibitem{Krempasky2024}
\bibinfo{author}{Krempask{\'{y}}, J.} \emph{et~al.}
\newblock \bibinfo{title}{{Altermagnetic lifting of Kramers spin degeneracy}}.
\newblock \emph{\bibinfo{journal}{Nature}} \textbf{\bibinfo{volume}{626}},
  \bibinfo{pages}{517--522} (\bibinfo{year}{2024}).
%\newblock \urlprefix\url{https://doi.org/10.1038/s41586-023-06907-7
%  https://www.nature.com/articles/s41586-023-06907-7
%  http://arxiv.org/abs/2308.10681}.
%\newblock \eprint{2308.10681}.

\newblock{\bf This paper and Refs. 99-101 reported experimental spectroscopic observations of a g-wave altermagnetic band structure of MnTe.} 

\bibitem{Lee2024}
\bibinfo{author}{Lee, S.~S.} \emph{et~al.}
\newblock \bibinfo{title}{{Broken Kramers Degeneracy in Altermagnetic MnTe}}.
\newblock \emph{\bibinfo{journal}{Physical Review Letters}}
  \textbf{\bibinfo{volume}{132}}, \bibinfo{pages}{036702}
  (\bibinfo{year}{2024}).
%\newblock \urlprefix\url{https://arxiv.org/abs/2308.11180v1
%  http://arxiv.org/abs/2308.11180
%  https://link.aps.org/doi/10.1103/PhysRevLett.132.036702}.
%\newblock \eprint{2308.11180}.

\bibitem{Osumi2024}
\bibinfo{author}{Osumi, T.} \emph{et~al.}
\newblock \bibinfo{title}{{Observation of a giant band splitting in
  altermagnetic MnTe}}.
\newblock \emph{\bibinfo{journal}{Physical Review B}}
  \textbf{\bibinfo{volume}{109}}, \bibinfo{pages}{115102}
  (\bibinfo{year}{2024}).
%\newblock \urlprefix\url{https://arxiv.org/abs/2308.10117v1
%  http://arxiv.org/abs/2308.10117
%  https://link.aps.org/doi/10.1103/PhysRevB.109.115102}.
%\newblock \eprint{2308.10117}.

\bibitem{Hajlaoui2024}
\bibinfo{author}{Hajlaoui, M.} \emph{et~al.}
\newblock \bibinfo{title}{{Temperature Dependence of Relativistic Valence Band
  Splitting Induced by an Altermagnetic Phase Transition}}.
\newblock \emph{\bibinfo{journal}{Advanced Materials}}
  \textbf{\bibinfo{volume}{36}}, \bibinfo{pages}{2314076}
  (\bibinfo{year}{2024}).
%\newblock
%  \urlprefix\url{https://onlinelibrary.wiley.com/doi/10.1002/adma.202314076}.

\bibitem{Aoyama2024}
\bibinfo{author}{Aoyama, T.} \& \bibinfo{author}{Ohgushi, K.}
\newblock \bibinfo{title}{{Piezomagnetic properties in altermagnetic MnTe}}.
\newblock \emph{\bibinfo{journal}{Physical Review Materials}}
  \textbf{\bibinfo{volume}{8}}, \bibinfo{pages}{L041402}
  (\bibinfo{year}{2024}).
%\newblock \urlprefix\url{http://arxiv.org/abs/2305.14786
%  https://link.aps.org/doi/10.1103/PhysRevMaterials.8.L041402}.
%\newblock \eprint{2305.14786}.

\bibitem{Kluczyk2024}
\bibinfo{author}{Kluczyk, K.~P.} \emph{et~al.}
\newblock \bibinfo{title}{{Coexistence of anomalous Hall effect and weak
  magnetization in a nominally collinear antiferromagnet MnTe}}.
\newblock \emph{\bibinfo{journal}{Physical Review B}}
  \textbf{\bibinfo{volume}{110}}, \bibinfo{pages}{155201}
  (\bibinfo{year}{2024}).
%\newblock \urlprefix\url{https://link.aps.org/doi/10.1103/PhysRevB.110.155201}.

\bibitem{Reimers2024}
\bibinfo{author}{Reimers, S.} \emph{et~al.}
\newblock \bibinfo{title}{{Direct observation of altermagnetic band splitting
  in CrSb thin films}}.
\newblock \emph{\bibinfo{journal}{Nature Communications}}
  \textbf{\bibinfo{volume}{15}}, \bibinfo{pages}{2116} (\bibinfo{year}{2024}).
%\newblock \urlprefix\url{https://www.nature.com/articles/s41467-024-46476-5
%  http://arxiv.org/abs/2310.17280}.
%\newblock \eprint{2310.17280}.

\bibitem{Yang2024}
\bibinfo{author}{Yang, G.} \emph{et~al.}
\newblock \bibinfo{title}{{Three-dimensional mapping of the altermagnetic spin
  splitting in CrSb}}.
\newblock \emph{\bibinfo{journal}{Nature Communications}}
  \textbf{\bibinfo{volume}{16}}, \bibinfo{pages}{1442} (\bibinfo{year}{2025}).
%\newblock \urlprefix\url{http://arxiv.org/abs/2405.12575
%  https://www.nature.com/articles/s41467-025-56647-7}.
%\newblock \eprint{2405.12575}.

\bibitem{Ding2024}
\bibinfo{author}{Ding, J.} \emph{et~al.}
\newblock \bibinfo{title}{{Large Band Splitting in g-Wave Altermagnet CrSb}}.
\newblock \emph{\bibinfo{journal}{Physical Review Letters}}
  \textbf{\bibinfo{volume}{133}}, \bibinfo{pages}{206401}
  (\bibinfo{year}{2024}).
%\newblock \urlprefix\url{Phys.Rev.Lett.133,206401(2024)
%  https://link.aps.org/doi/10.1103/PhysRevLett.133.206401}.
%\newblock \eprint{2405.12687}.

\bibitem{Li2024}
\bibinfo{author}{Li, C.} \emph{et~al.}
\newblock \bibinfo{title}{{Topological Weyl Altermagnetism in CrSb}}.
\newblock \emph{\bibinfo{journal}{Arxiv Preprint}}  (\bibinfo{year}{2024}).
\newblock \urlprefix\url{https://arxiv.org/abs/2405.14777v1
%  http://arxiv.org/abs/2405.14777}.
%\newblock \eprint{2405.14777}.

\bibitem{Lu2025}
\bibinfo{author}{Lu, W.} \emph{et~al.}
\newblock \bibinfo{title}{{Signature of Topological Surface Bands in
  Altermagnetic Weyl Semimetal CrSb}}.
\newblock \emph{\bibinfo{journal}{Nano Letters}} \textbf{\bibinfo{volume}{25}},
  \bibinfo{pages}{7343--7350} (\bibinfo{year}{2025}).

\bibitem{Zeng2024}
\bibinfo{author}{Zeng, M.} \emph{et~al.}
\newblock \bibinfo{title}{{Observation of Spin Splitting in Room‐Temperature
  Metallic Antiferromagnet CrSb}}.
\newblock \emph{\bibinfo{journal}{Advanced Science}}
  \textbf{\bibinfo{volume}{11}}, \bibinfo{pages}{2406529}
  (\bibinfo{year}{2024}).
%\newblock \urlprefix\url{https://arxiv.org/abs/2405.12679v1
%  http://arxiv.org/abs/2405.12679
%  https://onlinelibrary.wiley.com/doi/10.1002/advs.202406529}.
%\newblock \eprint{2405.12679}.

\bibitem{Dale2024}
\bibinfo{author}{Dale, N.} \emph{et~al.}
\newblock \bibinfo{title}{{Non-relativistic spin splitting above and below the
  Fermi level in a g-wave altermagnet}}  (\bibinfo{year}{2024}).
\newblock \urlprefix\url{http://arxiv.org/abs/2411.18761}.
%\newblock \eprint{2411.18761}.

\bibitem{Liu2024b}
\bibinfo{author}{Liu, Z.}, \bibinfo{author}{Ozeki, M.}, \bibinfo{author}{Asai,
  S.}, \bibinfo{author}{Itoh, S.} \& \bibinfo{author}{Masuda, T.}
\newblock \bibinfo{title}{{Chiral-Split Magnon in Altermagnetic MnTe}}.
\newblock \emph{\bibinfo{journal}{Physical Review Letters}}
  \textbf{\bibinfo{volume}{133}}, \bibinfo{pages}{156702}
  (\bibinfo{year}{2024}).
%\newblock \urlprefix\url{http://arxiv.org/abs/2408.16490}.
%\newblock \eprint{2408.16490}.

\bibitem{Jiang2025}
\bibinfo{author}{Jiang, B.} \emph{et~al.}
\newblock \bibinfo{title}{{A metallic room-temperature d-wave altermagnet}}.
\newblock \emph{\bibinfo{journal}{Nature Physics}}
  \textbf{\bibinfo{volume}{21}}, \bibinfo{pages}{754--759}
  (\bibinfo{year}{2025}).
%\newblock \urlprefix\url{https://www.nature.com/articles/s41567-025-02822-y}.

\newblock{\bf This paper and Refs. 113 reported experimental spectroscopic observations of a d-wave altermagnetic band structure.} 

\bibitem{Zhang2025a}
\bibinfo{author}{Zhang, F.} \emph{et~al.}
\newblock \bibinfo{title}{{Crystal-symmetry-paired spin–valley locking in a
  layered room-temperature metallic altermagnet candidate}}.
\newblock \emph{\bibinfo{journal}{Nature Physics}}
  \textbf{\bibinfo{volume}{21}}, \bibinfo{pages}{760--767}
  (\bibinfo{year}{2025}).
%\newblock \urlprefix\url{https://www.nature.com/articles/s41567-025-02864-2}.

\bibitem{Antonenko2025}
\bibinfo{author}{Antonenko, D.~S.}, \bibinfo{author}{Fernandes, R.~M.} \&
  \bibinfo{author}{Venderbos, J. W.~F.}
\newblock \bibinfo{title}{{Mirror Chern Bands and Weyl Nodal Loops in
  Altermagnets}}.
\newblock \emph{\bibinfo{journal}{Physical Review Letters}}
  \textbf{\bibinfo{volume}{134}}, \bibinfo{pages}{096703}
  (\bibinfo{year}{2025}).
%\newblock
%  \urlprefix\url{https://link.aps.org/doi/10.1103/PhysRevLett.134.096703}.

\newblock{\bf This paper reported a theoretical study of topological phenomena in the band structure of a 2D Lieb lattice.} 

\bibitem{Kaushal2024}
\bibinfo{author}{Kaushal, N.} \& \bibinfo{author}{Franz, M.}
\newblock \bibinfo{title}{{Altermagnetism in modified Lieb lattice Hubbard
  model}}.
\newblock \emph{\bibinfo{journal}{Arxiv Preprint}}  (\bibinfo{year}{2024}).
\newblock \urlprefix\url{http://arxiv.org/abs/2412.16421}.
%\newblock \eprint{2412.16421}.

\bibitem{Wei2025}
\bibinfo{author}{Wei, C.~C.} \emph{et~al.}
\newblock \bibinfo{title}{{La$_2$O$_3$Mn$_2$Se$_2$: A correlated insulating layered $d$-wave altermagnet}}.
\newblock \emph{\bibinfo{journal}{Physical Review Materials}}
  \textbf{\bibinfo{volume}{9}}, \bibinfo{pages}{24402} (\bibinfo{year}{2025}).
%\newblock \urlprefix\url{https://doi.org/10.1103/PhysRevMaterials.9.024402}.

\bibitem{Maharaj2020}
\bibinfo{author}{Maharaj, D.~D.} \emph{et~al.}
\newblock \bibinfo{title}{{Octupolar versus N{\'{e}}el Order in Cubic 5d$^2$
  Double Perovskites}}.
\newblock \emph{\bibinfo{journal}{Physical Review Letters}}
  \textbf{\bibinfo{volume}{124}}, \bibinfo{pages}{87206}
  (\bibinfo{year}{2020}).
%\newblock \urlprefix\url{https://doi.org/10.1103/PhysRevLett.124.087206}.
%\newblock \eprint{1909.03113}.

\bibitem{Neel1971}
\bibinfo{author}{N{\'{e}}el, L.}
\newblock \bibinfo{title}{{Magnetism and Local Molecular Field}}.
\newblock \emph{\bibinfo{journal}{Science}} \textbf{\bibinfo{volume}{174}},
  \bibinfo{pages}{985--992} (\bibinfo{year}{1971}).
%\newblock \urlprefix\url{http://www.ncbi.nlm.nih.gov/pubmed/17757022
%  http://www.sciencemag.org/cgi/doi/10.1126/science.174.4013.985}.

\bibitem{Chakraborty2024}
\bibinfo{author}{Chakraborty, A.}, \bibinfo{author}{{Gonz{\'{a}}lez
  Hern{\'{a}}ndez}, R.}, \bibinfo{author}{{\v{S}}mejkal, L.} \&
  \bibinfo{author}{Sinova, J.}
\newblock \bibinfo{title}{{Strain-induced phase transition from antiferromagnet
  to altermagnet}}.
\newblock \emph{\bibinfo{journal}{Physical Review B}}
  \textbf{\bibinfo{volume}{109}}, \bibinfo{pages}{144421}
  (\bibinfo{year}{2024}).
%\newblock \urlprefix\url{http://arxiv.org/abs/2402.00151
%  https://link.aps.org/doi/10.1103/PhysRevB.109.144421}.
%\newblock \eprint{2402.00151}.

\bibitem{Liu2024}
\bibinfo{author}{Liu, Y.}, \bibinfo{author}{Yu, J.} \& \bibinfo{author}{Liu,
  C.-C.}
\newblock \bibinfo{title}{{Twisted Magnetic Van der Waals Bilayers: An Ideal
  Platform for Altermagnetism}}.
\newblock \emph{\bibinfo{journal}{Physical Review Letters}}
  \textbf{\bibinfo{volume}{133}}, \bibinfo{pages}{206702}
  (\bibinfo{year}{2024}).
%\newblock \urlprefix\url{http://arxiv.org/abs/2404.17146
%  https://link.aps.org/doi/10.1103/PhysRevLett.133.206702}.
%\newblock \eprint{2404.17146}.

\bibitem{Leeb2023}
\bibinfo{author}{Leeb, V.}, \bibinfo{author}{Mook, A.},
  \bibinfo{author}{{\v{S}}mejkal, L.} \& \bibinfo{author}{Knolle, J.}
\newblock \bibinfo{title}{{Spontaneous Formation of Altermagnetism from Orbital
  Ordering}}.
\newblock \emph{\bibinfo{journal}{Physical Review Letters}}
  \textbf{\bibinfo{volume}{132}}, \bibinfo{pages}{236701}
  (\bibinfo{year}{2024}).
%\newblock \urlprefix\url{http://arxiv.org/abs/2312.10839
%  https://journals.aps.org/prl/abstract/10.1103/PhysRevLett.132.236701
%  https://link.aps.org/doi/10.1103/PhysRevLett.132.236701}.
%\newblock \eprint{2312.10839}.

\bibitem{Fernandes2023}
\bibinfo{author}{Fernandes, R.~M.}, \bibinfo{author}{de~Carvalho, V.~S.},
  \bibinfo{author}{Birol, T.} \& \bibinfo{author}{Pereira, R.~G.}
\newblock \bibinfo{title}{{Topological transition from nodal to nodeless Zeeman
  splitting in altermagnets}}.
\newblock \emph{\bibinfo{journal}{Physical Review B}}
  \textbf{\bibinfo{volume}{109}}, \bibinfo{pages}{024404}
  (\bibinfo{year}{2024}).
%\newblock \urlprefix\url{http://arxiv.org/abs/2307.12380
%  https://link.aps.org/doi/10.1103/PhysRevB.109.024404}.
%\newblock \eprint{2307.12380}.

\bibitem{Das2023}
\bibinfo{author}{Das, S.}, \bibinfo{author}{Suri, D.} \&
  \bibinfo{author}{Soori, A.}
\newblock \bibinfo{title}{{Transport across junctions of altermagnets with
  normal metals and ferromagnets}}.
\newblock \emph{\bibinfo{journal}{Journal of Physics: Condensed Matter}}
  \textbf{\bibinfo{volume}{35}}, \bibinfo{pages}{435302}
  (\bibinfo{year}{2023}).
%\newblock \urlprefix\url{http://arxiv.org/abs/2305.06680
%  http://dx.doi.org/10.1088/1361-648X/acea12
%  https://iopscience.iop.org/article/10.1088/1361-648X/acea12}.
%\newblock \eprint{2305.06680}.

\bibitem{Maier2023}
\bibinfo{author}{Maier, T.~A.} \& \bibinfo{author}{Okamoto, S.}
\newblock \bibinfo{title}{{Weak-coupling theory of neutron scattering as a
  probe of altermagnetism}}.
\newblock \emph{\bibinfo{journal}{Physical Review B}}
  \textbf{\bibinfo{volume}{108}}, \bibinfo{pages}{L100402}
  (\bibinfo{year}{2023}).
%\newblock \urlprefix\url{http://arxiv.org/abs/2307.03793
%  http://dx.doi.org/10.1103/PhysRevB.108.L100402
%  https://link.aps.org/doi/10.1103/PhysRevB.108.L100402}.
%\newblock \eprint{2307.03793}.

\bibitem{Sato2024}
\bibinfo{author}{Sato, T.}, \bibinfo{author}{Haddad, S.},
  \bibinfo{author}{Fulga, I.~C.}, \bibinfo{author}{Assaad, F.~F.} \&
  \bibinfo{author}{van~den Brink, J.}
\newblock \bibinfo{title}{{Altermagnetic Anomalous Hall Effect Emerging from
  Electronic Correlations}}.
\newblock \emph{\bibinfo{journal}{Physical Review Letters}}
  \textbf{\bibinfo{volume}{133}}, \bibinfo{pages}{086503}
  (\bibinfo{year}{2024}).
%\newblock
%  \urlprefix\url{https://link.aps.org/doi/10.1103/PhysRevLett.133.086503}.

\bibitem{Roig2024}
\bibinfo{author}{Roig, M.}, \bibinfo{author}{Kreisel, A.}, \bibinfo{author}{Yu,
  Y.}, \bibinfo{author}{Andersen, B.~M.} \& \bibinfo{author}{Agterberg, D.~F.}
\newblock \bibinfo{title}{{Minimal models for altermagnetism}}.
\newblock \emph{\bibinfo{journal}{Physical Review B}}
  \textbf{\bibinfo{volume}{110}}, \bibinfo{pages}{144412}
  (\bibinfo{year}{2024}).
%\newblock \urlprefix\url{https://arxiv.org/abs/2402.15616v2
%  http://arxiv.org/abs/2402.15616
%  https://link.aps.org/doi/10.1103/PhysRevB.110.144412}.
%\newblock \eprint{2402.15616}.

\bibitem{Bose2024}
\bibinfo{author}{Bose, A.}, \bibinfo{author}{Vadnais, S.} \&
  \bibinfo{author}{Paramekanti, A.}
\newblock \bibinfo{title}{{Altermagnetism and superconductivity in a
  multiorbital t-J model}}.
\newblock \emph{\bibinfo{journal}{Physical Review B}}
  \textbf{\bibinfo{volume}{110}}, \bibinfo{pages}{205120}
  (\bibinfo{year}{2024}).
%\newblock \urlprefix\url{http://arxiv.org/abs/2403.17050
%  https://link.aps.org/doi/10.1103/PhysRevB.110.205120}.
%\newblock \eprint{2403.17050}.

\bibitem{Ferrari2024}
\bibinfo{author}{Ferrari, F.} \& \bibinfo{author}{Valent{\'{i}}, R.}
\newblock \bibinfo{title}{{Altermagnetism on the Shastry-Sutherland lattice}}.
\newblock \emph{\bibinfo{journal}{Physical Review B}}
  \textbf{\bibinfo{volume}{110}}, \bibinfo{pages}{205140}
  (\bibinfo{year}{2024}).
%\newblock \urlprefix\url{http://arxiv.org/abs/2408.00841
%  https://link.aps.org/doi/10.1103/PhysRevB.110.205140}.
%\newblock \eprint{2408.00841}.

\bibitem{Rooj2025}
\bibinfo{author}{Rooj, S.}, \bibinfo{author}{Saxena, S.} \&
  \bibinfo{author}{Ganguli, N.}
\newblock \bibinfo{title}{{Altermagnetism in the orthorhombic Pnm structure
  through group theory and DFT calculations}}.
\newblock \emph{\bibinfo{journal}{Physical Review B}}
  \textbf{\bibinfo{volume}{111}}, \bibinfo{pages}{014434}
  (\bibinfo{year}{2025}).
%\newblock \urlprefix\url{http://arxiv.org/abs/2406.06232
%  https://link.aps.org/doi/10.1103/PhysRevB.111.014434}.
%\newblock \eprint{2406.06232}.

\bibitem{Topfer1997}
\bibinfo{author}{T{\"{o}}pfer, J.} \& \bibinfo{author}{Goodenough, J.}
\newblock \bibinfo{title}{{LaMnO3+$\delta$Revisited}}.
\newblock \emph{\bibinfo{journal}{Journal of Solid State Chemistry}}
  \textbf{\bibinfo{volume}{130}}, \bibinfo{pages}{117--128}
  (\bibinfo{year}{1997}).
%\newblock
%  \urlprefix\url{https://linkinghub.elsevier.com/retrieve/pii/S002245969797287X}.

\bibitem{Chen2014}
\bibinfo{author}{Chen, H.}, \bibinfo{author}{Niu, Q.} \&
  \bibinfo{author}{Macdonald, A.~H.}
\newblock \bibinfo{title}{{Anomalous hall effect arising from noncollinear
  antiferromagnetism}}.
\newblock \emph{\bibinfo{journal}{Physical Review Letters}}
  \textbf{\bibinfo{volume}{112}}, \bibinfo{pages}{017205}
  (\bibinfo{year}{2014}).
%\newblock \eprint{1309.4041}.

\newblock {\bf This paper predicted time-reversal symmetry breaking spin-split electronic structure and the anomalous Hall effect in a non-collinear compensated magnet.} 

\bibitem{Zhou2016}
\bibinfo{author}{Zhou, J.} \emph{et~al.}
\newblock \bibinfo{title}{{Predicted Quantum Topological Hall Effect and
  Noncoplanar Antiferromagnetism in K$_{0.5}$RhO$_2$}}.
\newblock \emph{\bibinfo{journal}{Physical Review Letters}}
  \textbf{\bibinfo{volume}{116}}, \bibinfo{pages}{256601}
  (\bibinfo{year}{2016}).
%\newblock \eprint{1602.08553}.

\bibitem{Feng2020}
\bibinfo{author}{Feng, W.} \emph{et~al.}
\newblock \bibinfo{title}{{Topological magneto-optical effects and their
  quantization in noncoplanar antiferromagnets}}.
\newblock \emph{\bibinfo{journal}{Nature Communications}}
  \textbf{\bibinfo{volume}{11}}, \bibinfo{pages}{118} (\bibinfo{year}{2020}).
%\newblock \urlprefix\url{https://arxiv.org/pdf/1811.05803.pdf
%  https://www.nature.com/articles/s41467-019-13968-8
%  http://www.nature.com/articles/s41467-019-13968-8}.
%\newblock \eprint{1811.05803}.

\bibitem{Ahn2019}
\bibinfo{author}{Ahn, K.-H.}, \bibinfo{author}{Hariki, A.},
  \bibinfo{author}{Lee, K.-W.} \& \bibinfo{author}{Kune{\v{s}}, J.}
\newblock \bibinfo{title}{{Antiferromagnetism in RuO$_2$ as $d$-wave Pomeranchuk
  instability}}.
\newblock \emph{\bibinfo{journal}{Physical Review B}}
  \textbf{\bibinfo{volume}{99}}, \bibinfo{pages}{184432}
  (\bibinfo{year}{2019}).
%\newblock \urlprefix\url{https://link.aps.org/doi/10.1103/PhysRevB.99.184432}.
%\newblock \eprint{1902.04436}.

\bibitem{Naka2019}
\bibinfo{author}{Naka, M.} \emph{et~al.}
\newblock \bibinfo{title}{{Spin current generation in organic
  antiferromagnets}}.
\newblock \emph{\bibinfo{journal}{Nature Communications}}
  \textbf{\bibinfo{volume}{10}}, \bibinfo{pages}{4305} (\bibinfo{year}{2019}).
%\newblock \urlprefix\url{http://dx.doi.org/10.1038/s41467-019-12229-y
%  http://www.nature.com/articles/s41467-019-12229-y
%  https://doi.org/10.1038/s41467-019-12229-y}.
%\newblock \eprint{1902.02506}.

\bibitem{Guo2023b}
\bibinfo{author}{Guo, Y.} \emph{et~al.}
\newblock \bibinfo{title}{{Spin-split collinear antiferromagnets: A large-scale
  ab-initio study}}.
\newblock \emph{\bibinfo{journal}{Materials Today Physics}}
  \textbf{\bibinfo{volume}{32}}, \bibinfo{pages}{100991}
  (\bibinfo{year}{2023}).
%\newblock
%  \urlprefix\url{https://linkinghub.elsevier.com/retrieve/pii/S2542529323000275}.
%\newblock \eprint{2207.07592}.

\bibitem{Ma2021}
\bibinfo{author}{Ma, H.-Y.} \emph{et~al.}
\newblock \bibinfo{title}{{Multifunctional antiferromagnetic materials with
  giant piezomagnetism and noncollinear spin current}}.
\newblock \emph{\bibinfo{journal}{Nature Communications}}
  \textbf{\bibinfo{volume}{12}}, \bibinfo{pages}{2846} (\bibinfo{year}{2021}).
%\newblock \urlprefix\url{https://doi.org/10.1038/s41467-021-23127-7
%  http://www.nature.com/articles/s41467-021-23127-7}.
%\newblock \eprint{arXiv:2104.00561}.

\bibitem{Egorov2021}
\bibinfo{author}{Egorov, S.~A.} \& \bibinfo{author}{Evarestov, R.~A.}
\newblock \bibinfo{title}{{Colossal Spin Splitting in the Monolayer of the
  Collinear Antiferromagnet MnF$_2$}}.
\newblock \emph{\bibinfo{journal}{Journal of Physical Chemistry Letters}}
  \textbf{\bibinfo{volume}{12}}, \bibinfo{pages}{2363--2369}
  (\bibinfo{year}{2021}).
%\newblock \urlprefix\url{https://pubs.acs.org/doi/10.1021/acs.jpclett.1c00282}.

\bibitem{Cui2023}
\bibinfo{author}{Cui, Q.}, \bibinfo{author}{Zhu, Y.}, \bibinfo{author}{Yao,
  X.}, \bibinfo{author}{Cui, P.} \& \bibinfo{author}{Yang, H.}
\newblock \bibinfo{title}{{Giant spin-Hall and tunneling magnetoresistance
  effects based on a two-dimensional nonrelativistic antiferromagnetic metal}}.
\newblock \emph{\bibinfo{journal}{Physical Review B}}
  \textbf{\bibinfo{volume}{108}}, \bibinfo{pages}{024410}
  (\bibinfo{year}{2023}).
%\newblock
%  \urlprefix\url{https://journals.aps.org/prb/abstract/10.1103/PhysRevB.108.024410
 % https://link.aps.org/doi/10.1103/PhysRevB.108.024410}.

\bibitem{Chen2023b}
\bibinfo{author}{Chen, X.}, \bibinfo{author}{Wang, D.}, \bibinfo{author}{Li,
  L.} \& \bibinfo{author}{Sanyal, B.}
\newblock \bibinfo{title}{{Giant spin-splitting and tunable spin-momentum
  locked transport in room temperature collinear antiferromagnetic semimetallic
  CrO monolayer}}.
\newblock \emph{\bibinfo{journal}{Applied Physics Letters}}
  \textbf{\bibinfo{volume}{123}} (\bibinfo{year}{2023}).
%\newblock
%  \urlprefix\url{https://pubs.aip.org/apl/article/123/2/022402/2901960/Giant-spin-splitting-and-tunable-spin-momentum}.

\bibitem{Sodequist2024}
\bibinfo{author}{S{\o}dequist, J.} \& \bibinfo{author}{Olsen, T.}
\newblock \bibinfo{title}{{Two-dimensional altermagnets from high throughput
  computational screening: Symmetry requirements, chiral magnons, and
  spin-orbit effects}}.
\newblock \emph{\bibinfo{journal}{Applied Physics Letters}}
  \textbf{\bibinfo{volume}{124}}, \bibinfo{pages}{1--7} (\bibinfo{year}{2024}).
%\newblock \urlprefix\url{http://arxiv.org/abs/2401.05992
%  https://pubs.aip.org/apl/article/124/18/182409/3288014/Two-dimensional-altermagnets-from-high-throughput}.
%\newblock \eprint{2401.05992}.

\bibitem{Zhu2024}
\bibinfo{author}{Zhu, Y.-P.} \emph{et~al.}
\newblock \bibinfo{title}{{Observation of plaid-like spin splitting in a
  noncoplanar antiferromagnet}}.
\newblock \emph{\bibinfo{journal}{Nature}} \textbf{\bibinfo{volume}{626}},
  \bibinfo{pages}{523--528} (\bibinfo{year}{2024}).
%\newblock \urlprefix\url{https://www.nature.com/articles/s41586-024-07023-w
%  https://arxiv.org/abs/2303.04549v2%0Ahttp://arxiv.org/abs/2303.04549}.
%\newblock \eprint{2303.04549}.

\bibitem{Zhu2024a}
\bibinfo{author}{Zhu, Y.} \emph{et~al.}
\newblock \bibinfo{title}{{Multipiezo Effect in Altermagnetic V$_2$SeTeO
  Monolayer}}.
\newblock \emph{\bibinfo{journal}{Nano Letters}} \textbf{\bibinfo{volume}{24}},
  \bibinfo{pages}{472--478} (\bibinfo{year}{2024}).
%\newblock
%  \urlprefix\url{https://pubs.acs.org/doi/10.1021/acs.nanolett.3c04330}.

\bibitem{Wigner1932}
\bibinfo{author}{Wigner, E.}
\newblock \bibinfo{title}{{Ueber die Operation der Zeitumkehr in der
  Quantenmechanik}}.
\newblock \emph{\bibinfo{journal}{Nachrichten von der Gesellschaft der
  Wissenschaften zu G{\"{o}}ttingen, Mathematisch-Physikalische Klasse}}
  \textbf{\bibinfo{volume}{1932}}, \bibinfo{pages}{546--559}
  (\bibinfo{year}{1932}).
%\newblock
%  \urlprefix\url{http://www.digizeitschriften.de/dms/img/?PPN=GDZPPN002509032}.

\bibitem{Tang2016}
\bibinfo{author}{Tang, P.}, \bibinfo{author}{Zhou, Q.}, \bibinfo{author}{Xu,
  G.} \& \bibinfo{author}{Zhang, S.-C.}
\newblock \bibinfo{title}{{Dirac fermions in an antiferromagnetic semimetal}}.
\newblock \emph{\bibinfo{journal}{Nature Physics}}
  \textbf{\bibinfo{volume}{12}}, \bibinfo{pages}{1100--1104}
  (\bibinfo{year}{2016}).
%\newblock \urlprefix\url{http://www.nature.com/articles/nphys3839}.
%\newblock \eprint{1603.08060}.

\bibitem{Smejkal2017c}
\bibinfo{author}{{\v{S}}mejkal, L.}, \bibinfo{author}{{\v{Z}}elezn{\'{y}}, J.},
  \bibinfo{author}{Sinova, J.} \& \bibinfo{author}{Jungwirth, T.}
\newblock \bibinfo{title}{{Electric Control of Dirac Quasiparticles by
  Spin-Orbit Torque in an Antiferromagnet}}.
\newblock \emph{\bibinfo{journal}{Physical Review Letters}}
  \textbf{\bibinfo{volume}{118}}, \bibinfo{pages}{106402}
  (\bibinfo{year}{2017}).
%\newblock
%  \urlprefix\url{https://link.aps.org/doi/10.1103/PhysRevLett.118.106402}.
%\newblock \eprint{1610.08107}.

\bibitem{Smejkal2018}
\bibinfo{author}{{\v{S}}mejkal, L.}, \bibinfo{author}{Mokrousov, Y.},
  \bibinfo{author}{Yan, B.} \& \bibinfo{author}{MacDonald, A.~H.}
\newblock \bibinfo{title}{{Topological antiferromagnetic spintronics}}.
\newblock \emph{\bibinfo{journal}{Nature Physics}}
  \textbf{\bibinfo{volume}{14}}, \bibinfo{pages}{242--251}
  (\bibinfo{year}{2018}).
%\newblock \urlprefix\url{http://arxiv.org/abs/1706.00670
%  http://www.nature.com/articles/s41567-018-0064-5}.
%\newblock \eprint{1706.00670}.

\bibitem{Kubler2014}
\bibinfo{author}{K{\"{u}}bler, J.} \& \bibinfo{author}{Felser, C.}
\newblock \bibinfo{title}{{Non-collinear antiferromagnets and the anomalous
  Hall effect}}.
\newblock \emph{\bibinfo{journal}{Epl}} \textbf{\bibinfo{volume}{108}},
  \bibinfo{pages}{67001} (\bibinfo{year}{2014}).
%\newblock \eprint{1410.5985}.

\bibitem{Nakatsuji2015}
\bibinfo{author}{Nakatsuji, S.}, \bibinfo{author}{Kiyohara, N.} \&
  \bibinfo{author}{Higo, T.}
\newblock \bibinfo{title}{{Large anomalous Hall effect in a non-collinear
  antiferromagnet at room temperature}}.
\newblock \emph{\bibinfo{journal}{Nature}} \textbf{\bibinfo{volume}{527}},
  \bibinfo{pages}{212--215} (\bibinfo{year}{2015}).
%\newblock \urlprefix\url{http://www.nature.com/articles/nature15723;}.

\newblock {\bf This paper reported the experimental observation of  the anomalous Hall effect in a non-collinear compensated magnet.} 

\bibitem{Kiyohara2015}
\bibinfo{author}{Kiyohara, N.}, \bibinfo{author}{Tomita, T.} \&
  \bibinfo{author}{Nakatsuji, S.}
\newblock \bibinfo{title}{{Giant Anomalous Hall Effect in the Chiral
  Antiferromagnet Mn3Ge}}.
\newblock \emph{\bibinfo{journal}{Physical Review Applied}}
  \textbf{\bibinfo{volume}{5}}, \bibinfo{pages}{064009} (\bibinfo{year}{2016}).
%\newblock \eprint{1511.04619}.

\bibitem{Nayak2016}
\bibinfo{author}{Nayak, A.~K.} \emph{et~al.}
\newblock \bibinfo{title}{{Large anomalous Hall effect driven by a nonvanishing
  Berry curvature in the noncolinear antiferromagnet Mn 3 Ge}}.
\newblock \emph{\bibinfo{journal}{Science Advances}}
  \textbf{\bibinfo{volume}{2}}, \bibinfo{pages}{e1501870--e1501870}
  (\bibinfo{year}{2016}).
%\newblock \urlprefix\url{https://www.science.org/doi/10.1126/sciadv.1501870}.
%\newblock \eprint{1511.03128}.

\bibitem{Takagi2023}
\bibinfo{author}{Takagi, H.} \emph{et~al.}
\newblock \bibinfo{title}{{Spontaneous topological Hall effect induced by
  non-coplanar antiferromagnetic order in intercalated van der Waals
  materials}}.
\newblock \emph{\bibinfo{journal}{Nature Physics}}
  \textbf{\bibinfo{volume}{19}}, \bibinfo{pages}{961--968}
  (\bibinfo{year}{2023}).
%\newblock \urlprefix\url{https://www.nature.com/articles/s41567-023-02017-3}.

\bibitem{Ikhlas2017}
\bibinfo{author}{Ikhlas, M.} \emph{et~al.}
\newblock \bibinfo{title}{{Large anomalous Nernst effect at room temperature in
  a chiral antiferromagnet}}.
\newblock \emph{\bibinfo{journal}{Nature Physics}}
  \textbf{\bibinfo{volume}{13}}, \bibinfo{pages}{1085--1090}
  (\bibinfo{year}{2017}).
%\newblock \urlprefix\url{http://www.nature.com/doifinder/10.1038/nphys4181}.
%\newblock \eprint{1710.00062}.

\bibitem{Li2017a}
\bibinfo{author}{Li, X.} \emph{et~al.}
\newblock \bibinfo{title}{{Anomalous Nernst and Righi-Leduc Effects in Mn$_3$Sn:
  Berry Curvature and Entropy Flow}}.
\newblock \emph{\bibinfo{journal}{Physical Review Letters}}
  \textbf{\bibinfo{volume}{119}}, \bibinfo{pages}{056601}
  (\bibinfo{year}{2017}).
%\newblock
%  \urlprefix\url{http://link.aps.org/doi/10.1103/PhysRevLett.119.056601}.
%\newblock \eprint{1612.06128}.

\bibitem{Higo2018c}
\bibinfo{author}{Higo, T.} \emph{et~al.}
\newblock \bibinfo{title}{{Large magneto-optical Kerr effect and imaging of
  magnetic octupole domains in an antiferromagnetic metal}}.
\newblock \emph{\bibinfo{journal}{Nature Photonics}}
  \textbf{\bibinfo{volume}{12}}, \bibinfo{pages}{73--78}
  (\bibinfo{year}{2018}).
%\newblock \urlprefix\url{http://www.nature.com/articles/s41566-017-0086-z}.
%\newblock \eprint{1805.06758}.

\bibitem{Matsuda2020}
\bibinfo{author}{Matsuda, T.}, \bibinfo{author}{Kanda, N.},
  \bibinfo{author}{Higo, T.} \& \bibinfo{author}{Matsunaga, R.}
\newblock \bibinfo{title}{{Room-temperature terahertz anomalous Hall effect in
  Weyl antiferromagnet Mn$_3$Sn thin films}}.
\newblock \emph{\bibinfo{journal}{Nature Communications}}
  \textbf{\bibinfo{volume}{11}}, \bibinfo{pages}{909} (\bibinfo{year}{2020}).
%\newblock \urlprefix\url{http://dx.doi.org/10.1038/s41467-020-14690-6}.

\bibitem{Kimata2021}
\bibinfo{author}{Kimata, M.} \emph{et~al.}
\newblock \bibinfo{title}{{X-ray study of ferroic octupole order producing
  anomalous Hall effect}}.
\newblock \emph{\bibinfo{journal}{Nature Communications}}
  \textbf{\bibinfo{volume}{12}}, \bibinfo{pages}{5582} (\bibinfo{year}{2021}).
%\newblock \urlprefix\url{http://dx.doi.org/10.1038/s41467-021-25834-7
%  https://www.nature.com/articles/s41467-021-25834-7}.

\bibitem{Sakamoto2021}
\bibinfo{author}{Sakamoto, S.} \emph{et~al.}
\newblock \bibinfo{title}{{Observation of spontaneous x-ray magnetic circular
  dichroism in a chiral antiferromagnet}}.
\newblock \emph{\bibinfo{journal}{Physical Review B}}
  \textbf{\bibinfo{volume}{104}}, \bibinfo{pages}{134431}
  (\bibinfo{year}{2021}).
%\newblock \urlprefix\url{https://link.aps.org/doi/10.1103/PhysRevB.104.134431}.

\bibitem{Zelezny2017a}
\bibinfo{author}{{\v{Z}}elezn{\'{y}}, J.}, \bibinfo{author}{Zhang, Y.},
  \bibinfo{author}{Felser, C.} \& \bibinfo{author}{Yan, B.}
\newblock \bibinfo{title}{{Spin-Polarized Current in Noncollinear
  Antiferromagnets}}.
\newblock \emph{\bibinfo{journal}{Physical Review Letters}}
  \textbf{\bibinfo{volume}{119}}, \bibinfo{pages}{187204}
  (\bibinfo{year}{2017}).
%\newblock
%  \urlprefix\url{https://link.aps.org/doi/10.1103/PhysRevLett.119.187204
%  http://arxiv.org/abs/1702.00295%0Ahttp://dx.doi.org/10.1103/PhysRevLett.119.187204}.
%\newblock \eprint{1702.00295}.

\bibitem{Zhang2018h}
\bibinfo{author}{Zhang, Y.}, \bibinfo{author}{{\v{Z}}elezn{\'{y}}, J.},
  \bibinfo{author}{Sun, Y.}, \bibinfo{author}{van~den Brink, J.} \&
  \bibinfo{author}{Yan, B.}
\newblock \bibinfo{title}{{Spin Hall effect emerging from a noncollinear
  magnetic lattice without spin–orbit coupling}}.
\newblock \emph{\bibinfo{journal}{New Journal of Physics}}
  \textbf{\bibinfo{volume}{20}}, \bibinfo{pages}{073028}
  (\bibinfo{year}{2018}).
%\newblock \urlprefix\url{http://arxiv.org/abs/1704.03917
%  http://stacks.iop.org/1367-2630/20/i=7/a=073028?key=crossref.767fedc8eb66d00618bb699a08ae7db3
%  https://iopscience.iop.org/article/10.1088/1367-2630/aad1eb}.
%\newblock \eprint{1704.03917}.

\bibitem{Kimata2019a}
\bibinfo{author}{Kimata, M.} \emph{et~al.}
\newblock \bibinfo{title}{{Magnetic and magnetic inverse spin Hall effects in a
  non-collinear antiferromagnet}}.
\newblock \emph{\bibinfo{journal}{Nature}} \textbf{\bibinfo{volume}{565}},
  \bibinfo{pages}{627--630} (\bibinfo{year}{2019}).
%\newblock \urlprefix\url{https://doi.org/10.1038/s41586-018-0853-0
%  http://www.nature.com/articles/s41586-018-0853-0}.

\bibitem{Hu2022}
\bibinfo{author}{Hu, S.} \emph{et~al.}
\newblock \bibinfo{title}{{Efficient perpendicular magnetization switching by a
  magnetic spin Hall effect in a noncollinear antiferromagnet}}.
\newblock \emph{\bibinfo{journal}{Nature Communications}}
  \textbf{\bibinfo{volume}{13}}, \bibinfo{pages}{4447} (\bibinfo{year}{2022}).
%\newblock \urlprefix\url{https://www.nature.com/articles/s41467-022-32179-2}.

\bibitem{Tsai2020}
\bibinfo{author}{Tsai, H.} \emph{et~al.}
\newblock \bibinfo{title}{{Electrical manipulation of a topological
  antiferromagnetic state}}.
\newblock \emph{\bibinfo{journal}{Nature}} \textbf{\bibinfo{volume}{580}},
  \bibinfo{pages}{608--613} (\bibinfo{year}{2020}).
%\newblock \urlprefix\url{http://www.nature.com/articles/s41586-020-2211-2}.

\bibitem{Takeuchi2021}
\bibinfo{author}{Takeuchi, Y.} \emph{et~al.}
\newblock \bibinfo{title}{{Chiral-spin rotation of non-collinear
  antiferromagnet by spin–orbit torque}}.
\newblock \emph{\bibinfo{journal}{Nature Materials}}
  \textbf{\bibinfo{volume}{20}}, \bibinfo{pages}{1364--1370}
  (\bibinfo{year}{2021}).
%\newblock \urlprefix\url{https://www.nature.com/articles/s41563-021-01005-3
%  http://www.nature.com/articles/s41563-021-01005-3}.

\bibitem{Higo2022}
\bibinfo{author}{Higo, T.} \emph{et~al.}
\newblock \bibinfo{title}{{Perpendicular full switching of chiral
  antiferromagnetic order by current}}.
\newblock \emph{\bibinfo{journal}{Nature}} \textbf{\bibinfo{volume}{607}},
  \bibinfo{pages}{474--479} (\bibinfo{year}{2022}).
%\newblock \urlprefix\url{https://www.nature.com/articles/s41586-022-04864-1}.

\bibitem{Pal2022}
\bibinfo{author}{Pal, B.} \emph{et~al.}
\newblock \bibinfo{title}{{Setting of the magnetic structure of chiral kagome
  antiferromagnets by a seeded spin-orbit torque}}.
\newblock \emph{\bibinfo{journal}{Science Advances}}
  \textbf{\bibinfo{volume}{8}}, \bibinfo{pages}{eabo5930}
  (\bibinfo{year}{2022}).
%\newblock \urlprefix\url{https://www.science.org/doi/10.1126/sciadv.abo5930}.

\bibitem{Chen2023}
\bibinfo{author}{Chen, X.} \emph{et~al.}
\newblock \bibinfo{title}{{Octupole-driven magnetoresistance in an
  antiferromagnetic tunnel junction}}.
\newblock \emph{\bibinfo{journal}{Nature}} \textbf{\bibinfo{volume}{613}},
  \bibinfo{pages}{490--495} (\bibinfo{year}{2023}).
%\newblock \urlprefix\url{https://doi.org/10.1038/s41586-022-05463-w
%  https://www.nature.com/articles/s41586-022-05463-w}.

\bibitem{Qin2023}
\bibinfo{author}{Qin, P.} \emph{et~al.}
\newblock \bibinfo{title}{{Room-temperature magnetoresistance in an
  all-antiferromagnetic tunnel junction}}.
\newblock \emph{\bibinfo{journal}{Nature}} \textbf{\bibinfo{volume}{613}},
  \bibinfo{pages}{485--489} (\bibinfo{year}{2023}).
%\newblock \urlprefix\url{https://doi.org/10.1038/s41586-022-05461-y
%  https://www.nature.com/articles/s41586-022-05461-y}.

\bibitem{Nakatsuji2022}
\bibinfo{author}{Nakatsuji, S.} \& \bibinfo{author}{Arita, R.}
\newblock \bibinfo{title}{{Topological Magnets: Functions Based on Berry Phase
  and Multipoles}}.
\newblock \emph{\bibinfo{journal}{Annual Review of Condensed Matter Physics}}
  \textbf{\bibinfo{volume}{13}} (\bibinfo{year}{2022}).

\bibitem{Rimmler2024}
\bibinfo{author}{Rimmler, B.~H.}, \bibinfo{author}{Pal, B.} \&
  \bibinfo{author}{Parkin, S. S.~P.}
\newblock \bibinfo{title}{{Non-collinear antiferromagnetic spintronics}}.
\newblock \emph{\bibinfo{journal}{Nature Reviews Materials}}
  \bibinfo{pages}{1--19} (\bibinfo{year}{2024}).
%\newblock \urlprefix\url{https://www.nature.com/articles/s41578-024-00706-w}.

\bibitem{Han2025}
\bibinfo{author}{Han, J.}, \bibinfo{author}{Yoon, J.-Y.},
  \bibinfo{author}{Ohno, H.} \& \bibinfo{author}{Fukami, S.}
\newblock \bibinfo{title}{{Unconventional responses in non-collinear
  antiferromagnets}}.
\newblock \emph{\bibinfo{journal}{Newton}} \textbf{\bibinfo{volume}{1}},
  \bibinfo{pages}{100012} (\bibinfo{year}{2025}).
%\newblock
%  \urlprefix\url{https://linkinghub.elsevier.com/retrieve/pii/S2950636025000040}.

\bibitem{Brekke2024}
\bibinfo{author}{Brekke, B.}, \bibinfo{author}{Sukhachov, P.},
  \bibinfo{author}{Giil, H.~G.}, \bibinfo{author}{Brataas, A.} \&
  \bibinfo{author}{Linder, J.}
\newblock \bibinfo{title}{{Minimal Models and Transport Properties of
  Unconventional p-Wave Magnets}}.
\newblock \emph{\bibinfo{journal}{Physical Review Letters}}
  \textbf{\bibinfo{volume}{133}}, \bibinfo{pages}{236703}
  (\bibinfo{year}{2024}).
%\newblock \urlprefix\url{http://arxiv.org/abs/2405.15823
%  https://link.aps.org/doi/10.1103/PhysRevLett.133.236703}.
%\newblock \eprint{2405.15823}.

\bibitem{Ezawa2024d}
\bibinfo{author}{Ezawa, M.}
\newblock \bibinfo{title}{{Topological insulators and superconductors based on
  p-wave magnets: Electrical control and detection of a domain wall}}.
\newblock \emph{\bibinfo{journal}{Physical Review B}}
  \textbf{\bibinfo{volume}{110}}, \bibinfo{pages}{165429}
  (\bibinfo{year}{2024}).
%\newblock
%  \urlprefix\url{https://journals.aps.org/prb/abstract/10.1103/PhysRevB.110.165429
%  https://link.aps.org/doi/10.1103/PhysRevB.110.165429}.

\bibitem{Sivianes2025a}
\bibinfo{author}{Sivianes, J.}, \bibinfo{author}{dos Santos, F.~J.} \&
  \bibinfo{author}{Iba{\~{n}}ez-Azpiroz, J.}
\newblock \bibinfo{title}{{Optical Signatures of Spin Symmetries in
  Unconventional Magnets}}.
\newblock \emph{\bibinfo{journal}{Physical Review Letters}}
  \textbf{\bibinfo{volume}{134}}, \bibinfo{pages}{196907}
  (\bibinfo{year}{2025}).
%\newblock
%  \urlprefix\url{https://link.aps.org/doi/10.1103/PhysRevLett.134.196907}.

\bibitem{Chakraborty2024c}
\bibinfo{author}{Chakraborty, A.} \emph{et~al.}
\newblock \bibinfo{title}{{Highly Efficient Non-relativistic Edelstein effect
  in p-wave magnets}}.
\newblock \emph{\bibinfo{journal}{Nature Communications}}
  \textbf{\bibinfo{volume}{16}}, \bibinfo{pages}{7270} (\bibinfo{year}{2023}).
%\newblock \urlprefix\url{http://arxiv.org/abs/2411.16378}.
%\newblock \eprint{2411.16378}.

\bibitem{Yu2025}
\bibinfo{author}{Yu, Y.} \emph{et~al.}
\newblock \bibinfo{title}{{Odd-parity Magnetism Driven by Antiferromagnetic
  Exchange}}.
\newblock \emph{\bibinfo{journal}{Arxiv Preprint}}  (\bibinfo{year}{2025}).
\newblock \urlprefix\url{http://arxiv.org/abs/2501.02057}.
%\newblock \eprint{2501.02057}.

\bibitem{Wang2023a}
\bibinfo{author}{Wang, M.} \emph{et~al.}
\newblock \bibinfo{title}{{Emergent zero-field anomalous Hall effect in a
  reconstructed rutile antiferromagnetic metal}}.
\newblock \emph{\bibinfo{journal}{Nature Communications}}
  \textbf{\bibinfo{volume}{14}}, \bibinfo{pages}{8240} (\bibinfo{year}{2023}).
%\newblock \urlprefix\url{https://www.nature.com/articles/s41467-023-43962-0}.

\bibitem{Han2024}
\bibinfo{author}{Han, L.} \emph{et~al.}
\newblock \bibinfo{title}{{Electrical 180° switching of N{\'{e}}el vector in
  spin-splitting antiferromagnet}}.
\newblock \emph{\bibinfo{journal}{Science Advances}}
  \textbf{\bibinfo{volume}{10}}, \bibinfo{pages}{eadn0479}
  (\bibinfo{year}{2024}).
%\newblock \urlprefix\url{https://www.science.org/doi/10.1126/sciadv.adn0479}.

\bibitem{Takagi2025}
\bibinfo{author}{Takagi, R.} \emph{et~al.}
\newblock \bibinfo{title}{{Spontaneous Hall effect induced by collinear
  antiferromagnetic order at room temperature}}.
\newblock \emph{\bibinfo{journal}{Nature Materials}}
  \textbf{\bibinfo{volume}{24}}, \bibinfo{pages}{63--68}
  (\bibinfo{year}{2025}).
%\newblock \urlprefix\url{https://www.nature.com/articles/s41563-024-02058-w}.

\bibitem{Ray2025}
\bibinfo{author}{Ray, M.~K.} \emph{et~al.}
\newblock \bibinfo{title}{{Zero-field Hall effect emerging from a non-Fermi
  liquid in a collinear antiferromagnet V$_{1/3}$NbS$_2$}}.
\newblock \emph{\bibinfo{journal}{Nature Communications}}
  \textbf{\bibinfo{volume}{16}}, \bibinfo{pages}{3532} (\bibinfo{year}{2025}).
%\newblock \urlprefix\url{https://www.nature.com/articles/s41467-025-58476-0}.

\bibitem{Fakhredine2023}
\bibinfo{author}{Fakhredine, A.}, \bibinfo{author}{Sattigeri, R.~M.},
  \bibinfo{author}{Cuono, G.} \& \bibinfo{author}{Autieri, C.}
\newblock \bibinfo{title}{{Interplay between altermagnetism and nonsymmorphic
  symmetries generating large anomalous Hall conductivity by semi-Dirac points
  induced anticrossings}}.
\newblock \emph{\bibinfo{journal}{Physical Review B}}
  \textbf{\bibinfo{volume}{108}}, \bibinfo{pages}{115138}
  (\bibinfo{year}{2023}).
%\newblock \urlprefix\url{http://arxiv.org/abs/2308.08416
%  http://dx.doi.org/10.1103/PhysRevB.108.115138
%  https://link.aps.org/doi/10.1103/PhysRevB.108.115138}.
%\newblock \eprint{2308.08416}.

\bibitem{Fang2024}
\bibinfo{author}{Fang, Y.}, \bibinfo{author}{Cano, J.} \&
  \bibinfo{author}{Ghorashi, S. A.~A.}
\newblock \bibinfo{title}{{Quantum Geometry Induced Nonlinear Transport in
  Altermagnets}}.
\newblock \emph{\bibinfo{journal}{Physical Review Letters}}
  \textbf{\bibinfo{volume}{133}}, \bibinfo{pages}{106701}
  (\bibinfo{year}{2024}).
%\newblock
%  \urlprefix\url{https://link.aps.org/doi/10.1103/PhysRevLett.133.106701}.

\bibitem{Li2024b}
\bibinfo{author}{Li, Y.-X.}, \bibinfo{author}{Liu, Y.} \& \bibinfo{author}{Liu,
  C.-C.}
\newblock \bibinfo{title}{{Creation and manipulation of higher-order
  topological states by altermagnets}}.
\newblock \emph{\bibinfo{journal}{Physical Review B}}
  \textbf{\bibinfo{volume}{109}}, \bibinfo{pages}{L201109}
  (\bibinfo{year}{2024}).
%\newblock
%  \urlprefix\url{https://link.aps.org/doi/10.1103/PhysRevB.109.L201109}.

\bibitem{Zhou2023}
\bibinfo{author}{Zhou, X.} \emph{et~al.}
\newblock \bibinfo{title}{{Crystal Thermal Transport in Altermagnetic RuO$_2$}}.
\newblock \emph{\bibinfo{journal}{Physical Review Letters}}
  \textbf{\bibinfo{volume}{132}}, \bibinfo{pages}{056701}
  (\bibinfo{year}{2024}).
%\newblock \urlprefix\url{http://arxiv.org/abs/2305.01410
%  https://link.aps.org/doi/10.1103/PhysRevLett.132.056701}.
%\newblock \eprint{2305.01410}.

\bibitem{Zhan2023}
\bibinfo{author}{Zhan, J.}, \bibinfo{author}{Li, J.}, \bibinfo{author}{Shi,
  W.}, \bibinfo{author}{Chen, X.~Q.} \& \bibinfo{author}{Sun, Y.}
\newblock \bibinfo{title}{{Coexistence of Weyl semimetal and Weyl nodal loop
  semimetal phases in a collinear antiferromagnet}}.
\newblock \emph{\bibinfo{journal}{Physical Review B}}
  \textbf{\bibinfo{volume}{107}}, \bibinfo{pages}{224402}
  (\bibinfo{year}{2023}).
%\newblock \urlprefix\url{https://doi.org/10.1103/PhysRevB.107.224402}.
%\newblock \eprint{2207.11472}.

\bibitem{Nag2024}
\bibinfo{author}{Nag, J.} \emph{et~al.}
\newblock \bibinfo{title}{{GdAlSi: An antiferromagnetic topological Weyl
  semimetal with nonrelativistic spin splitting}}.
\newblock \emph{\bibinfo{journal}{Physical Review B}}
  \textbf{\bibinfo{volume}{110}}, \bibinfo{pages}{224436}
  (\bibinfo{year}{2024}).
%\newblock \urlprefix\url{https://link.aps.org/doi/10.1103/PhysRevB.110.224436}.

\bibitem{Parshukov2024}
\bibinfo{author}{Parshukov, K.}, \bibinfo{author}{Wiedmann, R.} \&
  \bibinfo{author}{Schnyder, A.~P.}
\newblock \bibinfo{title}{{Topological responses from gapped Weyl points in 2D
  altermagnets}}.
\newblock \emph{\bibinfo{journal}{Arxiv Preprint}} \bibinfo{pages}{1--7}
  (\bibinfo{year}{2024}).
\newblock \urlprefix\url{http://arxiv.org/abs/2403.09520}.
%\newblock \eprint{2403.09520}.

\bibitem{Rao2024}
\bibinfo{author}{Rao, P.}, \bibinfo{author}{Mook, A.} \&
  \bibinfo{author}{Knolle, J.}
\newblock \bibinfo{title}{{Tunable band topology and optical conductivity in
  altermagnets}}.
\newblock \emph{\bibinfo{journal}{Physical Review B}}
  \textbf{\bibinfo{volume}{110}}, \bibinfo{pages}{024425}
  (\bibinfo{year}{2024}).
%\newblock \urlprefix\url{http://arxiv.org/abs/2403.10509
%  http://dx.doi.org/10.1103/PhysRevB.110.024425
%  https://link.aps.org/doi/10.1103/PhysRevB.110.024425}.
%\newblock \eprint{2403.10509}.

\bibitem{Tao2018}
\bibinfo{author}{Tao, L.~L.} \& \bibinfo{author}{Tsymbal, E.~Y.}
\newblock \bibinfo{title}{{Persistent spin texture enforced by symmetry}}.
\newblock \emph{\bibinfo{journal}{Nature Communications}}
  \textbf{\bibinfo{volume}{9}}, \bibinfo{pages}{2763} (\bibinfo{year}{2018}).
%\newblock \urlprefix\url{https://www.nature.com/articles/s41467-018-05137-0}.

\bibitem{Ji2022}
\bibinfo{author}{Ji, J.}, \bibinfo{author}{Lou, F.}, \bibinfo{author}{Yu, R.},
  \bibinfo{author}{Feng, J.~S.} \& \bibinfo{author}{Xiang, H.~J.}
\newblock \bibinfo{title}{{Symmetry-protected full-space persistent spin
  texture in two-dimensional materials}}.
\newblock \emph{\bibinfo{journal}{Physical Review B}}
  \textbf{\bibinfo{volume}{105}}, \bibinfo{pages}{L041404}
  (\bibinfo{year}{2022}).
%\newblock
%  \urlprefix\url{https://link.aps.org/doi/10.1103/PhysRevB.105.L041404}.

\bibitem{Bernevig2006}
\bibinfo{author}{Bernevig, B.~A.}, \bibinfo{author}{Orenstein, J.} \&
  \bibinfo{author}{Zhang, S.~C.}
\newblock \bibinfo{title}{{Exact SU(2) symmetry and persistent spin helix in a
  spin-orbit coupled system}}.
\newblock \emph{\bibinfo{journal}{Physical Review Letters}}
  \textbf{\bibinfo{volume}{97}}, \bibinfo{pages}{236601}
  (\bibinfo{year}{2006}).
%\newblock \eprint{0606196}.

\bibitem{Koralek2009}
\bibinfo{author}{Koralek, J.~D.} \emph{et~al.}
\newblock \bibinfo{title}{{Emergence of the persistent spin helix in
  semiconductor quantum wells}}.
\newblock \emph{\bibinfo{journal}{Nature}} \textbf{\bibinfo{volume}{458}},
  \bibinfo{pages}{610--613} (\bibinfo{year}{2009}).
%\newblock \eprint{0903.4709}.

\bibitem{Kuroda2017}
\bibinfo{author}{Kuroda, K.} \emph{et~al.}
\newblock \bibinfo{title}{{Evidence for magnetic Weyl fermions in a correlated
  metal}}.
\newblock \emph{\bibinfo{journal}{Nature Materials}}
  \textbf{\bibinfo{volume}{16}}, \bibinfo{pages}{1090--1095}
  (\bibinfo{year}{2017}).
%\newblock
%  \urlprefix\url{http://arxiv.org/abs/1710.06167%0Ahttp://dx.doi.org/10.1038/nmat4987
%  http://www.nature.com/doifinder/10.1038/nmat4987}.
%\newblock \eprint{1710.06167}.

\bibitem{Liu2018b}
\bibinfo{author}{Liu, Z.~Q.} \emph{et~al.}
\newblock \bibinfo{title}{{Electrical switching of the topological anomalous
  Hall effect in a non-collinear antiferromagnet above room temperature}}.
\newblock \emph{\bibinfo{journal}{Nature Electronics}}
  \textbf{\bibinfo{volume}{1}}, \bibinfo{pages}{172--177}
  (\bibinfo{year}{2018}).
%\newblock \urlprefix\url{http://www.nature.com/articles/s41928-018-0040-1}.

\bibitem{Xu2020b}
\bibinfo{author}{Xu, L.} \emph{et~al.}
\newblock \bibinfo{title}{{Finite-temperature violation of the anomalous
  transverse Wiedemann-Franz law}}.
\newblock \emph{\bibinfo{journal}{Science Advances}}
  \textbf{\bibinfo{volume}{6}}, \bibinfo{pages}{eaaz3522}
  (\bibinfo{year}{2020}).

\bibitem{Chen2021a}
\bibinfo{author}{Chen, T.} \emph{et~al.}
\newblock \bibinfo{title}{{Anomalous transport due to Weyl fermions in the
  chiral antiferromagnets Mn$_3$X, X = Sn, Ge}}.
\newblock \emph{\bibinfo{journal}{Nature Communications}}
  \textbf{\bibinfo{volume}{12}}, \bibinfo{pages}{572} (\bibinfo{year}{2021}).
%\newblock \urlprefix\url{http://dx.doi.org/10.1038/s41467-020-20838-1
%  http://www.nature.com/articles/s41467-020-20838-1}.
%\newblock \eprint{2011.10942}.

\bibitem{Ghosh2025}
\bibinfo{author}{Ghosh, S.} \emph{et~al.}
\newblock \bibinfo{title}{{Raman spectroscopic evidence for linearly dispersed
  nodes and magnetic ordering in the topological semimetal V$_{1/3}$NbS$_2$}}
  (\bibinfo{year}{2025}).
%\newblock \urlprefix\url{http://arxiv.org/abs/2504.04590}.
%\newblock \eprint{2504.04590}.

\bibitem{Sun2009}
\bibinfo{author}{Sun, K.}, \bibinfo{author}{Yao, H.}, \bibinfo{author}{Fradkin,
  E.} \& \bibinfo{author}{Kivelson, S.~A.}
\newblock \bibinfo{title}{{Topological Insulators and Nematic Phases from
  Spontaneous Symmetry Breaking in 2D Fermi Systems with a Quadratic Band
  Crossing}}.
\newblock \emph{\bibinfo{journal}{Physical Review Letters}}
  \textbf{\bibinfo{volume}{103}}, \bibinfo{pages}{046811}
  (\bibinfo{year}{2009}).
%\newblock
 % \urlprefix\url{https://link.aps.org/doi/10.1103/PhysRevLett.103.046811}.

\bibitem{Steward2023}
\bibinfo{author}{Steward, C. R.~W.}, \bibinfo{author}{Fernandes, R.~M.} \&
  \bibinfo{author}{Schmalian, J.}
\newblock \bibinfo{title}{{Dynamic paramagnon-polarons in altermagnets}}.
\newblock \emph{\bibinfo{journal}{Physical Review B}}
  \textbf{\bibinfo{volume}{108}}, \bibinfo{pages}{144418}
  (\bibinfo{year}{2023}).
%\newblock \urlprefix\url{http://arxiv.org/abs/2307.01855
 % https://link.aps.org/doi/10.1103/PhysRevB.108.144418}.
%\newblock \eprint{2307.01855}.

\bibitem{Martin2008}
\bibinfo{author}{Martin, I.} \& \bibinfo{author}{Batista, C.~D.}
\newblock \bibinfo{title}{{Itinerant electron-driven chiral magnetic ordering
  and spontaneous quantum hall effect in triangular lattice models}}.
\newblock \emph{\bibinfo{journal}{Physical Review Letters}}
  \textbf{\bibinfo{volume}{101}}, \bibinfo{pages}{156402}
  (\bibinfo{year}{2008}).
%\newblock
%  \urlprefix\url{https://journals.aps.org/prl/abstract/10.1103/PhysRevLett.101.156402}.

\end{thebibliography}

\onecolumngrid

\begin{figure}[h!]
%	\centering
	\includegraphics[width=.6\linewidth]{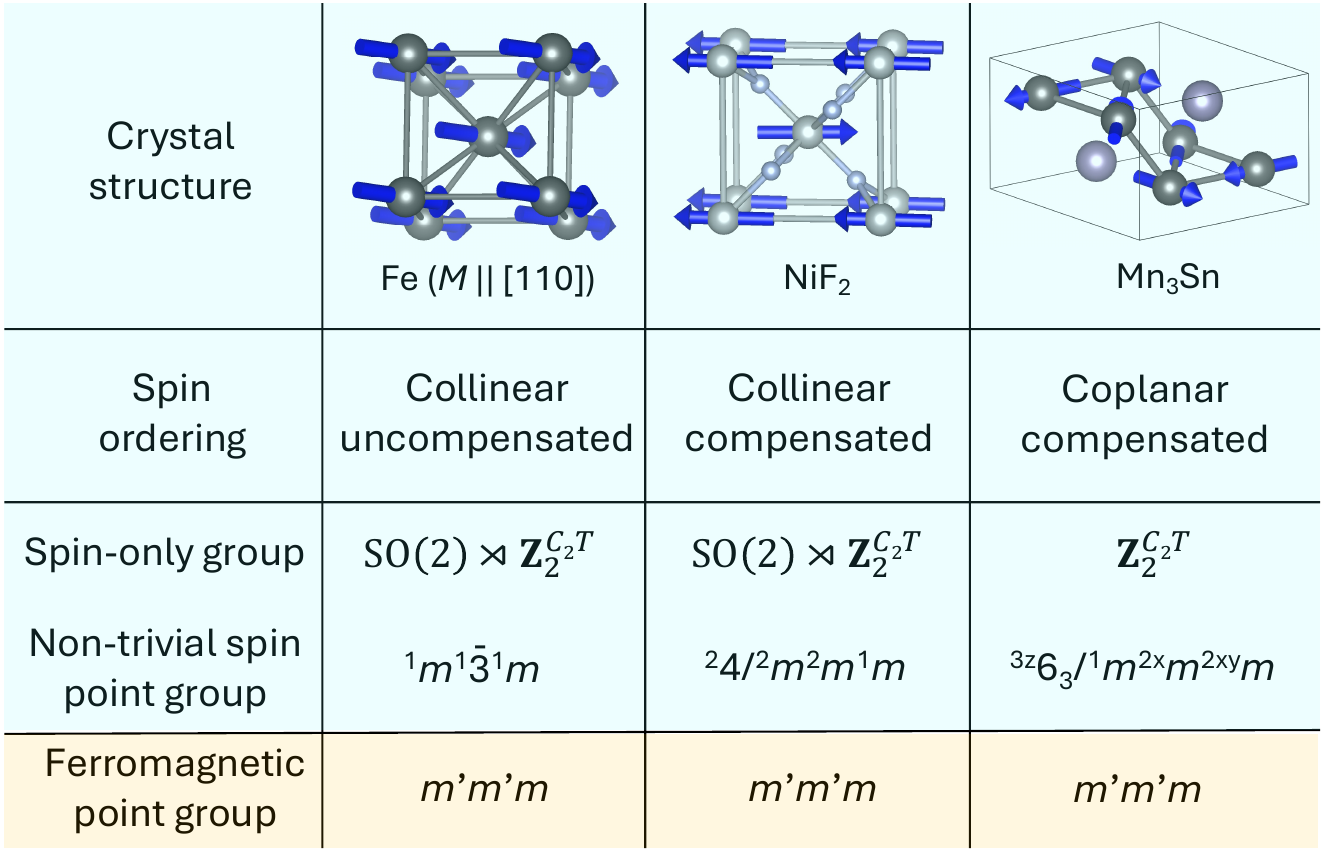}
	\vspace{.5cm}
%	\caption*{
%	\textbf{Spin groups and magnetic groups.}
		% 
%		Examples of collinear/non-collinear and compensated/uncompensated spin ordering with denoted corresponding spin-only group, non-trivial spin point group and magnetic point group. The middle column represents altermagnetism.
%}
%\label{fig:groups}

{\color{brown}\small FIG. Box. \textbf{Spin groups and magnetic groups.} Examples of collinear/non-collinear and compensated/uncompensated spin ordering with denoted corresponding spin-only group, non-trivial spin point group and magnetic point group. The middle column represents altermagnetism.}
\end{figure}

%\vspace{-0.5cm}

\end{document}